\documentclass[useAMS,usenatbib]{mn2e}
\usepackage[dvips]{graphicx}
\usepackage{aas_macros}

\title[A search for relic radio emission with the GMRT]
  {Deep low-frequency observations with the Giant Metrewave Radio Telescope:
   a search for relic radio emission}
\author[S. K. Sirothia et al.]
  {S.~K.~Sirothia,\thanks{Email: sirothia@ncra.tifr.res.in (SKS), djs@ncra.tifr.res.in (DJS), 
   ishwar@ncra.tifr.res.in (CHI) and ngk@ncra.tifr.res.in (NGK)}
   D.~J.~Saikia, C.~H. Ishwara-Chandra and N.~G.~Kantharia\\
National Centre for Radio Astrophysics,  Tata Institute of Fundamental Research, Post Bag 3, Ganeshkhind, Pune 411007, India\\
}
\date{Version 2008 Sep. 25}

\pagerange{\pageref{firstpage}--\pageref{lastpage}} \pubyear{2002}

\def\LaTeX{L\kern-.36em\raise.3ex\hbox{a}\kern-.15em
    T\kern-.1667em\lower.7ex\hbox{E}\kern-.125emX}

\newcounter{saveeqn}

\pubyear{2008}

\begin{document}
\label{firstpage}

\maketitle

\begin{abstract}
We present deep multifrequency observations using the Giant Metrewave Radio Telescope at 153, 244, 610 and 1260 MHz of a field centered on J0916+6348, to search for evidence of fossil radio lobes which could be due to an earlier cycle of episodic activity of the  parent galaxy, as well as halos and relics in clusters of galaxies. We do not find any unambiguous evidence of episodic activity in a list of 374 sources, suggesting that such activity is rare even in relatively deep low-frequency observations. We examine the spectra of all the sources by combining our observations with those from the Westerbork Northern Sky Survey, NRAO VLA Sky Survey and the Faint Images of the Radio Sky at Twenty-centimeters survey. Considering only those which have measurements at a minimum of three different frequencies, we find that almost all sources are consistent with a straight spectrum with a median spectral index,  $\alpha\sim$0.8 (S$(\nu)\propto\nu^{-\alpha}$), which appears steeper than theoretical expectations of the injection spectral index. We identify 14 very steep-spectrum sources with $\alpha\geq$1.3. We examine their optical fields and discuss the nature of some of these sources.
\end{abstract}

\begin{keywords}
 Radio continuum: galaxies -- galaxies: active -- galaxies: clusters: general 
\end{keywords}

\section{INTRODUCTION}
Low-frequency radio observations of radio galaxies and quasars provide us 
with an opportunity to probe episodic activity in active galactic nuclei (AGN)
and also constrain the injection spectral indices of the electron energy spectrum. 
The extended radio emission in radio galaxies and quasars could provide 
constraints on the duration and duty cycles of episodic activity in AGN via
the structural and spectral information of the lobes of emission. The most
striking examples of such episodic nuclear activity are the `double-double'
radio galaxies (DDRGs) where a young pair of radio lobes is seen closer to the nucleus
in addition to the older and more distant lobes of emission (e.g. 
\citeauthor*{1996MNRAS.279..257S}~\citeyear{1996MNRAS.279..257S};
\citeauthor{1999A+A...348..699L}~\citeyear{1999A+A...348..699L};
\citeauthor{2000MNRAS.315..371S}~\citeyear{2000MNRAS.315..371S};
\citeauthor*{2006MNRAS.366.1391S}~\citeyear{2006MNRAS.366.1391S} and references therein). In addition to the 
DDRGs, diffuse relic radio emission due to an earlier
cycle of activity has also been suggested for a number of sources from structural
and spectral information. Some of the examples are 3C338 and 3C388
(\citeauthor*{1983ApJ...271..575B}~\citeyear{1983ApJ...271..575B};
\citeauthor{1994ApJ...421L..23R}~\citeyear{1994ApJ...421L..23R}),
Her A \citep{2003MNRAS.342..399G}, 3C310 (\citeauthor{1984ApJ...282L..55V}~\citeyear{1984ApJ...282L..55V};
\citeauthor*{1986MNRAS.222..753L}~\citeyear{1986MNRAS.222..753L}),
Cen A (\citeauthor*{1983ApJ...273..128B}~\citeyear{1983ApJ...273..128B};
\citeauthor*{1992ApJ...395..444C}~\citeyear{1992ApJ...395..444C};
\citeauthor{1993A+A...274.1009J}~\citeyear{1993A+A...274.1009J};
\citeauthor{1999MNRAS.307..750M}~\citeyear{1999MNRAS.307..750M}),
and more recently 4C29.30 \citep{2007MNRAS.378..581J}. \cite{2000A+A...362...27J} have 
listed a few cases of small-scale
halos associated with the known milliarcsec-scale structures of compact steep spectrum and
giga-hertz peaked spectrum sources from interplanetary scintillation observations with 
the Ooty Radio Telescope at 327 MHz. However, examples of such features appear to
be relatively rare, and several searches for relic emission around bright radio 
sources have not yielded positive results 
(e.g. \citeauthor*{1980A+A....89..204R}~\citeyear{1980A+A....89..204R};
 \citeauthor*{1980A+AS...42..299S}~\citeyear{1980A+AS...42..299S};
 \citeauthor*{1982ApJ...255L..93P}~\citeyear{1982ApJ...255L..93P};
 \citeauthor*{1983A+A...125..146K}~\citeyear{1983A+A...125..146K};
 \citeauthor*{1984IAUS..110....9V}~\citeyear{1984IAUS..110....9V};
 \citeauthor*{2001AJ....122.2940J}~\citeyear{2001AJ....122.2940J}).
 
\begin{table*}
\begin{center}
\caption{Observation summary and image parameters}
\label{0916p6348_f0:table:obs_sum}
\begin{tabular}{| l | l | l | l | l | l |} \hline
  & Center Frequency  & 153 MHz &  244 MHz & 610 MHz & 1260 MHz\\
  & Date & 2005 Dec 12  & 2005 Nov 26 & 2005 Nov 26 & 2008 Apr 22\\
  & Working antennas & 29  & 26 & 27 & 27\\
General  & Bandwidth & 5.6 MHz & 32 MHz & 32 MHz & 32 MHz\\
  & Polarization & RR, LL  & LL & RR & RR, LL\\
  & Visibility integration time (s) & 16.78 & 16.78 & 16.78 & 4.19 \\
  & Total observation time (hr) & 7.61 & 8.57 & 8.57 & 6.50 \\ \hline
  & Source & 3C147 & 3C147 & 3C147 & 3C147\\
Flux       & Time (hr) & 1.01 & 0.25 & 0.25 & 0.26 \\
Calibrator & Flux density (Jy) & 68.29 & 59.16 & 38.27 & 15.58 \\
  & Scale & Perley-Taylor 99 & Perley-Taylor 99 & Perley-Taylor 99 & Perley-Taylor 99\\ \hline
  & Source & 3C147 & J0834+5534 & J0834+5534 & J0834+5534\\
Phase & Time (hr) & 1.0 & 1.72 & 1.72 & 0.74 \\
Calibrator & Flux density (Jy) & 68.29 & 8.60 & 8.22 & 7.39 \\ \hline
  &Phase center & J0916+6348 & J0916+6348 & J0916+6348 & J0916+6348\\
Target & Time (hr) & 6.60 & 6.60 & 6.60 & 5.34 \\
Field   & Image size & $6144\times6144$ & $6144\times6144$ & $6144\times6144$ & $6144\times6144$\\
  & Pixel size & $4^{\prime\prime}\times4^{\prime\prime}$ & $3^{\prime\prime}\times3^{\prime\prime}$ & $2^{\prime\prime}\times2^{\prime\prime}$ & $1^{\prime\prime}\times1^{\prime\prime}$\\
  & Rms noise$^\dagger$ $\mu$Jy beam$^{-1}$ & $\approx 513$ & $\approx 173$ & $\approx 32$ & $\approx 22$\\
  & Synthesized beam & $21.48^{\prime\prime}\times18.41^{\prime\prime}$ & $12.09^{\prime\prime}\times10.80^{\prime\prime}$ & $5.51^{\prime\prime}\times4.59^{\prime\prime}$ & $2.85^{\prime\prime}\times2.13^{\prime\prime}$\\
  & PA & 45.8$^\circ$ & 57.4$^\circ$ & 47.2$^\circ$ & -47.1$^\circ$\\ \hline
Primary Beam & Gaussian, FWHM & 173.8$^{\prime}$ & 117.0$^{\prime}$ & 42.7$^{\prime}$ & 26.4$^{\prime}$\\
correction parameter  &  Reference Frequency & 153 MHz & 235 MHz & 610 MHz & 1280 MHz\\
\hline
\end{tabular}

$^\dagger$ The rms noise given is the median value of all estimates across the primary
beam till 20 per cent of the peak response, before primary beam correction. 

\end{center}
\end{table*}

The diffuse relic emission is expected to have a steep radio spectrum due
to radiative losses and remain visible for approximately $\sim$10$^8$ yr
(e.g. \citeauthor*{2000ApJ...543..611O}~\citeyear{2000ApJ...543..611O};
\citeauthor*{2002MNRAS.336..649K}~\citeyear{2002MNRAS.336..649K};
\citeauthor{2004A+A...427...79J}~\citeyear{2004A+A...427...79J}).
Low-frequency observations with low rms noise values and short spacings sensitive
to the diffuse emission could provide better constraints on the frequency of 
occurrence or detection of such relic emission. We explore this possibility from 
observations of a field centered on RA: 09$^{\rm h}$ 16$^{\rm m}$ 30$^{\rm s}$,
DEC: 63$^{\circ}$ 48$^{\prime}$ 00$^{\prime\prime}$ at a number of frequencies, 
namely 150, 244, 610 and 1260 MHz, with the Giant Metrewave Radio Telescope (GMRT). 
This field was chosen when several interesting radio sources, including a
wide-angle tailed source, were noticed while studying the nearby group of galaxies
Holmberg 124 \citep{2005A+A...435..483K}. In order to identify diffuse cluster
sources, it is important that the imaged field contains galaxy clusters. In these
observations, the clusters Abell 764 at a redshift of 0.166 \citep{1989ApJS...70....1A} and 
MaxBCG J139.14754+63.8034 at a redshift of 0.1215 \citep{2007ApJ...660..239K} are both very close to the
pointing centre.

Deep low-frequency observations are also extremely useful for identifying radio halos,
relics and core-halos or mini halos associated with clusters of galaxies. While core-halos
are less than about 500 kpc in extent and associated with the dominant galaxy in cooling core
clusters, halos and relics are much larger in size and not associated with any 
particular galaxy. Radio halos are usually projected towards the cluster
center and have a regular morphology, while relics are seen towards the periphery with
a variety of shapes (cf. \citeauthor{2008A+A...484..327V}~\citeyear{2008A+A...484..327V} and references therein).
While relics are believed to arise due to cluster mergers and/or matter accretion
\citep{1999ApJ...520..529S, 2003ApJ...593..599R, 2006MNRAS.367..113P, 2008A+A...486..347G}, the 
models for halos range from re-acceleration due to turbulence 
(\citeauthor{2003ASPC..301..349B}~\citeyear{2003ASPC..301..349B}; 
\citeauthor{2004JKAS...37..433S}~\citeyear{2004JKAS...37..433S};
\citeauthor*{2008SSRv..134..207P}~\citeyear{2008SSRv..134..207P};
\citeauthor{2007ApJ...670L...5B}~\citeyear{2007ApJ...670L...5B} and references therein) 
to production of relativistic 
electrons by hadronic collisions (e.g. \citeauthor*{2000A+A...362..151D}~\citeyear{2000A+A...362..151D} 
and references therein).

\begin{figure*}
\begin{center}
\vbox{
 \hbox{
  \includegraphics[angle=0, totalheight=2.8in, viewport=19 212 573 627, clip]{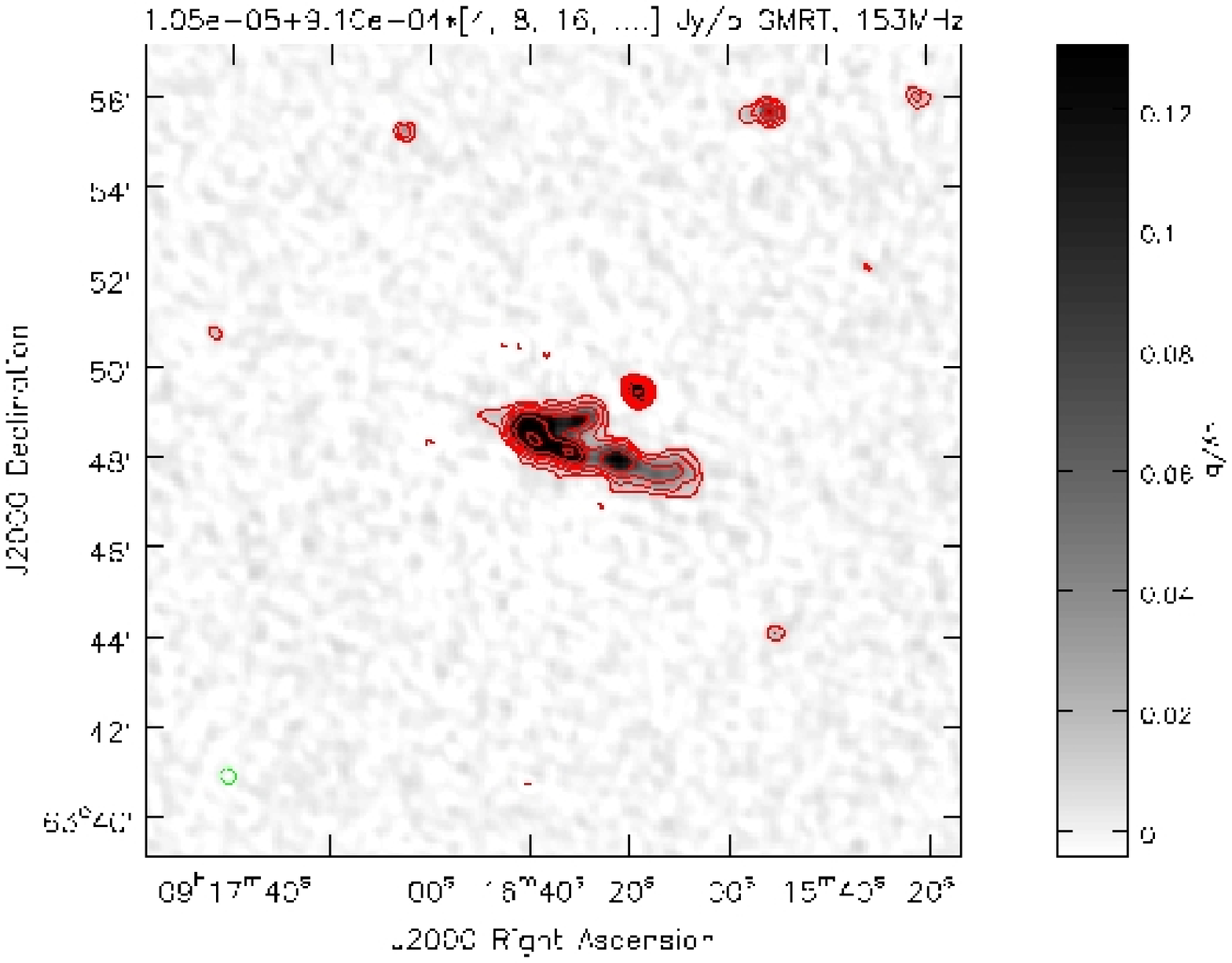}
  \includegraphics[angle=0, totalheight=2.8in, viewport=19 212 573 627, clip]{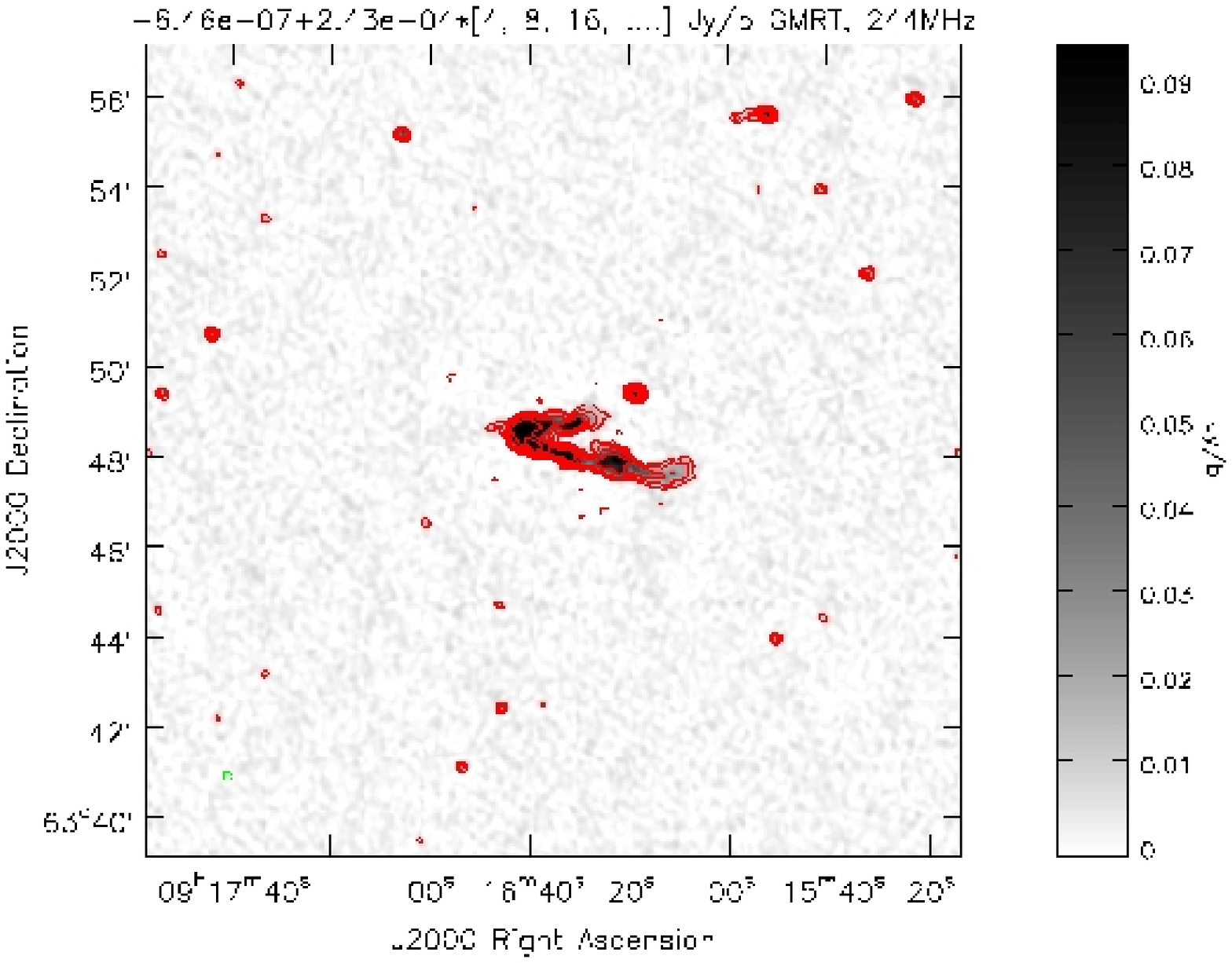}
 }
 \hbox{
  \includegraphics[angle=0, totalheight=2.8in, viewport=19 212 573 627, clip]{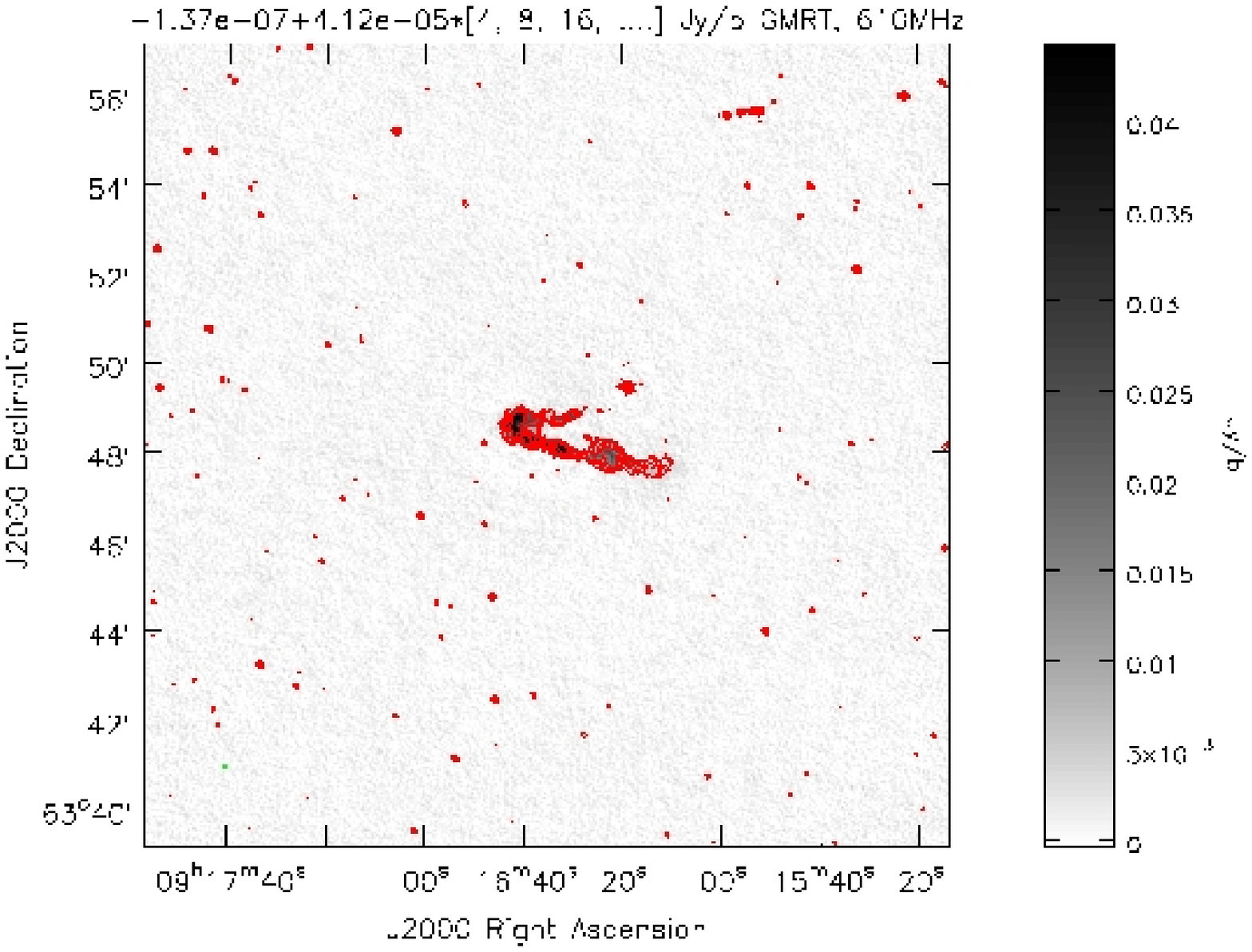}
  \includegraphics[angle=0, totalheight=2.8in, viewport=19 212 573 627, clip]{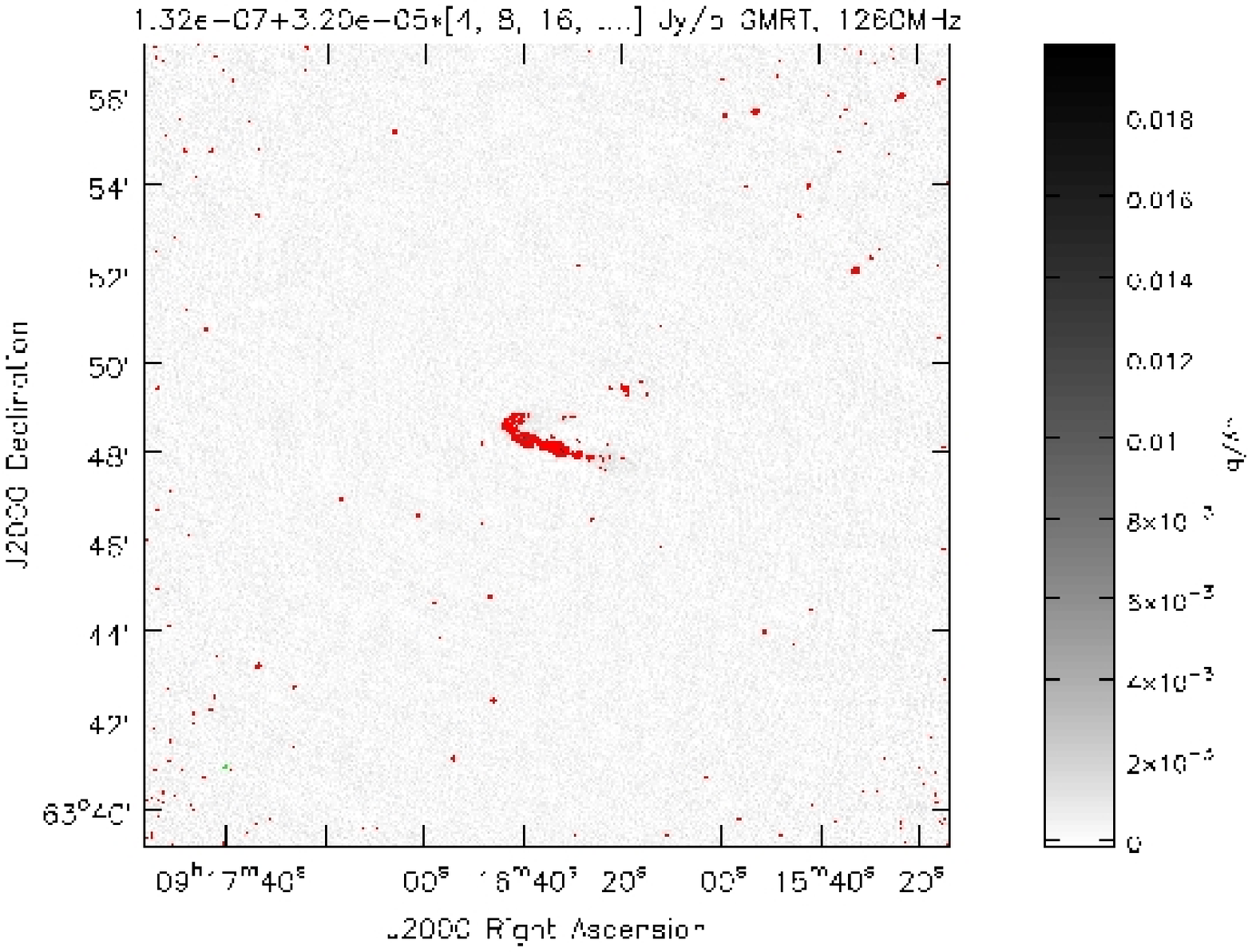}
 }
}
\caption{GMRT images of an area of 18$\times$18 arcmin$^2$ of the field J0916+6348 at 150, 244, 610 and 1260 MHz. The
resolutions and rms noise values are listed in Table~\ref{0916p6348_f0:table:obs_sum}. In this figure and in all the
images presented here, the contour levels in units of Jy beam$^{-1}$ are represented by mean$+$rms$\times$(n), where n
is the multiplication factor. The negative contours are shown as dashed lines. }
\label{0916p6348_f0:fig:multifreq}
\end{center}

\begin{center}
\vbox{
 \hbox{
  \includegraphics[angle=0, totalheight=2.8in, viewport=19 212 573 627, clip]{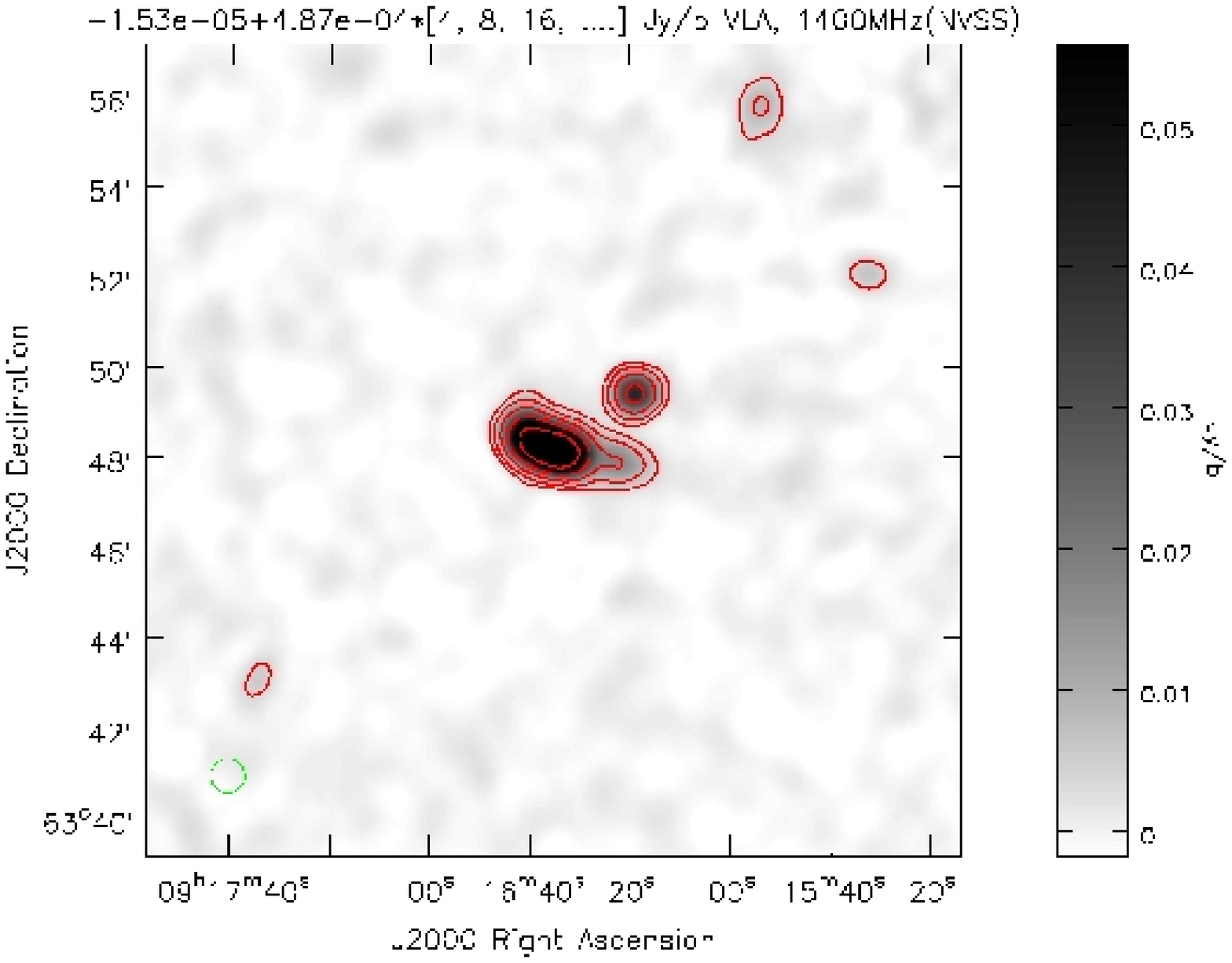}
  \includegraphics[angle=0, totalheight=2.8in, viewport=19 212 573 627, clip]{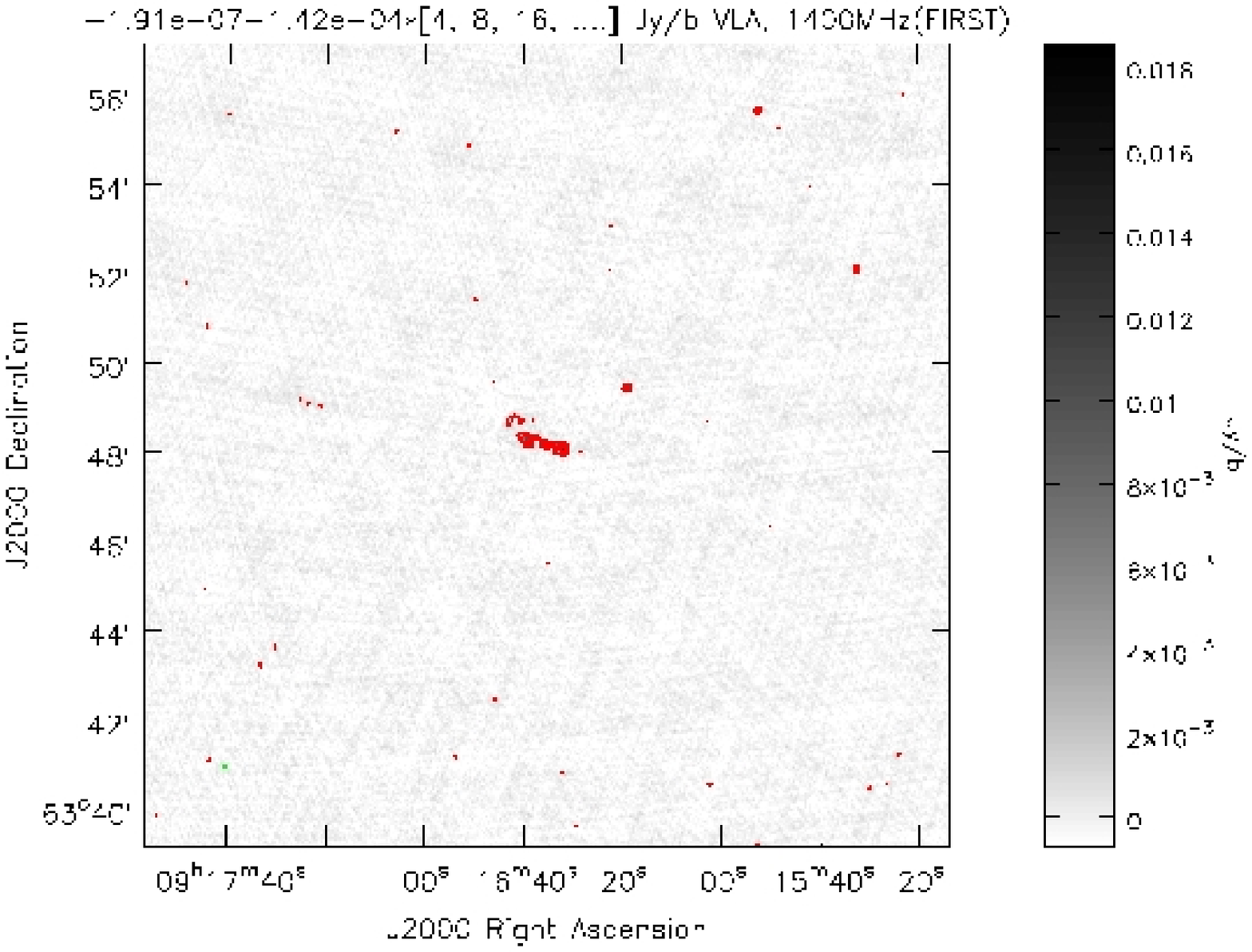}
 }
}
\caption{The NVSS (left) and FIRST (right) images of the same area shown in Fig.~\ref{0916p6348_f0:fig:multifreq} 
of the field J0916+6348 at 1400 MHz with angular resolutions of 45 and 5 arcsec respectively. 
Contour levels: mean$+$rms$\times$(n) in units of Jy beam$^{-1}$.}
\label{0916p6348_f0:fig:vla}
\end{center}
\end{figure*}

In addition to the search for fossil or relic radio emission of different forms, 
it is also useful to 
examine the low-frequency spectra of these weak sources, which could be close to
their injection spectral indices, $\alpha_{\rm inj}$ (flux density, S$\propto\nu^{-\alpha}$),
and compare these with values for stronger sources. Most of our sources are weak
and are not catalogued in the well-known low-frequency surveys such as the Cambridge surveys
at 151 MHz  (\citeauthor{2002IAUS..199...21G}~\citeyear{2002IAUS..199...21G} and references therein) and the
VLA Low-frequency Sky Survey (VLSS; \citeauthor{2007AJ....134.1245C}~\citeyear{2007AJ....134.1245C}; 
{\tt http://lwa.nrl.navy.mil/VLSS}).
In the case of the strong sources, $\alpha_{\rm inj}$ has been estimated to be 0.5
for Cygnus A. Similar values have been obtained by \cite{1989A+A...219...63M} for a
sample of powerful radio galaxies. From measurements between 38 MHz and 1 GHz, 
\cite*{1992MNRAS.257..545L} find a median value of 0.77. For a sample of 3CR sources \cite{1989MNRAS.239..401L}
estimate the hot spot spectral indices, which are similar to $\alpha_{\rm inj}$, 
to have a similar median value of $\sim$0.8. This is also similar to the estimate of
$\alpha_{\rm inj}$ for the compact steep spectrum source 3C190 by
\cite*{1997ApJ...479..258K}. For a sample of giant radio sources, using low-frequency
GMRT observations and higher frequency observations from the Very Large Array (VLA) as
well as data from the literature, \cite{2008MNRAS.385.1286J} find that  
$\alpha_{\rm inj}$ varies from $\sim$0.55 to 0.88 with a median value of $\sim$0.6.  
The theoretically expected values are somewhat smaller. For a strong, non-relativistic shock in a
Newtonian fluid $\alpha_{\rm inj}$ = 0.5 (\citeauthor{1978MNRAS.182..147B}~\citeyear{1978MNRAS.182..147B},~\citeyear{1978MNRAS.182..443B}; \citeauthor{1978ApJ...221L..29B}~\citeyear{1978ApJ...221L..29B}),
while the values range from 0.35 to 0.65 for relativistic shocks in different 
situations (\citeauthor{1989LNP...327..247H}~\citeyear{1989LNP...327..247H};
\citeauthor*{1987ApJ...315..425K}~\citeyear{1987ApJ...315..425K};
\citeauthor*{1981ApJ...248..344D}~\citeyear{1981ApJ...248..344D};
\citeauthor*{1982A+A...111..317A}~\citeyear{1982A+A...111..317A}). 

A summary of the observations and the parameters of the images obtained are presented
in Table~\ref{0916p6348_f0:table:obs_sum}, while the details of our observations and analysis 
are described in Section~\ref{0916p6348_f0:sec:obs}. The results are presented in 
Section~\ref{0916p6348_f0:sec:result}. The discussion and conclusions are summarized 
in Section~\ref{0916p6348_f0:sec:discussions}.

\section{OBSERVATIONS AND DATA REDUCTION}
\label{0916p6348_f0:sec:obs}
The observation summary along with the calibrators used, and the parameters of
the final images are  presented in Table~\ref{0916p6348_f0:table:obs_sum}. 
The data reduction was done mainly 
using AIPS++ (version: 1.9, build \#1556). 3C147 was the primary flux density and 
bandpass calibrator at all the four frequencies. After applying bandpass corrections on the
mentioned phase calibrator, gain and phase variations were estimated,
and then flux density, bandpass, gain and phase calibration from 
flux density and phase calibrator were applied on the target field.

While calibrating the data, bad data were flagged at various stages. 
The data for antennas with high errors in antenna-based solutions
were examined and flagged over certain time ranges. Some baselines
were flagged based on closure errors on the bandpass calibrator.
Channel and time-based flagging of data points corrupted by 
radio frequency interference (RFI) were done using a median filter 
with a $6\sigma$ threshold. Residual errors above $5\sigma$ were 
also flagged after a few rounds of imaging and self calibration.
The system temperature ($T_{sys}$) was found to vary with antenna, 
the ambient temperature and elevation \citep{sks2008}. 
In the absence of regular $T_{sys}$ measurements for GMRT antennas, 
this correction was estimated from the residuals of corrected data 
with respect to the model data. The corrections were then applied 
to the data. The final image was made after several rounds of phase
self calibration, and one round of amplitude self calibration, where
the data were normalized by the median gain for all the data. The
final image was also primary beam corrected using the Gaussian 
parameters as mentioned in Table~\ref{0916p6348_f0:table:obs_sum}.

\begin{figure*}
\begin{center}
\vbox{
 \hbox{\hskip 0.9cm GMRT091446+633757 \hskip 3.2cm GMRT091633+643528 \hskip 3.2cm  GMRT092234+640556}
 \hbox{
  \includegraphics[angle=0, totalheight=1.8in, viewport=19 212 573 627, clip]{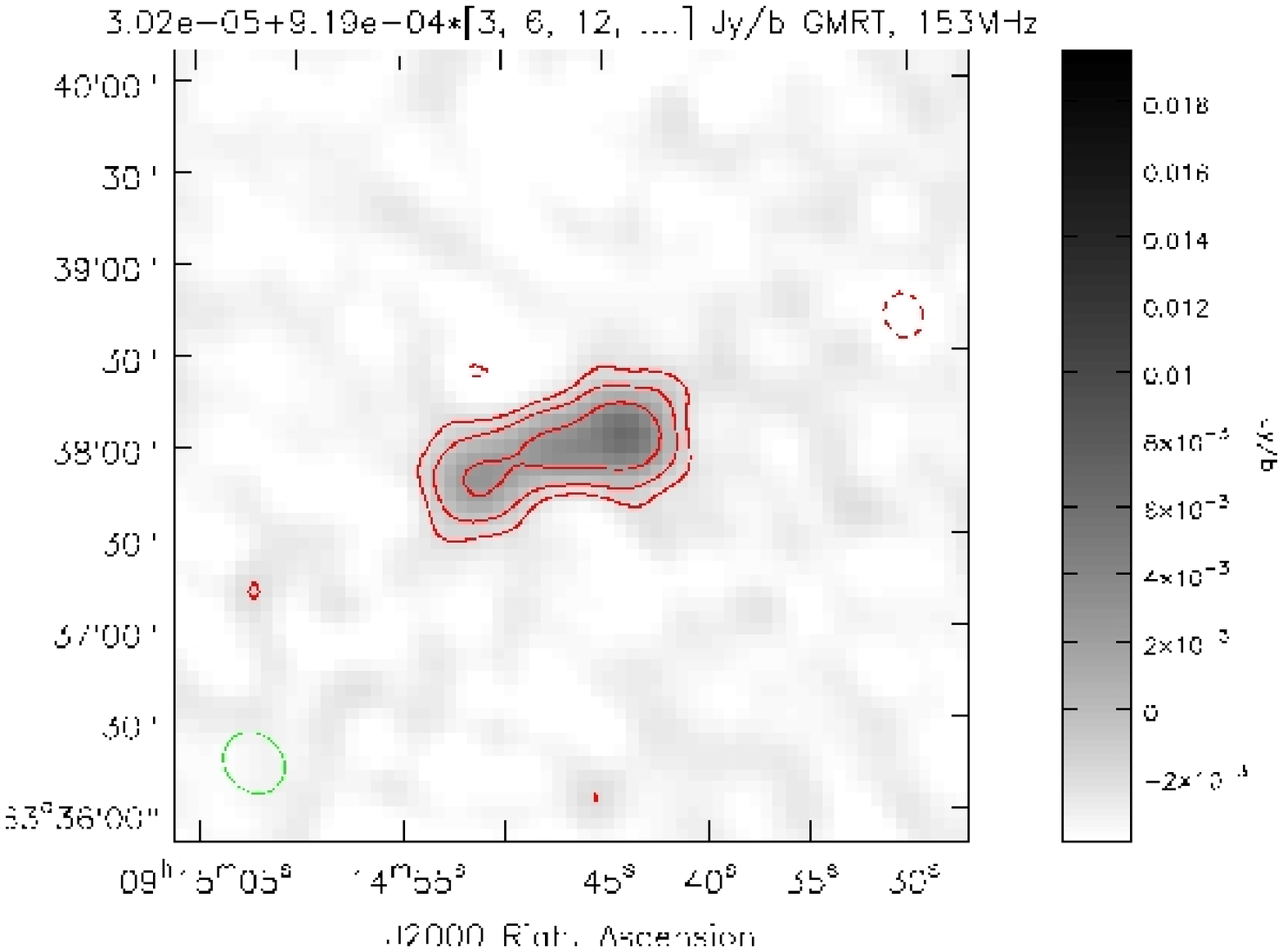}
  \includegraphics[angle=0, totalheight=1.8in, viewport=19 212 573 627, clip]{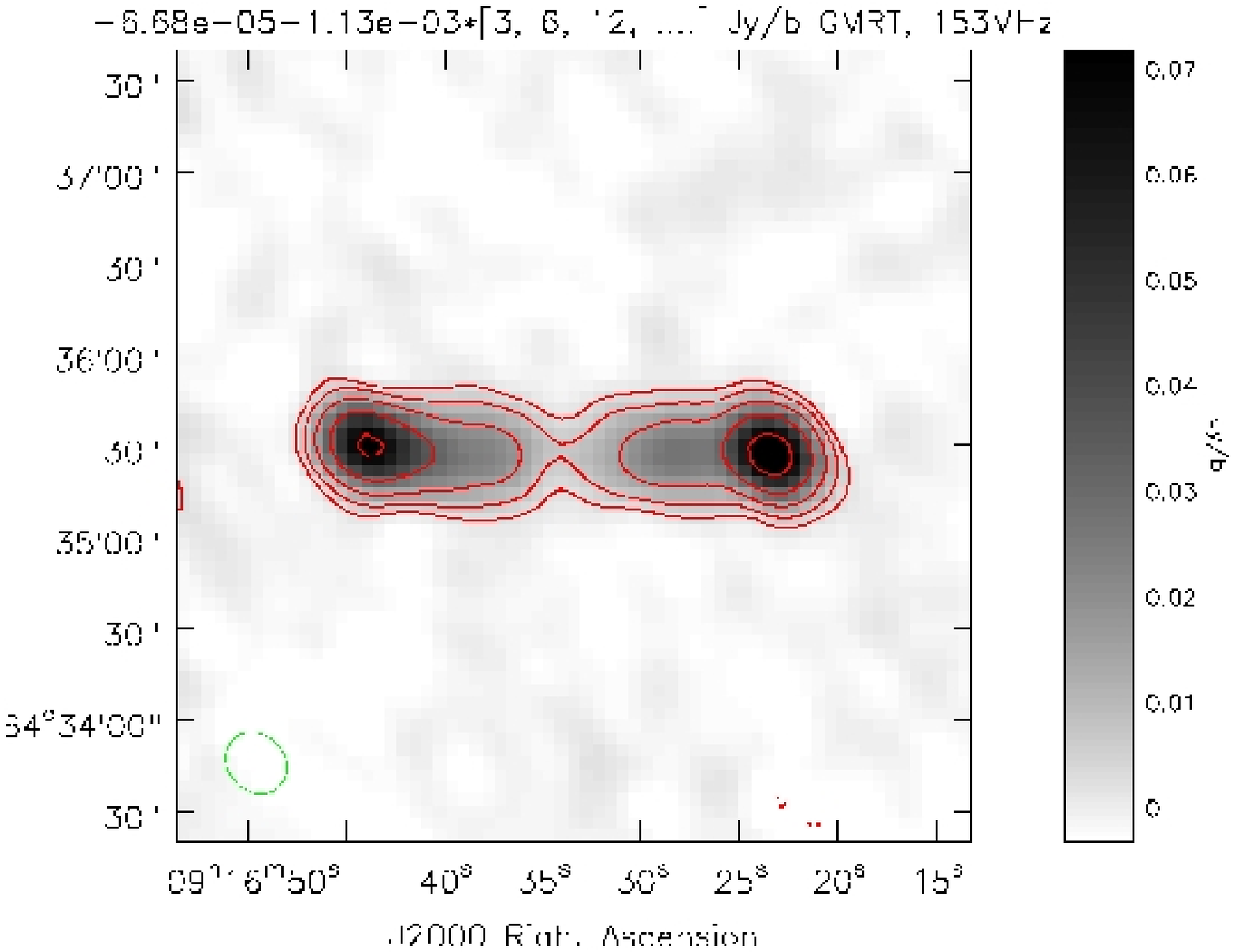}
  \includegraphics[angle=0, totalheight=1.8in, viewport=19 212 573 627, clip]{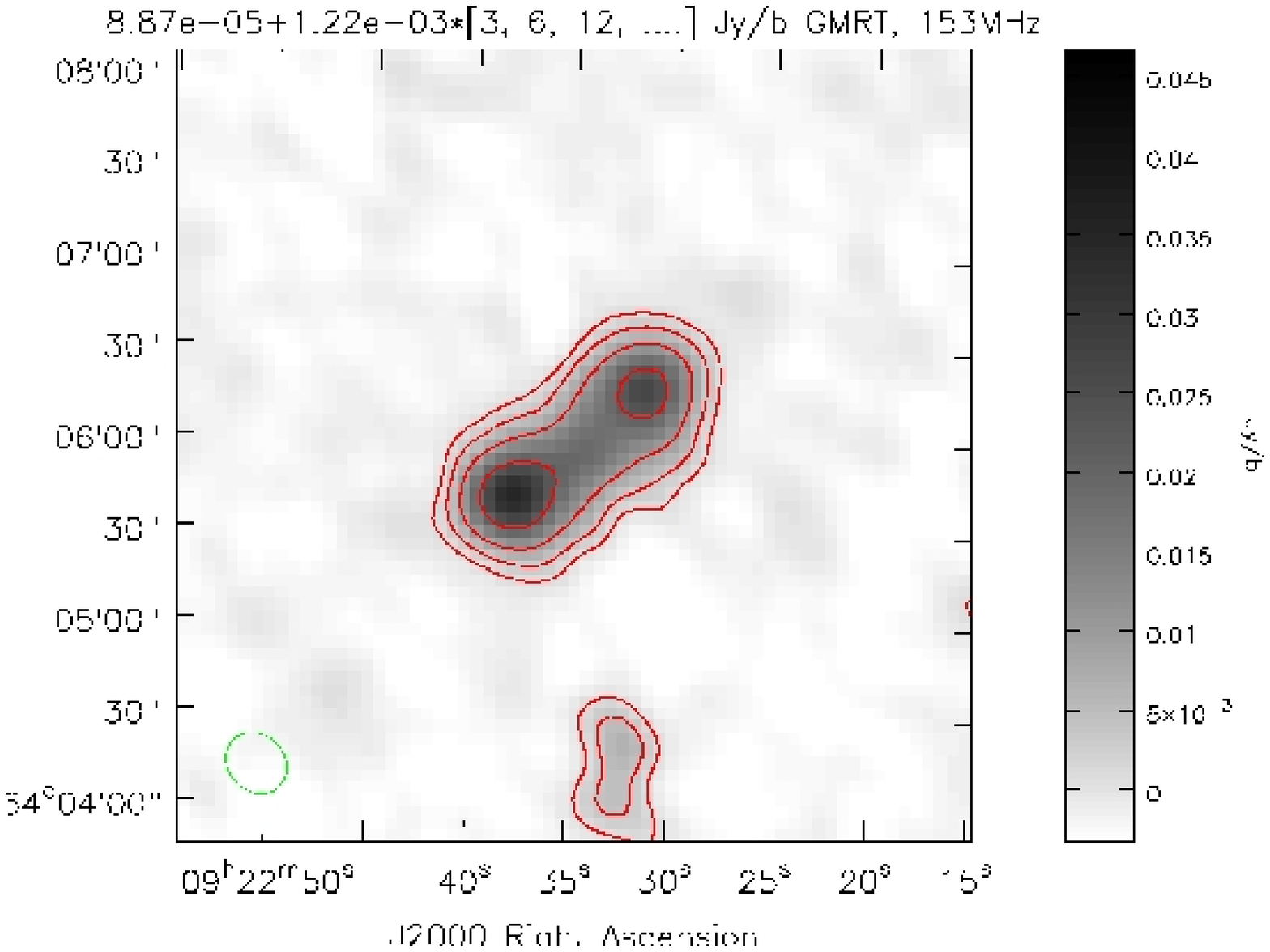}
 }
 \hbox{
  \includegraphics[angle=0, totalheight=1.8in, viewport=19 212 573 627, clip]{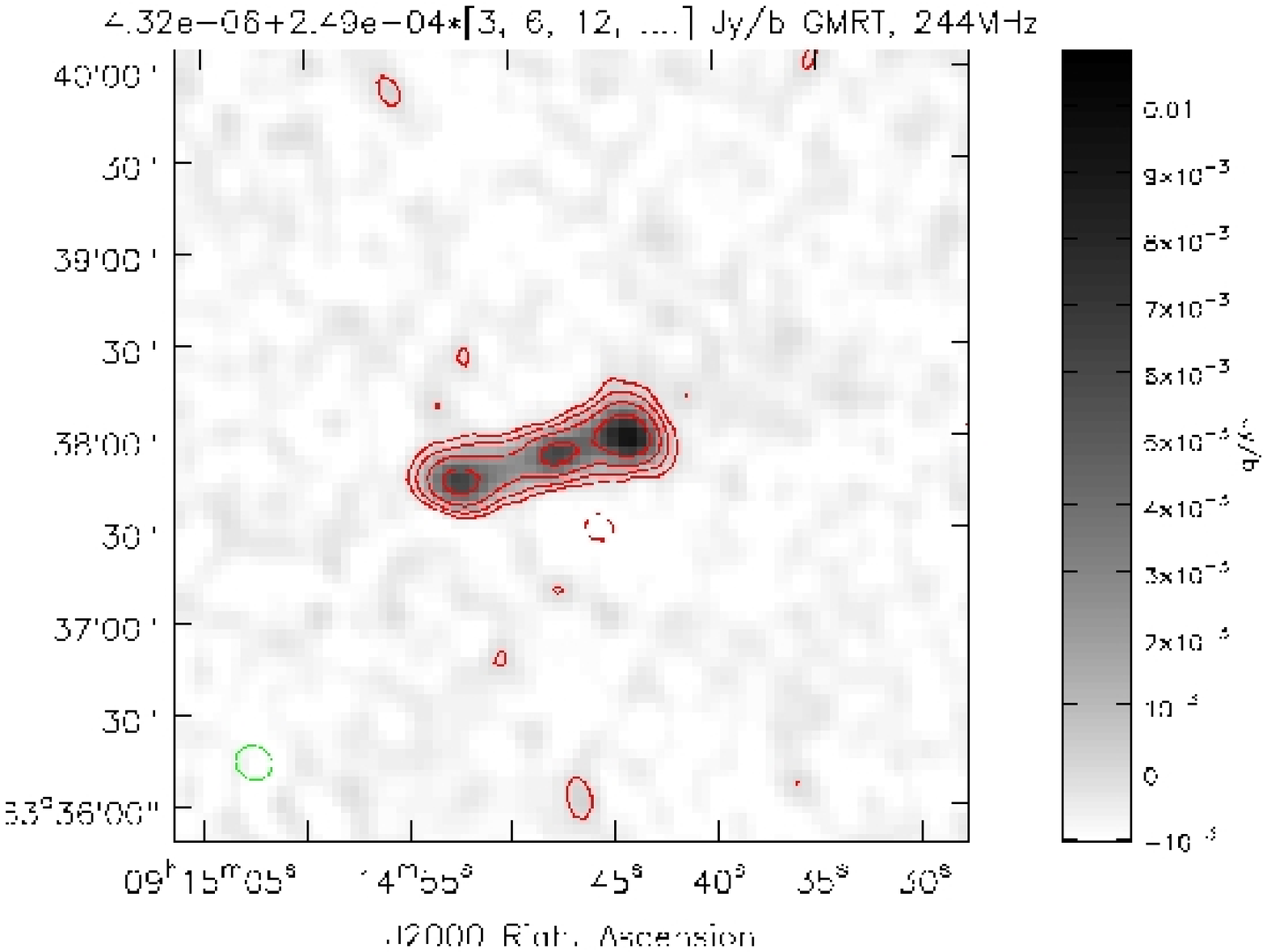}
  \includegraphics[angle=0, totalheight=1.8in, viewport=19 212 573 627, clip]{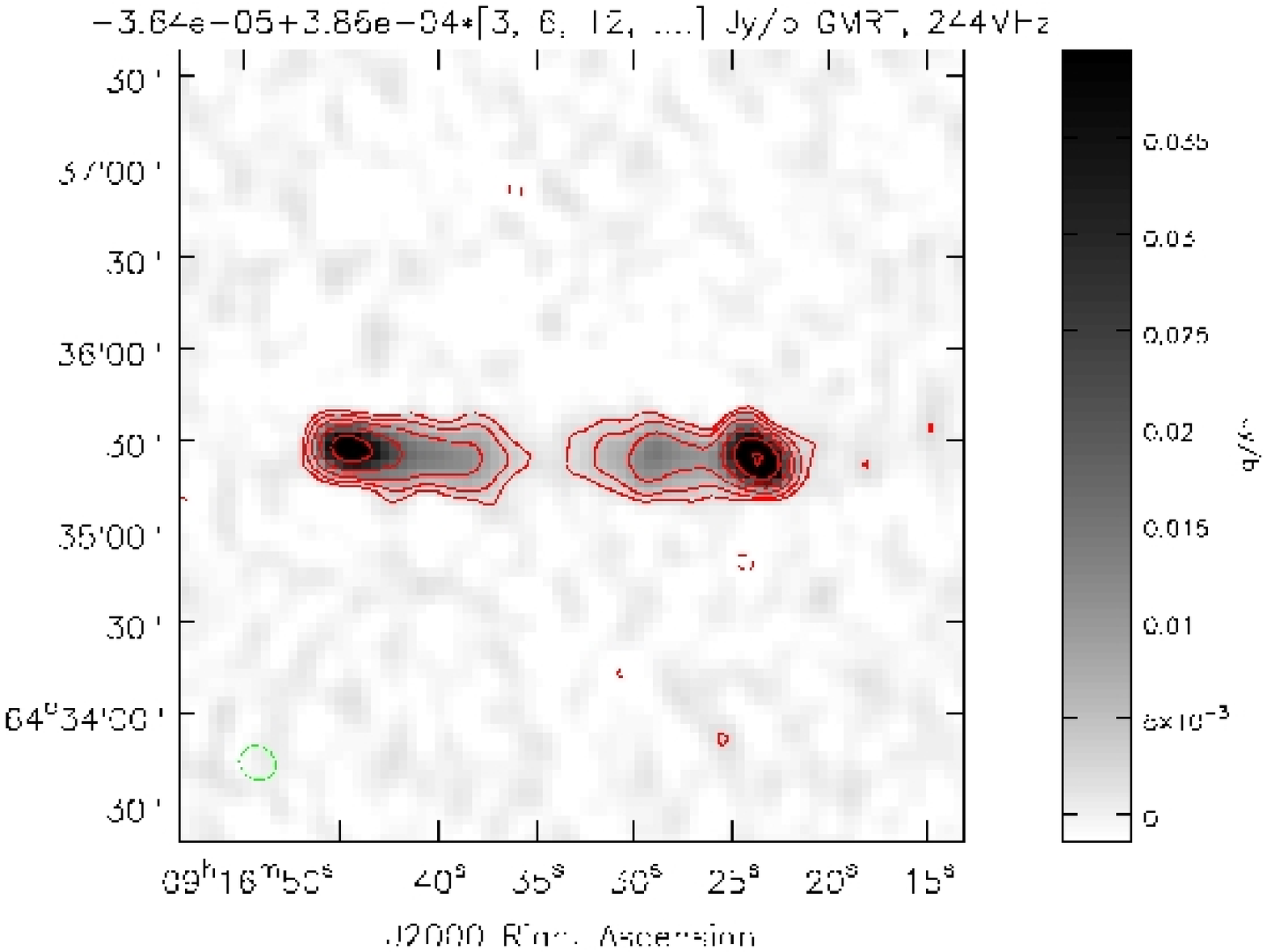}
  \includegraphics[angle=0, totalheight=1.8in, viewport=19 212 573 627, clip]{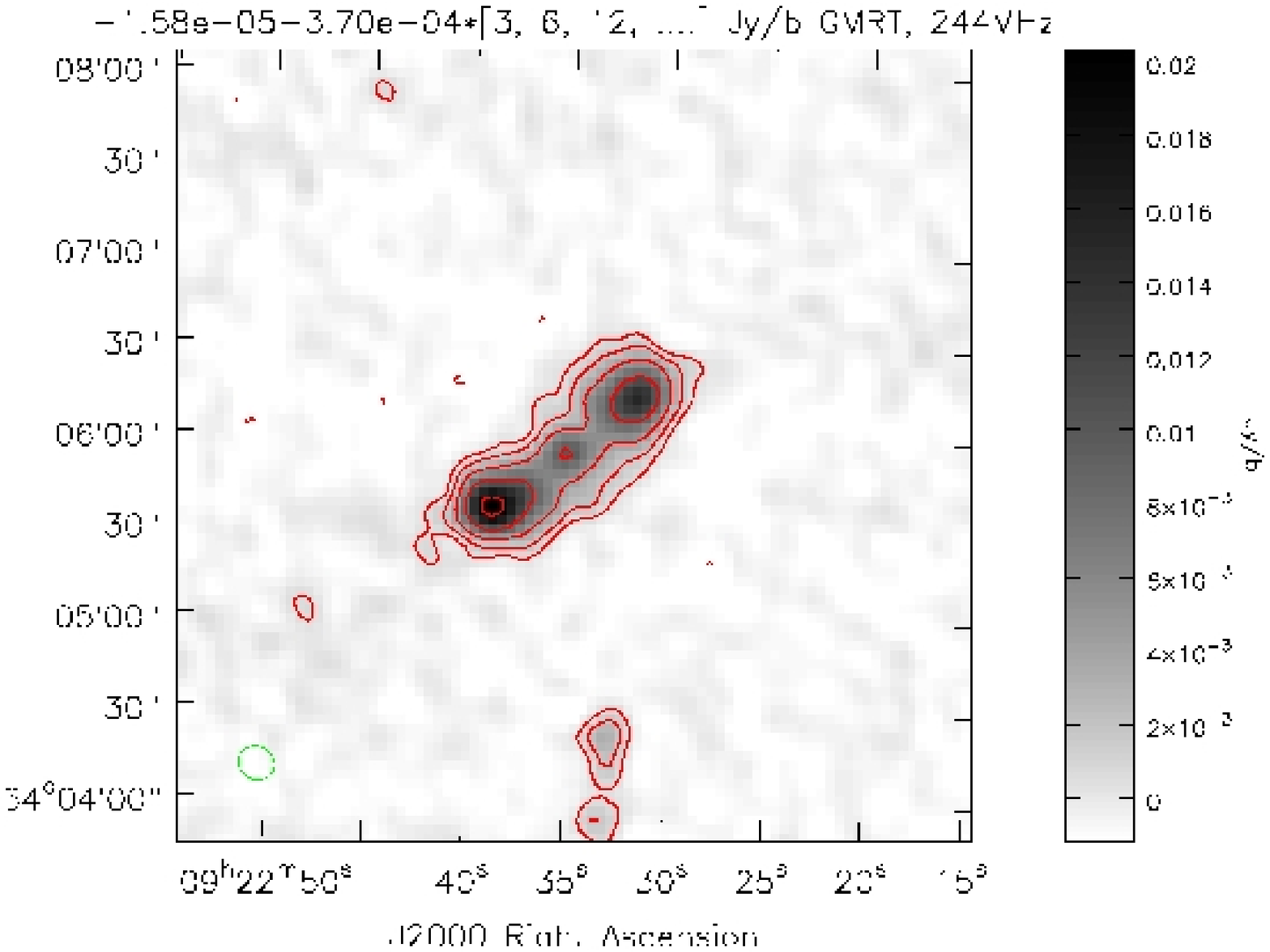}
 }
}
\caption{ The GMRT images of three double-lobed objects at 153 and 244 MHz. The images of the 
sources at 153 MHz are in the first row, while the corresponding images of the same source at
244 MHz are in the second row. Contour levels: mean$+$rms$\times$(n) in units of Jy beam$^{-1}$.}
\label{0916p6348_f0:fig:egobj}
\end{center}
\end{figure*}

\section{OBSERVATIONAL RESULTS}
\label{0916p6348_f0:sec:result}
The GMRT images of an area of $\sim$18$\times$18 arcmin$^2$ at all the four frequencies
are shown in Fig.~\ref{0916p6348_f0:fig:multifreq}, while the images from the NRAO VLA 
Sky Survey (NVSS; \citeauthor{1998AJ....115.1693C}~\citeyear{1998AJ....115.1693C}) and 
Faint Images of the Radio Sky at Twenty-centimeters survey
(FIRST; \citeauthor{1995ApJ...450..559B}~\citeyear{1995ApJ...450..559B})
are shown in Fig.~\ref{0916p6348_f0:fig:vla}. Amongst the GMRT images, the 1260-MHz
image has the lowest noise figure with an rms of about 22 $\mu$Jy beam$^{-1}$, while
the corresponding values for the 150-MHz, 244-MHz and 610-MHz images are 513, 173 
and 32 $\mu$Jy beam$^{-1}$ respectively.  Diffuse relic emission with a brightness of 
20 $\mu$Jy beam$^{-1}$ at 1260 MHz and a spectral index, $\alpha\sim$1.5, would 
have a surface brightness of $\sim$480 and 150 $\mu$Jy beam$^{-1}$ at 150 and 244
MHz respectively for a similar beam. Considering that our beam areas at these two
frequencies are larger than at 1260 MHz by $\sim$100 and 50 times respectively, we
should be able to detect diffuse relic emission associated with these sources at
the lower frequencies even if these were below the threshold at 1260 MHz. 

To examine the detection of such diffuse relic emission from either an earlier cycle of 
activity in a radio galaxy or from cluster relics and halos, and also estimate the 
low-frequency spectra of these sources, we compiled a list of
sources which are within 1.5$^\circ$ of the phase center of our 
observations at 153 MHz, and with a peak brightness which is at least 6 times 
larger than the local rms value. This yielded a list of 374 sources. 
For each of these sources we attempted to find counterparts at the other
frequencies of observations as well as in the Westerbork Northern Sky Survey
at 327 MHz (WENSS; \citeauthor{1997A+AS..124..259R}~\citeyear{1997A+AS..124..259R}; 
\citeauthor{2000yCat.8062....0D}~\citeyear{2000yCat.8062....0D}) when available, and from  NVSS and FIRST 
at 1400 MHz by searching within a radius of 30 arcsec of each source. The search radius was arrived at by
comparing the positions of compact sources in the different catalogues.

Of these 374 sources, 361 were found to have counterparts at 244 MHz. The 
remaining 13 sources are located between $\sim$1.06$^\circ$ and 1.46$^\circ$ from the pointing
center, the median value being $\sim$1.38$^\circ$. The primary beam response of the 244-MHz 
beam at 1.5$^\circ$ from the phase center being about 20 per cent of the peak value, the
rms noise is larger. We have checked the field of each of these 13 sources and 
find that these sources are visible in the 244-MHz image but do not satisfy our selection criterion
of the peak value being $\ge$6 times the rms value. Four of these sources are detected in
either the NVSS or FIRST catalogues. We have examined the spectra of these sources, but
given the uncertainties there is no clear example
of a very steep spectrum source amongst these objects.

The sources with a spectral index greater than 1.3 estimated from measurements at a minimum
of three frequencies are listed in Table~\ref{0916p6348_f0:table:ss} along with some of their 
observed properties.  The Table for the entire list of sources is 
available in the on-line version. The Tables are arranged as follows. 
Column 1: source name in J2000 co-ordinates where
hhmmss represents the hours, minutes and seconds of right ascension and
ddmmss represents the degrees, arcmin and arcsec of declination, estimated
from the flux-density weighted centroid of all the emission enclosed by
the 3$\sigma$ contour at 153 MHz;  columns 2 and 3: the right ascension and declination
of the source in J2000 co-ordinates; column 4: distance of the centroid of the source in
the 150-MHz image from
the pointing center in degrees; columns 5 to 10: the total flux densities
at 153, 244, 330, 610, 1260 and 1400 MHz in units of mJy; column 11: spectral index
from a linear least-squares fit for which the spectrum could be satisfactorily fitted
with a straight spectrum. For sources which show a strong departure from a straight-line fit,
a parabolic form log S = b(log$\nu$)$^2$+mlog$\nu$+c was fitted and the high frequency 
spectral indices between 610 and 1400 MHz have been quoted. The flux densities at 330 MHz are
from the WENSS catalogue while those at 1400 MHz are from the NVSS catalogue. The flux
densities from the GMRT observations at 610 and 1260 MHz have been listed for only those
sources which are 0.54$^\circ$ and 0.34$^\circ$ of the phase center, which corresponds to
about 20 per cent of the peak response of the primary beam. The flux densities of the
sources in all the GMRT images have been estimated by integrating the emission within the
3-$\sigma$ closed contour. We have also examined the
structure of each source at each frequency as well as in the FIRST and NVSS images.
We have not listed the flux densities of a few extended sources at specific frequencies
if significant flux density appears to be missing in the images at the given frequency 
even after convolving it to a lower resolution. This has been done
for a few sources at 610 and 1260 MHz.  Column 12: 
structural classification where S denotes a single source, D a double-lobed source, T a 
triple source with a possible core component and
Cplx a complex source. Diffuse extended emission without a clear double-lobed structure
has been marked as E. The structural classification has been usually done from the highest 
resolution image available, which is usually our GMRT image at either 610 or 1260 MHz or
the FIRST image at 1400 MHz.  The resolution of our 610-MHz
image is similar to that of the FIRST image although the noise in our image is significantly
better than that of FIRST.  We have estimated the errors in the flux densities to be 
$\sim$15 per cent at 150 and 244 MHz, 10 per cent at 610 MHz and 5 per cent at the
higher frequencies. For each source we have estimated the error in the value of the 
spectral index, and find the typical error to be $\sim$0.15.

\begin{figure*}
\begin{center}
\vbox{
 \hbox{
  \includegraphics[angle=270, totalheight=1.5in]{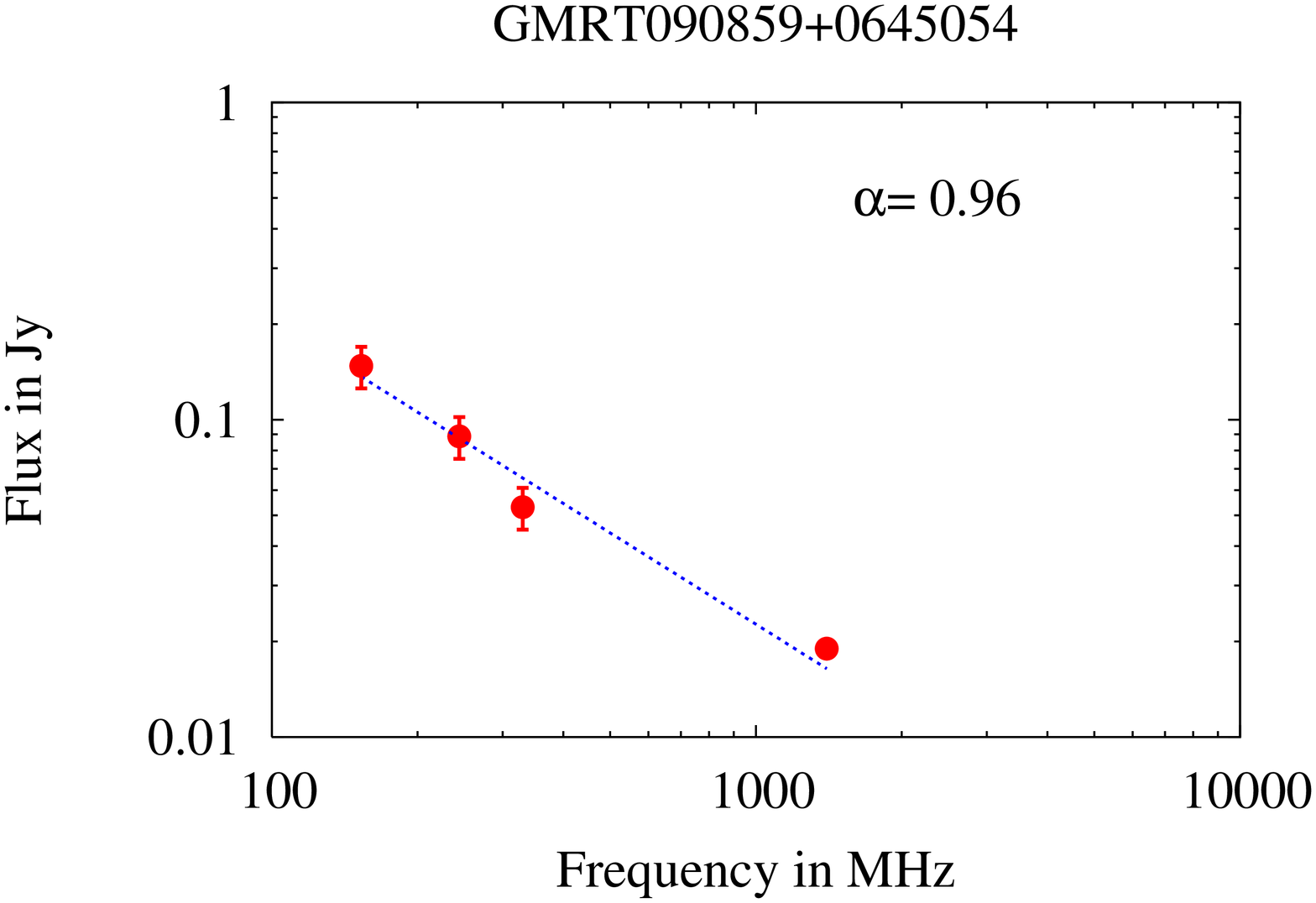}
  \includegraphics[angle=270, totalheight=1.5in]{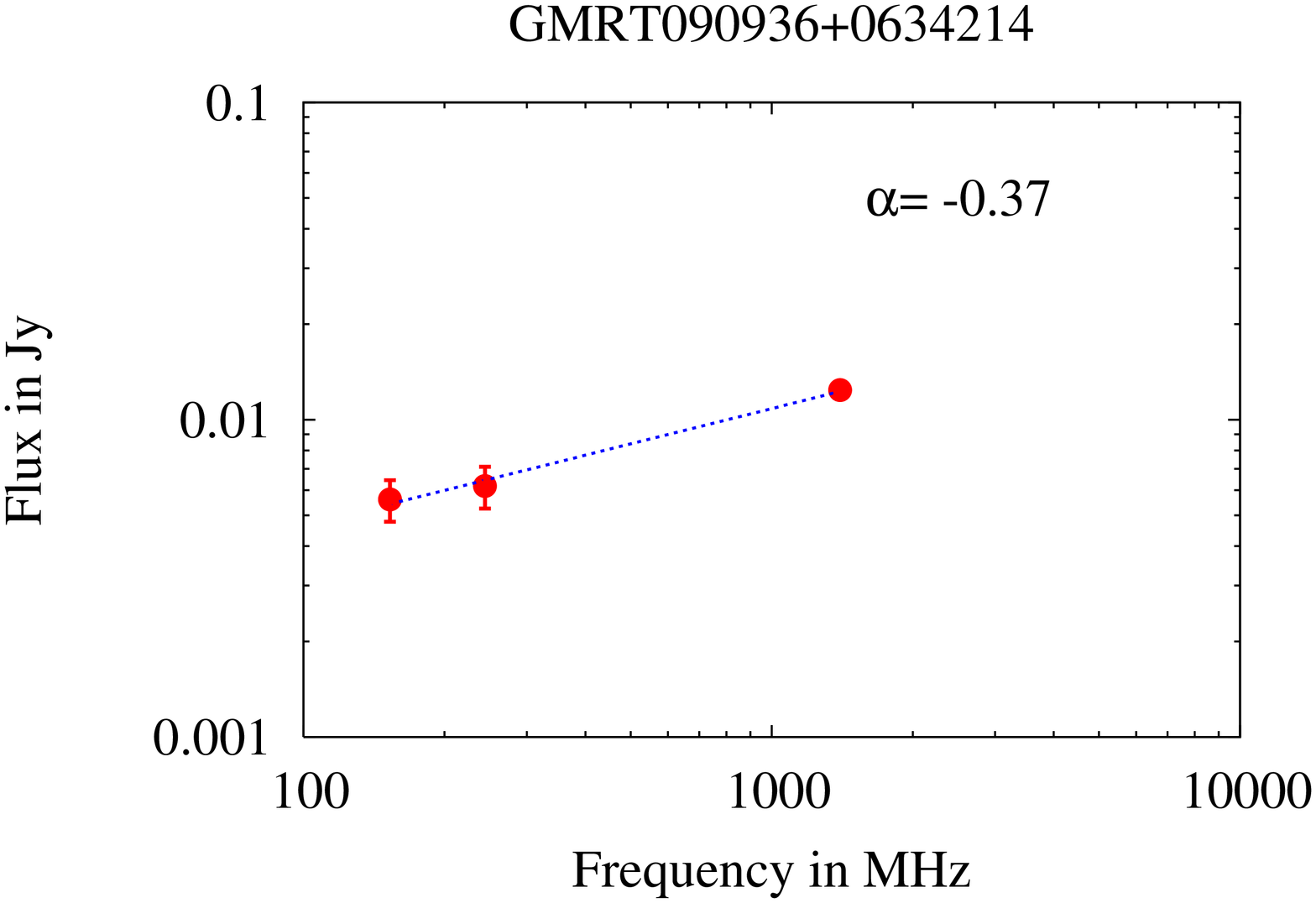}
  \includegraphics[angle=270, totalheight=1.5in]{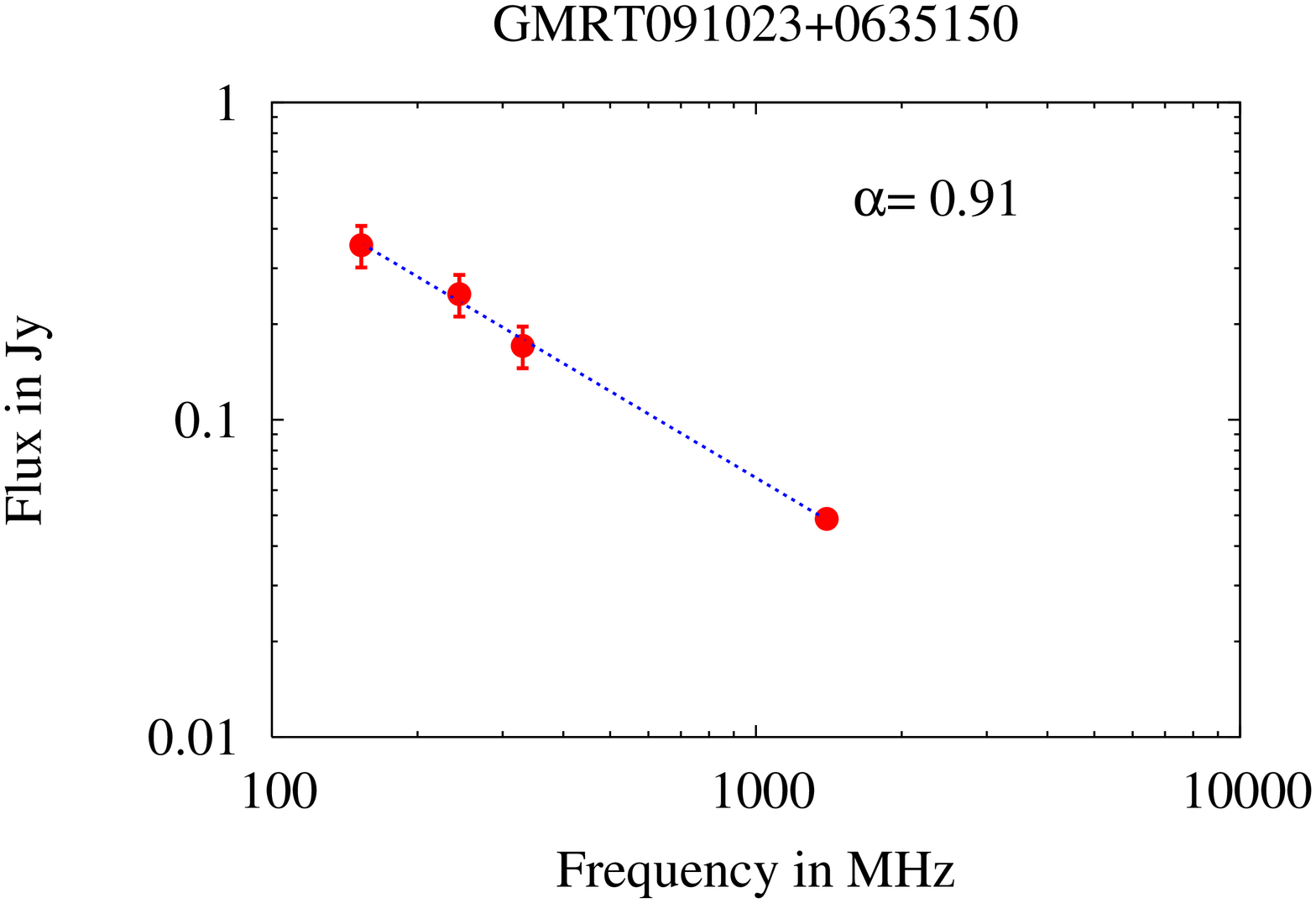}
 }
 \hbox{
  \includegraphics[angle=270, totalheight=1.5in]{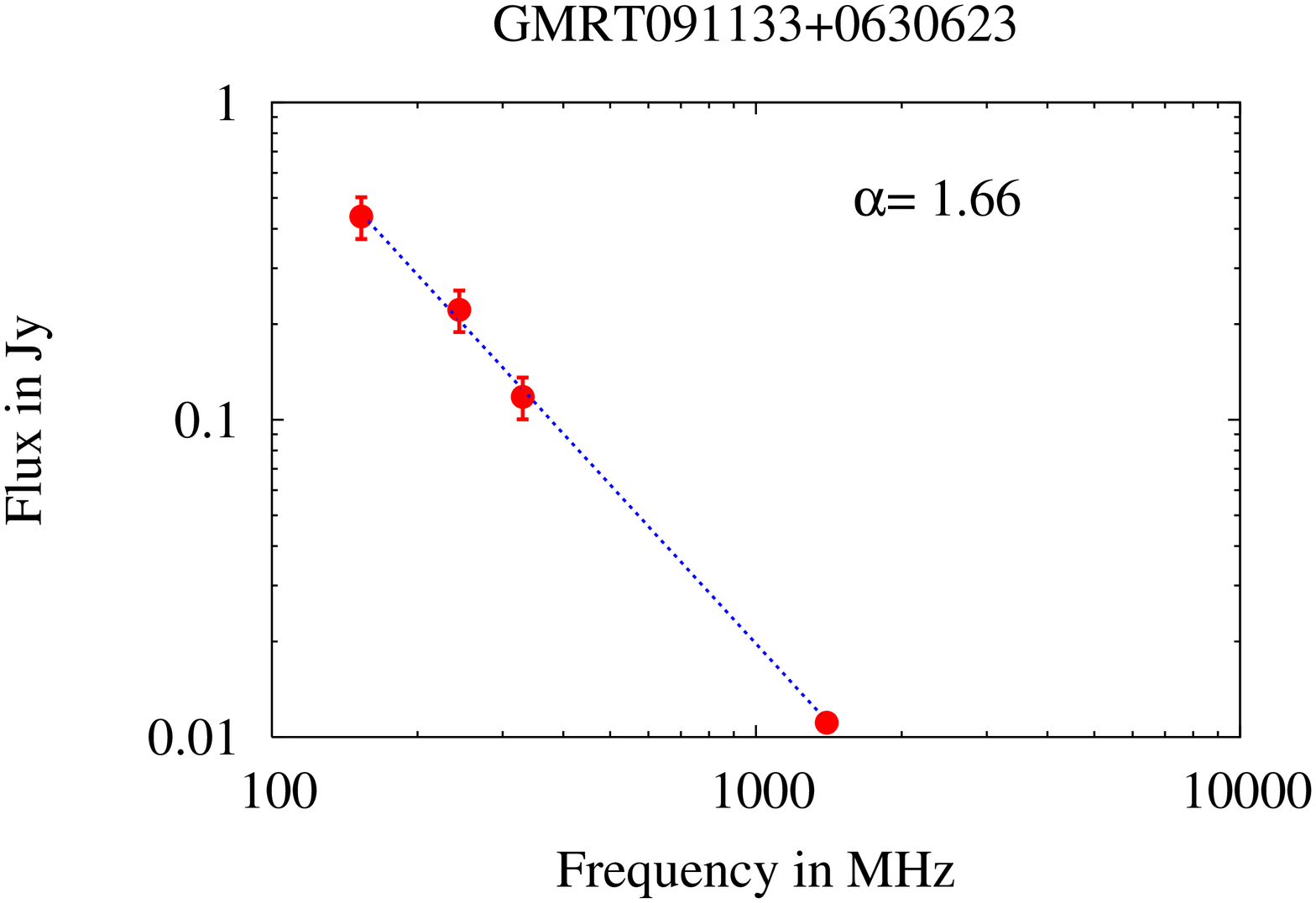}
  \includegraphics[angle=270, totalheight=1.5in]{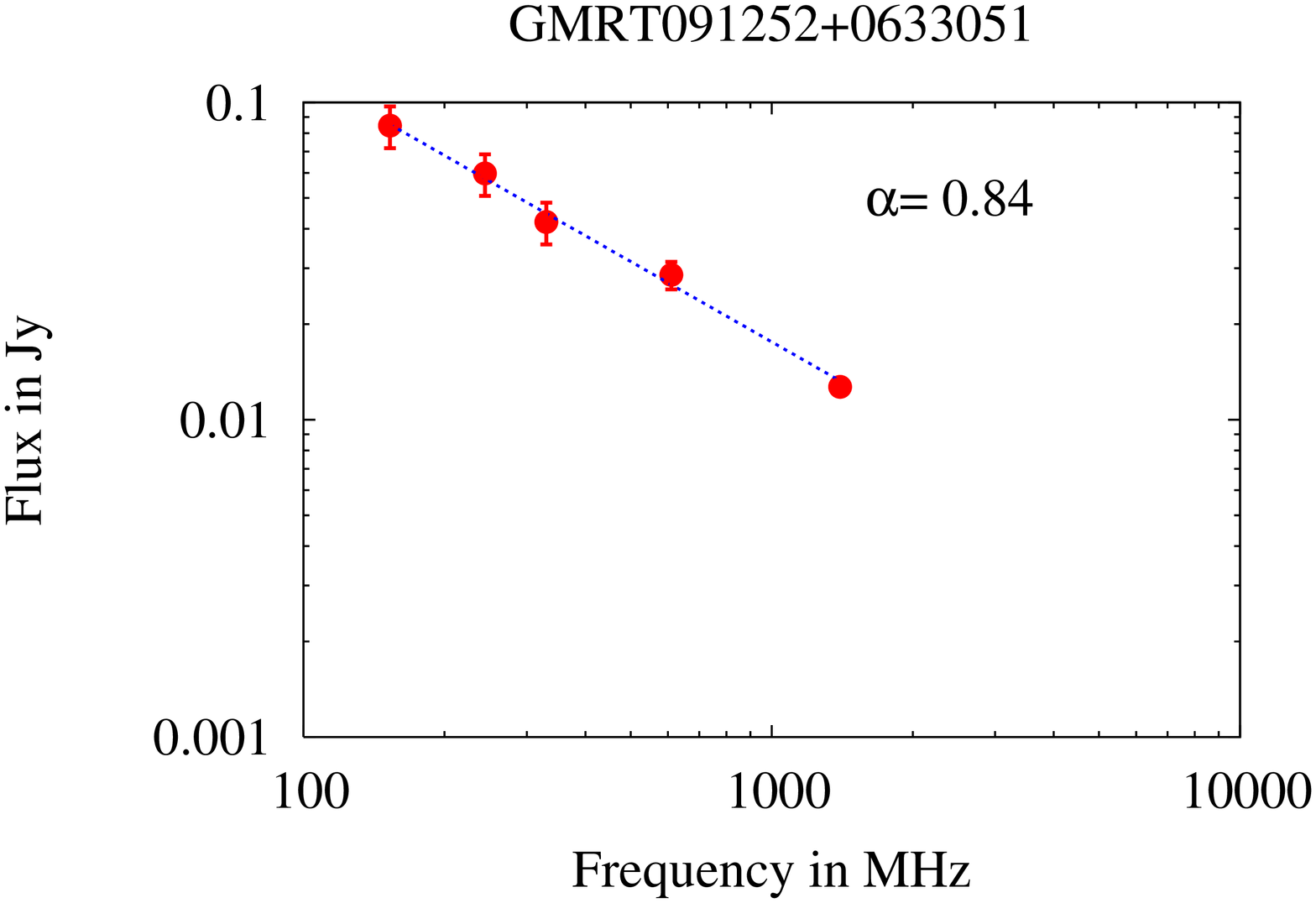}
  \includegraphics[angle=270, totalheight=1.5in]{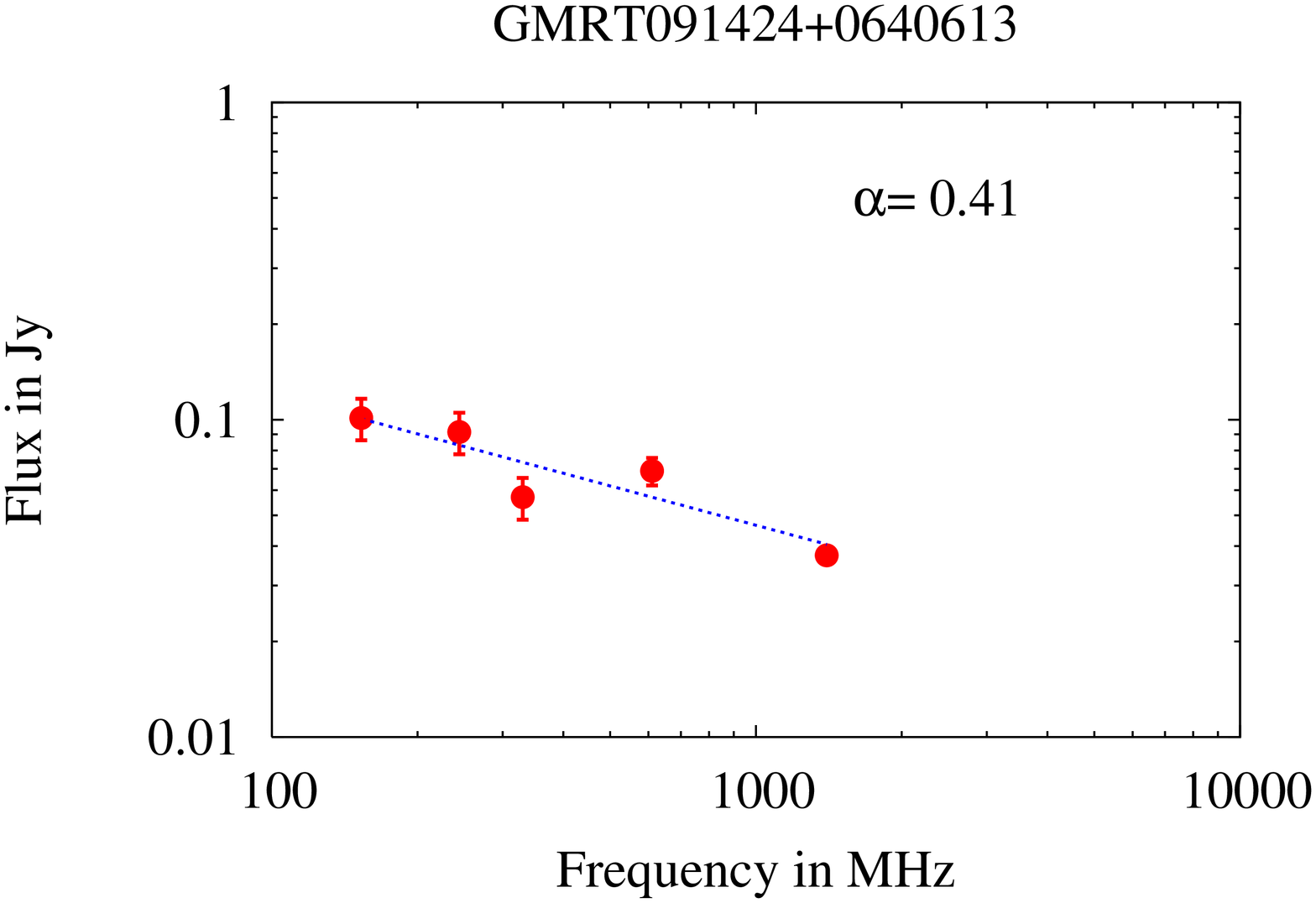}
 }
 \hbox{
  \includegraphics[angle=270, totalheight=1.5in]{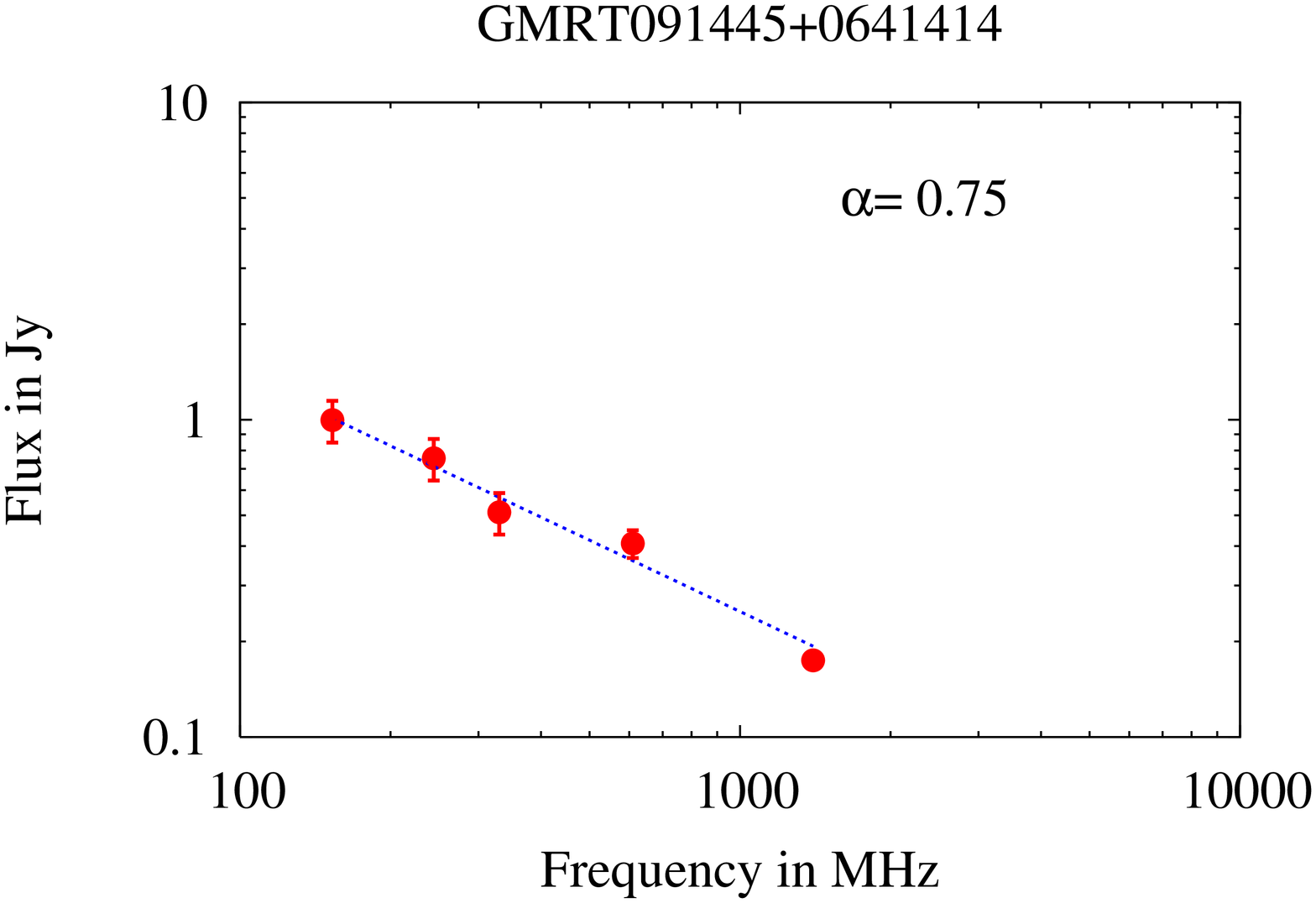}
  \includegraphics[angle=270, totalheight=1.5in]{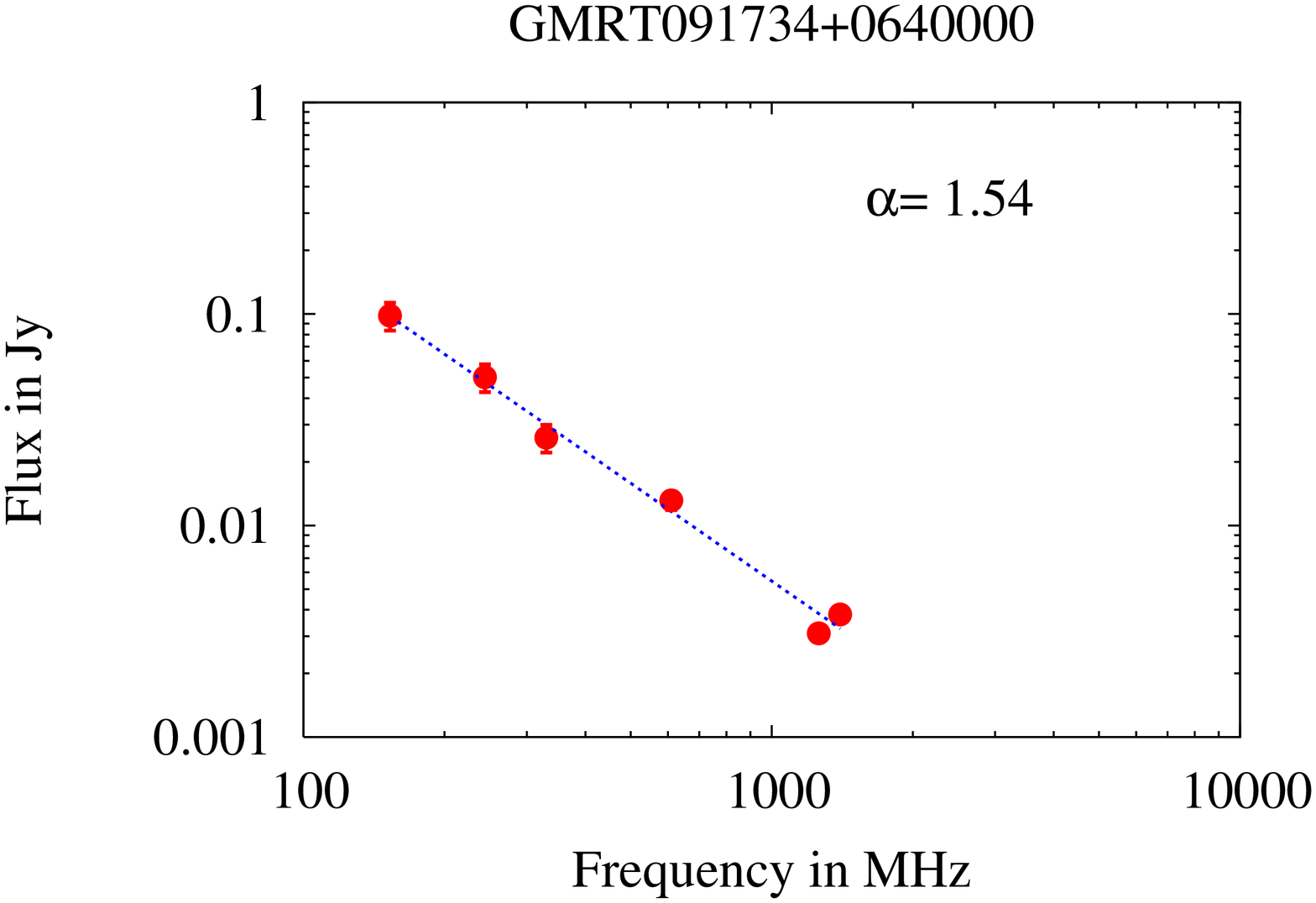}
  \includegraphics[angle=270, totalheight=1.5in]{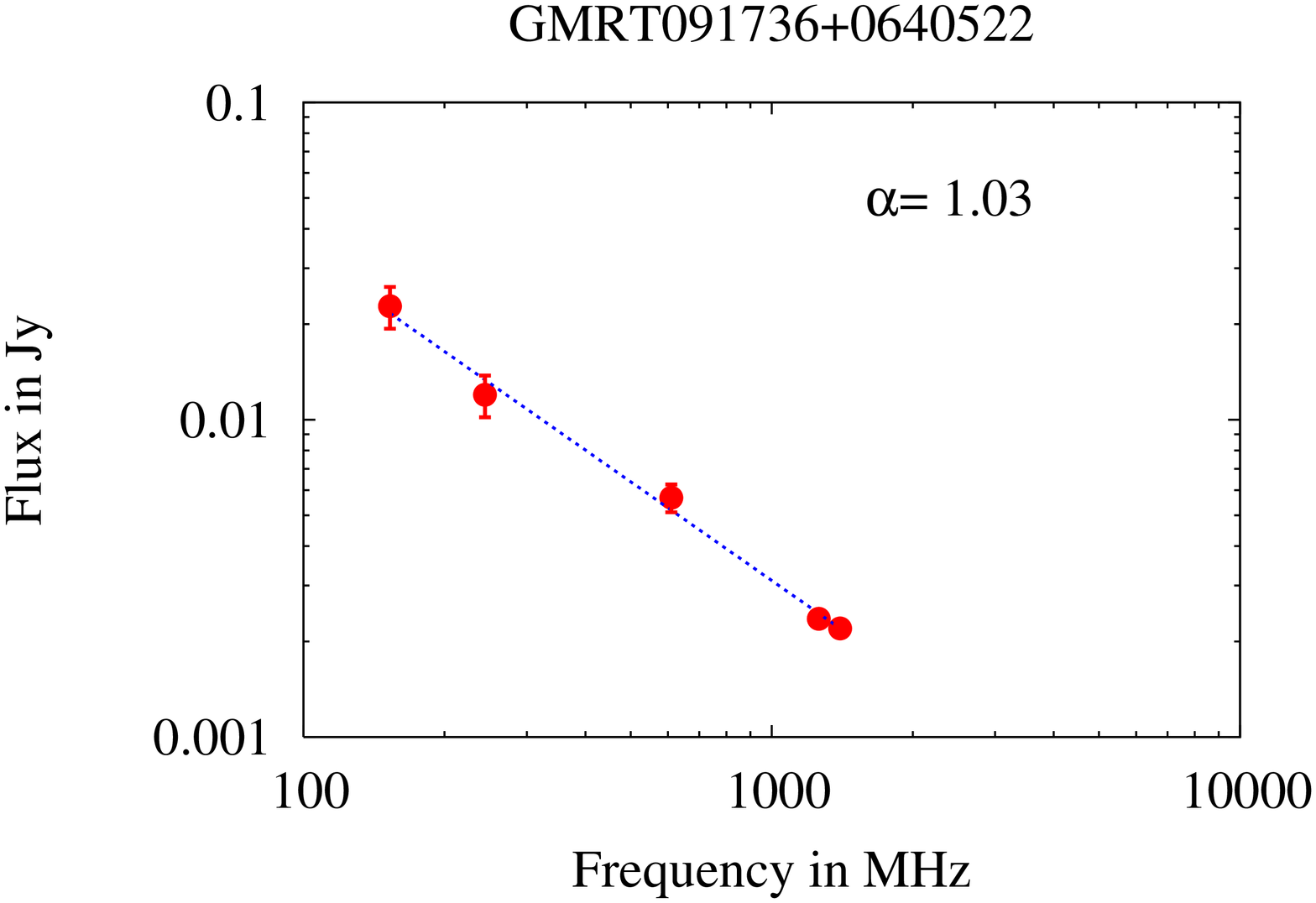}
 }
}
\caption{ Representative spectra of a few objects fitted with a linear least-squares fit.}
\label{0916p6348_f0:fig:spfits}
\end{center}
\begin{center}
\vbox{
 \hbox{
  \includegraphics[angle=270, totalheight=1.5in]{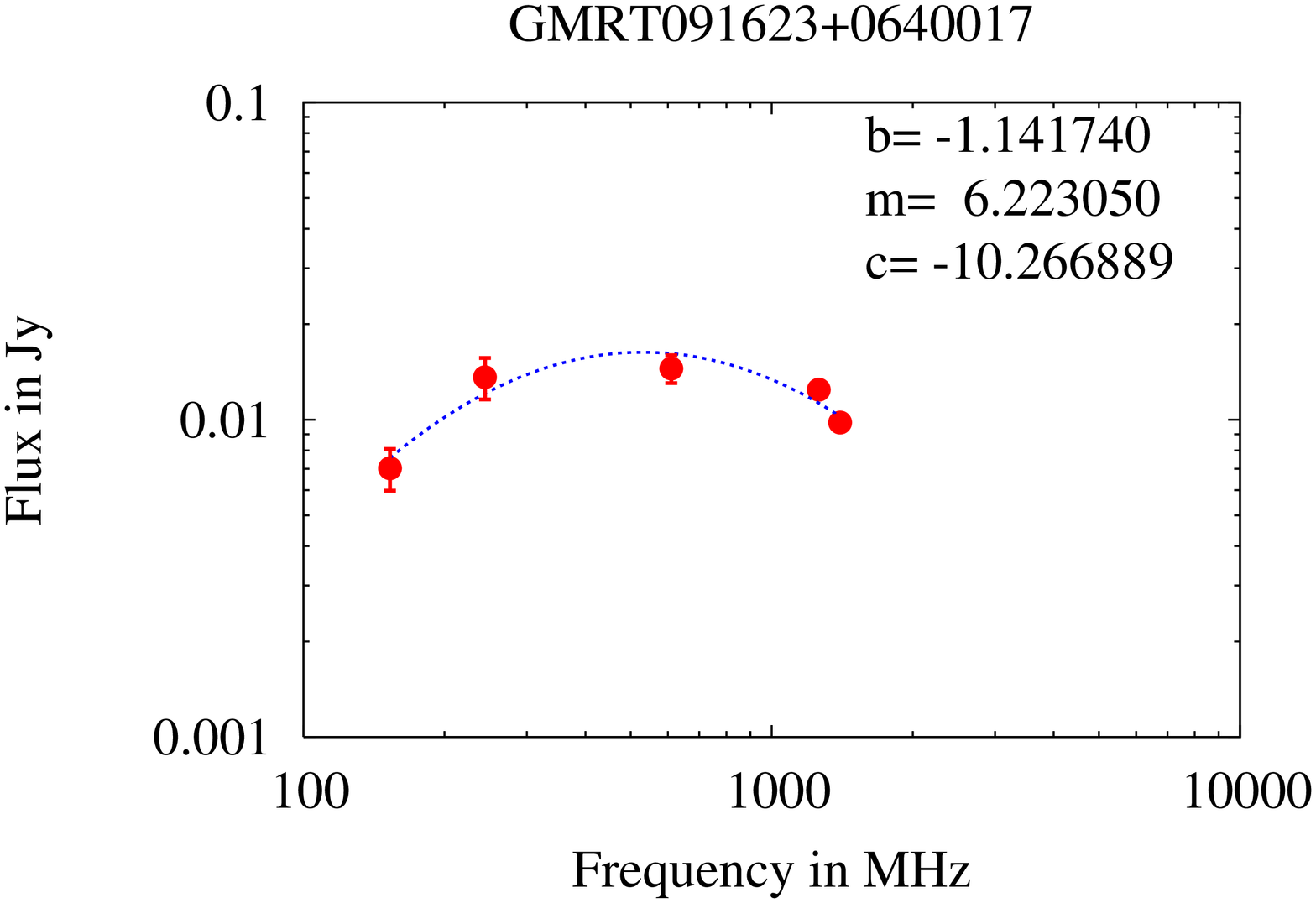}
  \includegraphics[angle=270, totalheight=1.5in]{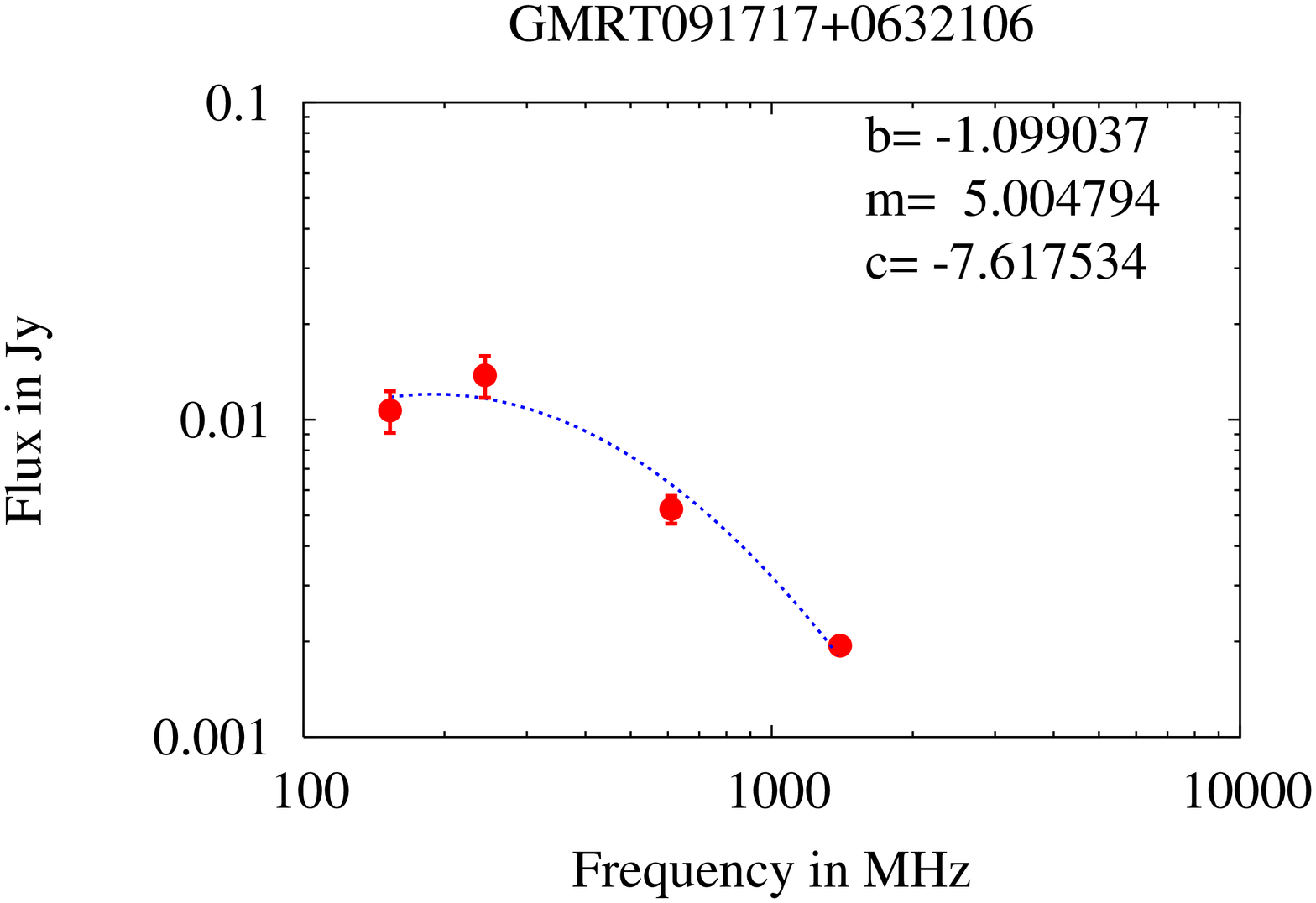}
  \includegraphics[angle=270, totalheight=1.5in]{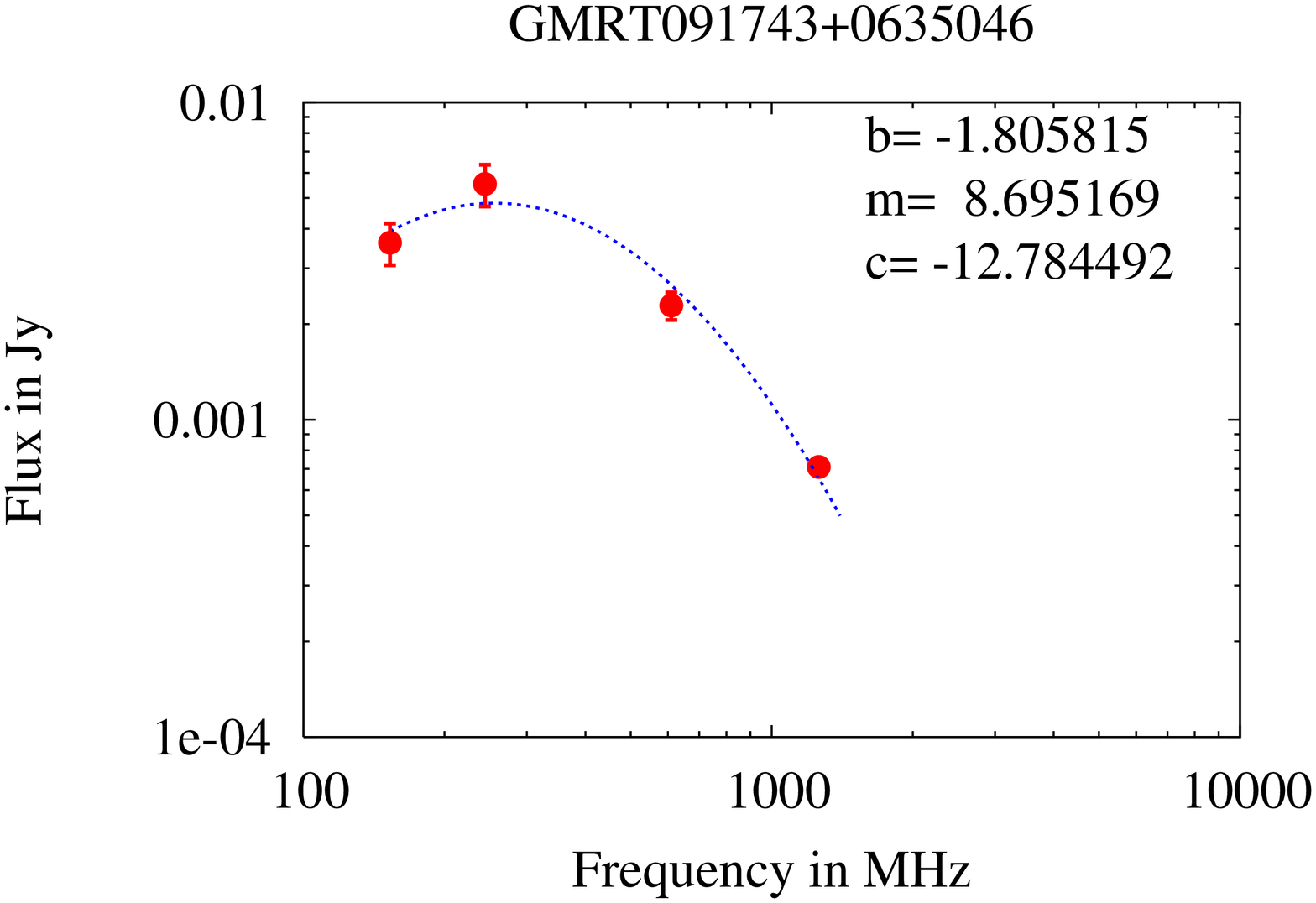}
 }
}
\caption{Spectra of the three objects fitted with a parabolic form.}
\label{0916p6348_f0:fig:spfitp}
\end{center}
\end{figure*}

\begin{table*}
\caption{The steep-spectrum sources with a spectral index steeper than 1.3. The Table for the entire list of sources with spectral index information is available in the on-line version.}
\label{0916p6348_f0:table:ss}
\begin{tabular}{l r r r r r r r r r r l}
\hline
\multicolumn{1}{c}{Source name} & \multicolumn{1}{c}{RA} & \multicolumn{1}{c}{DEC} & \multicolumn{1}{c}{Dis} & \multicolumn{1}{c}{S$_{153}$} & \multicolumn{1}{c}{S$_{244}$}
                    & \multicolumn{1}{c}{S$_{330}$} & \multicolumn{1}{c}{S$_{610}$} & \multicolumn{1}{c}{S$_{1260}$} & \multicolumn{1}{c}{S$_{1400}$}
                    & \multicolumn{1}{c}{$\alpha$} &\multicolumn{1}{c}{Class} \\
                    & \multicolumn{1}{c}{hh:mm:ss.s} & \multicolumn{1}{c}{dd:mm:ss.s} & \multicolumn{1}{c}{deg.} & \multicolumn{1}{c}{mJy} & \multicolumn{1}{c}{mJy}
                    & \multicolumn{1}{c}{mJy} & \multicolumn{1}{c}{mJy} & \multicolumn{1}{c}{mJy} & \multicolumn{1}{c}{mJy} &  &        \\
\multicolumn{1}{c}{(1)} & \multicolumn{1}{c}{(2)} & \multicolumn{1}{c}{(3)} & \multicolumn{1}{c}{(4)} &  \multicolumn{1}{c}{(5)} & \multicolumn{1}{c}{(6)} & \multicolumn{1}{c}{(7)} & \multicolumn{1}{c}{(8)} & \multicolumn{1}{c}{(9)} & \multicolumn{1}{c}{(10)} & \multicolumn{1}{c}{(11)} & \multicolumn{1}{c}{(12)} \\
\hline
GMRT090357+640848 & 09:03:57.8 & 64:08:48.2 & 1.43 &  427.0 &  202.4 &  139.0 &    $-$ &    $-$ &   25.1 &   1.31 & S \\
GMRT090740+642941 & 09:07:41.0 & 64:29:41.4 & 1.19 &   53.3 &   30.4 &    $-$ &    $-$ &    $-$ &    2.8 &   1.33 & D \\
GMRT090842+642154 & 09:08:42.4 & 64:21:54.2 & 1.03 &  102.3 &   57.2 &   32.0 &    $-$ &    $-$ &    6.0 &   1.31 & S \\
GMRT091133+630623 & 09:11:33.4 & 63:06:23.5 & 0.89 &  436.8 &  222.1 &  118.0 &    $-$ &    $-$ &   11.1 &   1.66 & D \\
GMRT091308+633945 & 09:13:08.5 & 63:39:45.8 & 0.41 &   19.1 &    4.9 &    $-$ &    0.7 &    $-$ &    $-$ &   2.37 & S \\
GMRT091328+640209 & 09:13:28.4 & 64:02:09.1 & 0.41 &    7.6 &    3.9 &    $-$ &    1.1 &    $-$ &    $-$ &   1.37 & S \\
GMRT091734+640001 & 09:17:34.2 & 64:00:01.0 & 0.23 &   98.2 &   50.2 &   26.0 &   13.1 &    3.1 &    3.8 &   1.54 & S \\
GMRT091744+630229 & 09:17:44.0 & 63:02:29.5 & 0.77 &   20.6 &   11.5 &    $-$ &    $-$ &    $-$ &    1.1 &   1.34 & S \\
GMRT091845+642854 & 09:18:45.5 & 64:28:54.4 & 0.72 &   62.4 &   32.9 &   17.0 &    $-$ &    $-$ &    2.5 &   1.48 & D \\
GMRT092004+633626 & 09:20:04.4 & 63:36:26.5 & 0.43 &   22.5 &    8.6 &    $-$ &    2.9 &    $-$ &    $-$ &   1.49 & S \\
GMRT092009+641550 & 09:20:09.6 & 64:15:50.1 & 0.60 &  256.5 &  145.9 &   94.0 &    $-$ &    $-$ &   14.4 &   1.30 & S \\
GMRT092131+633939 & 09:21:31.9 & 63:39:39.6 & 0.57 &  114.2 &   60.5 &   40.0 &    $-$ &    $-$ &    4.0 &   1.50 & D \\
GMRT092332+633611 & 09:23:32.4 & 63:36:11.9 & 0.80 &   89.0 &   48.6 &   38.0 &    $-$ &    $-$ &    4.6 &   1.31 & Cplx \\
GMRT092830+634748 & 09:28:30.7 & 63:47:48.8 & 1.32 &   96.8 &   33.1 &   23.0 &    $-$ &    $-$ &    $-$ &   1.95 & E \\
\hline \hline
\end{tabular}

\end{table*}
\section{DISCUSSION}
\label{0916p6348_f0:sec:discussions}

\subsection{Search for relic emission}
Relic radio emission associated with an earlier cycle of activity may be 
seen as either lobes of emission beyond the extent of the younger lobes
as in the case of the double-double radio galaxies, or just diffuse emission
associated with a smaller-scale double-lobed or triple radio source. If the
relic radio emission, which is likely to have a steep radio spectrum 
contributes a significant amount of the total flux density one might also
find signatures of it in the integrated spectra of these sources. There 
could also be relic sources without any recent activity; these sources
are likely to have a very steep radio spectrum due to radiative losses.

We have examined the structure and spectra of each source in our entire
list of 374 sources. 
Of these, 295 are classified as single, 61 as double,
9 as triple and the remaining 9 as complex sources. 
Although tails and bridges have been detected in many of the sources, there 
are no clear unambiguous signatures of episodic activity either in the 
form of DDRGs or diffuse emission associated with the double-lobed or single 
sources. The images of three of the large angular-sized, double-lobed
sources at both 153 and 244 MHz are shown in Fig.~\ref{0916p6348_f0:fig:egobj}.
This suggests that sources with clear signatures of episodic activity are 
rather rare even in fields selected and observed with high sensitivity 
at a low frequency. We made low-resolution images 
at the different frequencies to ensure that
we have not missed any diffuse steep-spectrum emission.

Most of the known dozen or so DDRGs are associated with large radio galaxies, 
with sizes typically over about a Mpc 
\citep[e.g.][]{2000MNRAS.315..371S, 2006MNRAS.366.1391S}.
This may be due to the typical time scales of episodic activity which 
have been estimated to be in the range of $\sim$10$^7$--10$^8$ yr
\citep{2000MNRAS.315..371S, 2006MNRAS.372..693K, 2008MNRAS.385.1286J, 2008MNRAS.385.2117S}.
About 79 per cent of our sources are single sources, 
with the median value of angular size for the entire
list of sources being less than 10 arcsec, which  corresponds to 
a linear size of only $\sim$60 and 80 kpc respectively at redshifts of
0.5 and 1 respectively.  The non-detection of unambiguous signatures of
episodic activity in these sources may possibly also be related to their small
linear sizes and hence younger ages. 

\subsection{Spectra}
\begin{figure}
\begin{center}
\includegraphics[angle=270, totalheight=2.3in]{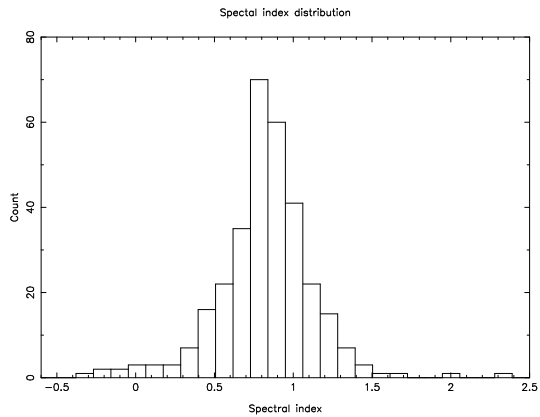}
\caption{The spectral index distribution for all the 317 sources with a minimum of three measurements
         at different frequencies. For the three sources fitted with a parabolic form, the high-frequency
         spectral index between 610 and 1400 MHz has been plotted.}
\label{0916p6348_f0:fig:spdis}
\end{center}
\end{figure}

We have fitted the spectra for 317 of the sources which have measurements at 
a minimum of three different frequencies. Since the measurements cover only about a 
decade in frequency and the number of independent flux density estimates
at different frequencies is also limited, we have fitted a straight 
spectrum to the sources unless the data show evidence of large deviations.
The representative spectra of a few sources with straight spectra are
shown in Fig.~\ref{0916p6348_f0:fig:spfits}, while those for the three 
sources with curved spectra
are shown in Fig.~\ref{0916p6348_f0:fig:spfitp}. The curved spectra appear 
to exhibit turn over frequencies ranging from $\sim$100  to 600 MHz.
The three sources with a possible low-frequency
turnover in their spectra also do not have any extended emission associated with it.
The distribution of spectral indices for all the sources is shown in 
Fig.~\ref{0916p6348_f0:fig:spdis}. For the three sources with curved spectra, we have plotted
the high-frequency spectral index. The median value of spectral index $\sim$0.82,
which is similar to estimates for the powerful 3CR sources
\citep{1992MNRAS.257..545L,1989MNRAS.239..401L,1997ApJ...479..258K},
but somewhat larger than the theoretically expected values of injection 
spectral indices discussed earlier. 
 
\begin{figure*}
\begin{center}
\vbox{
 \hbox{\hskip 0.9cm GMRT090357+640848 \hskip 3.2cm GMRT090740+642941 \hskip 3.2cm  GMRT090842+642154}
 \hbox{
  \includegraphics[angle=0, totalheight=1.8in, viewport=19 212 573 627, clip]{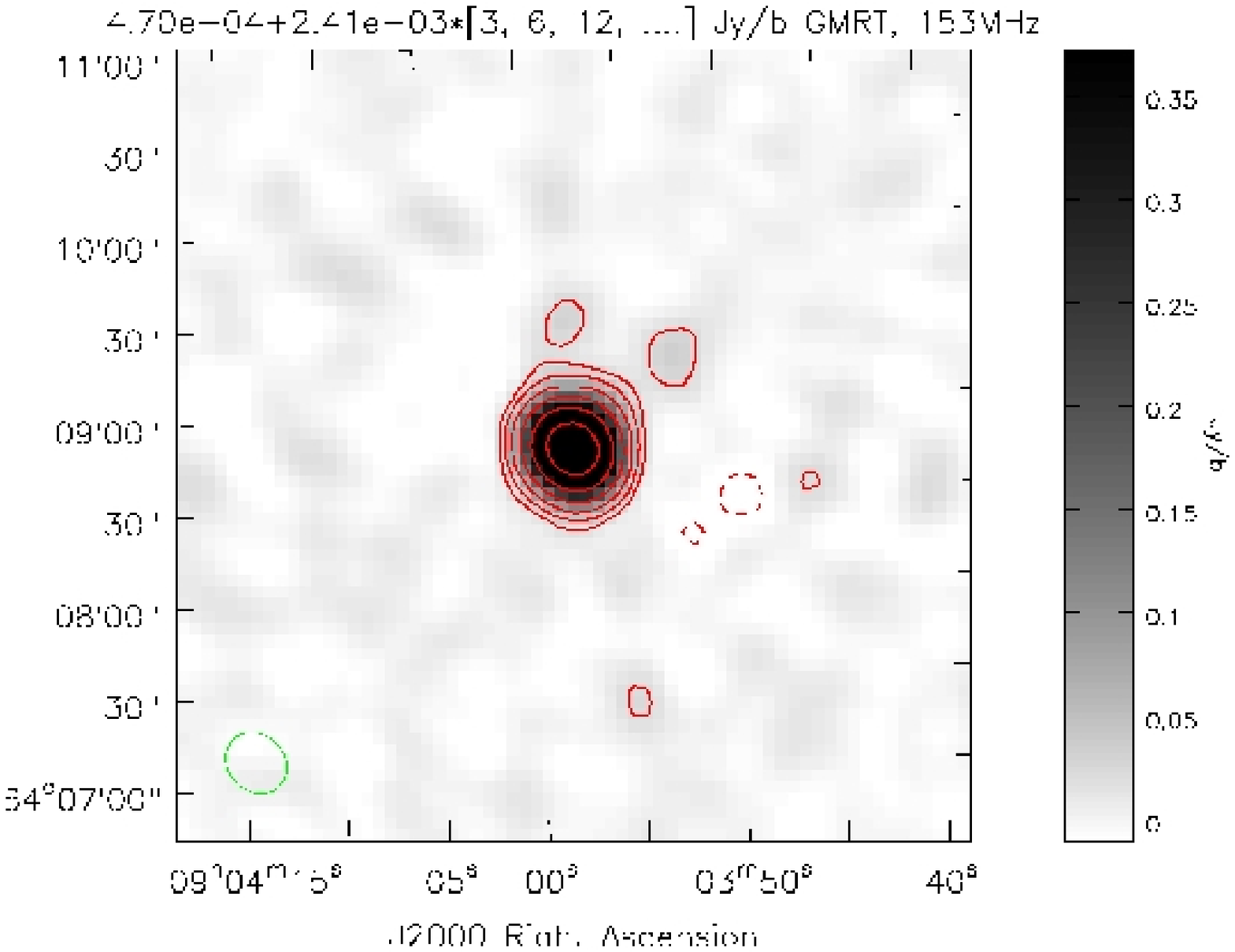}
  \includegraphics[angle=0, totalheight=1.8in, viewport=19 212 573 627, clip]{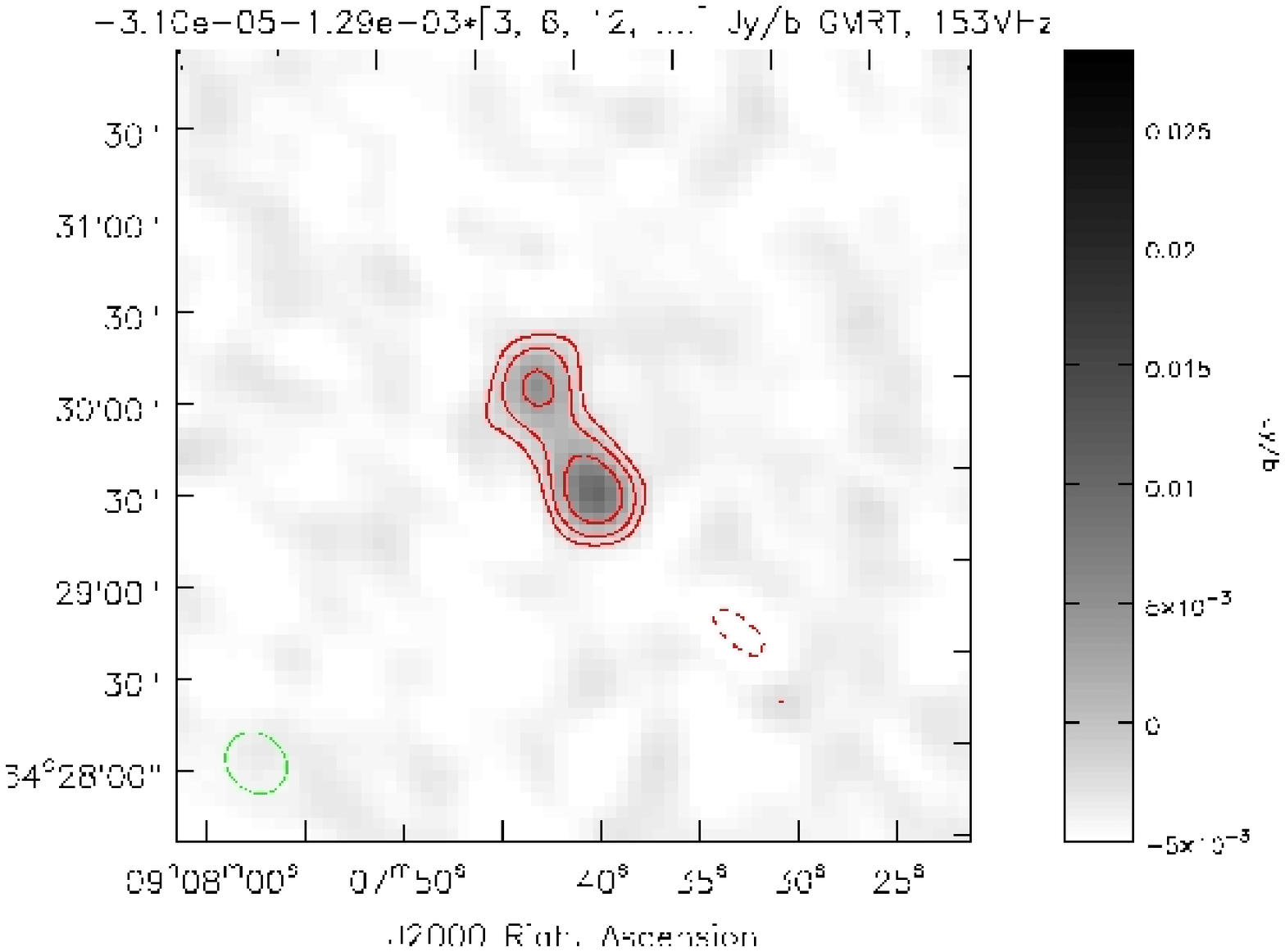}
  \includegraphics[angle=0, totalheight=1.8in, viewport=19 212 573 627, clip]{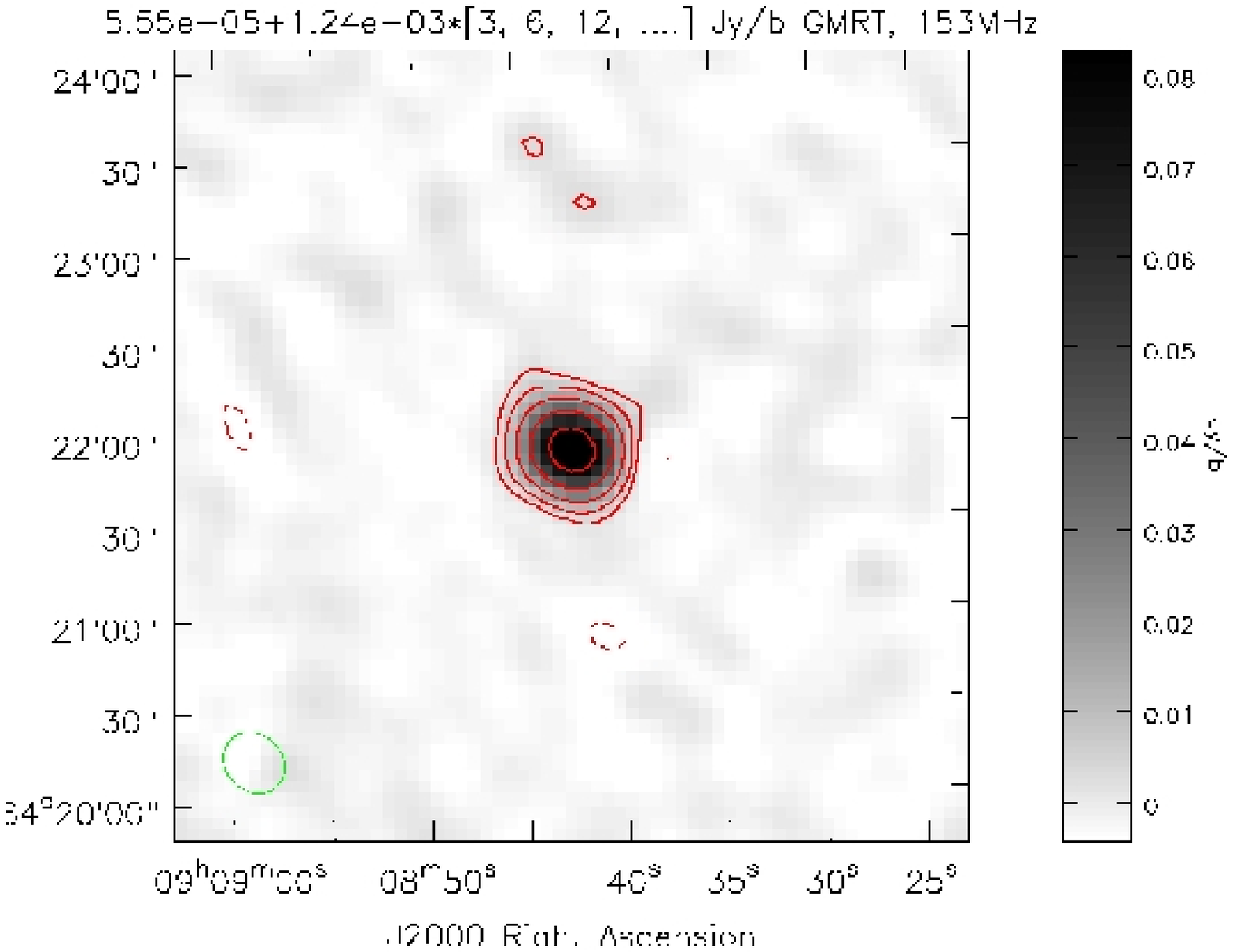}
 }
 \hbox{
  \includegraphics[angle=0, totalheight=1.8in, viewport=19 212 573 627, clip]{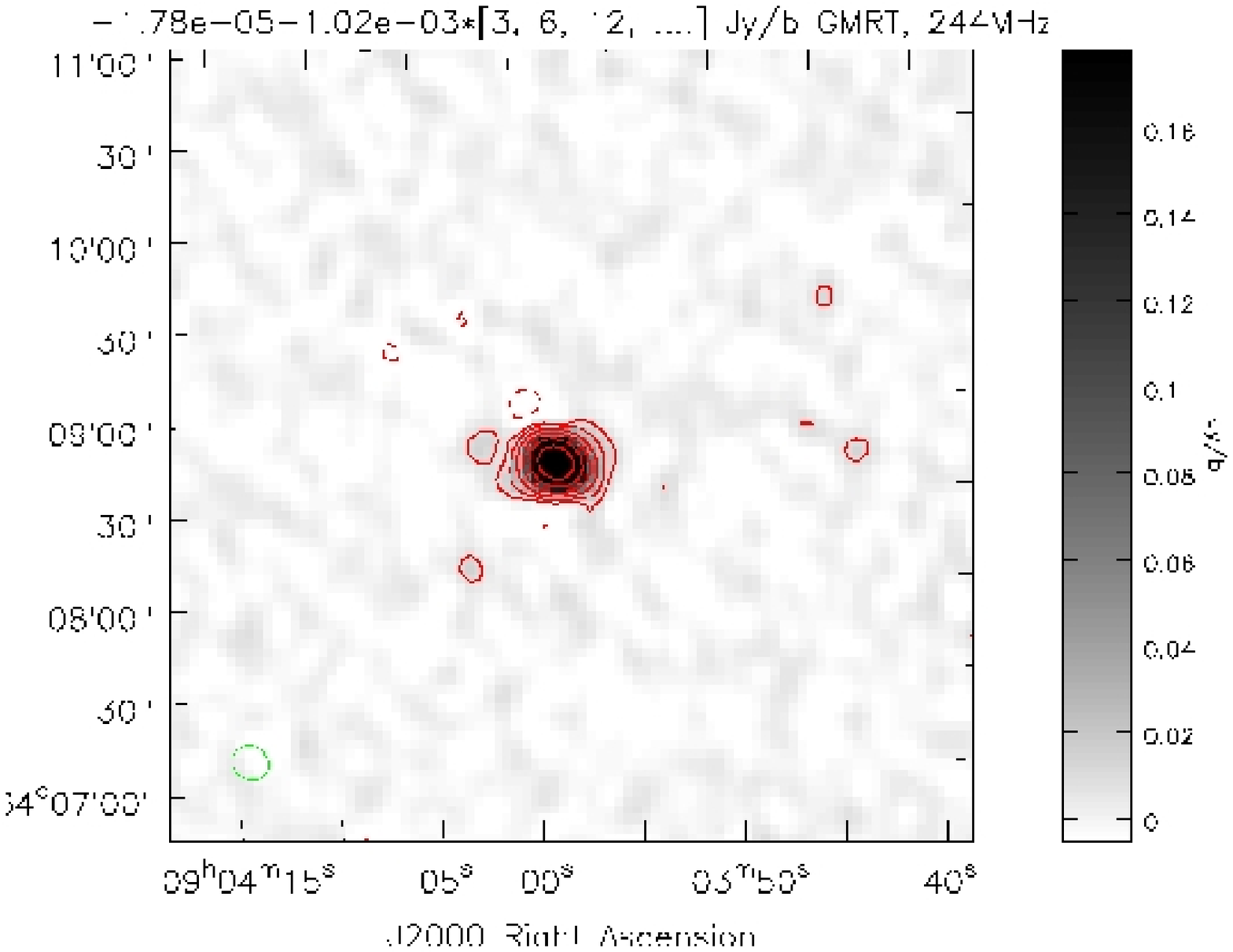}
  \includegraphics[angle=0, totalheight=1.8in, viewport=19 212 573 627, clip]{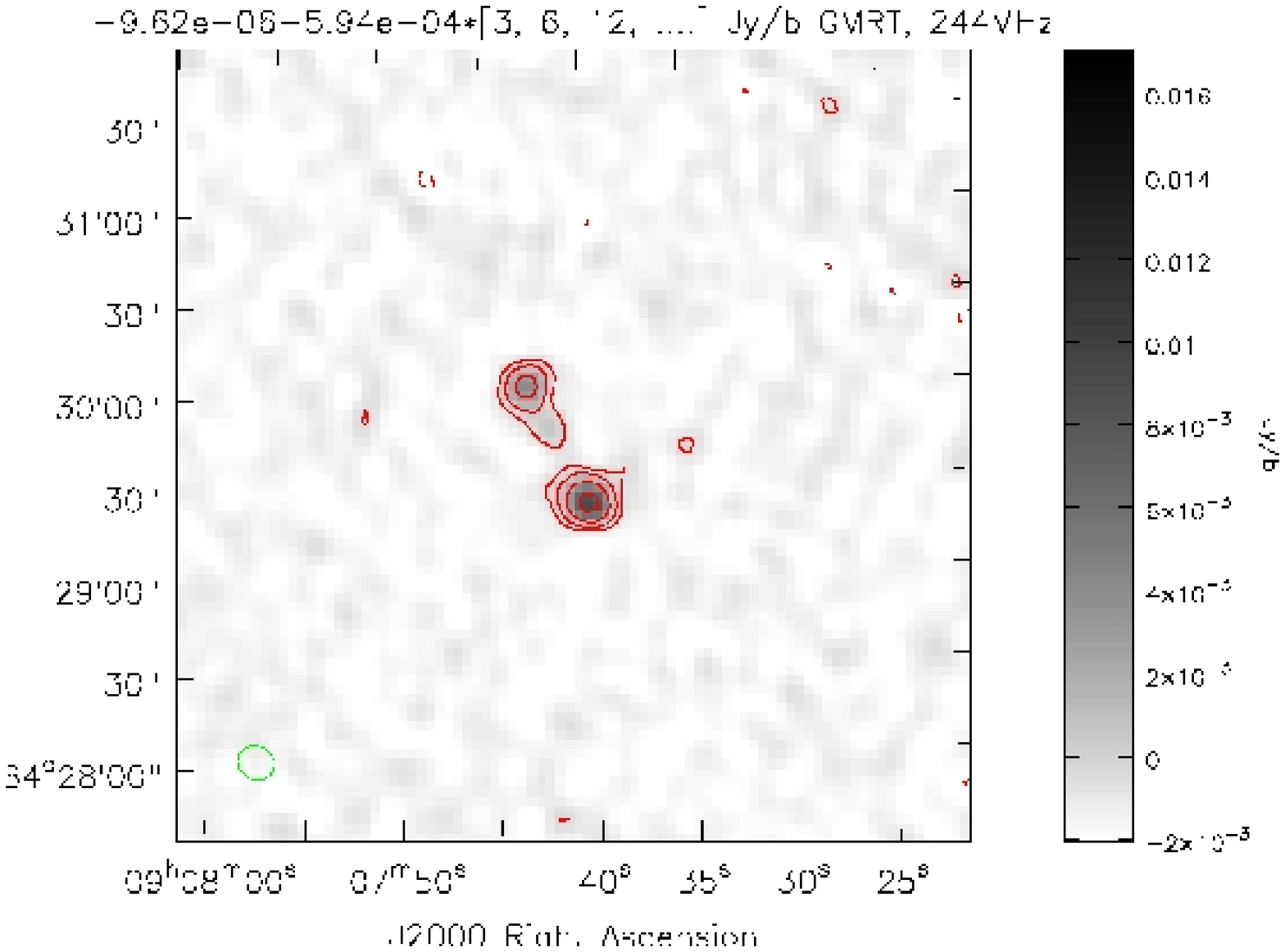}
  \includegraphics[angle=0, totalheight=1.8in, viewport=19 212 573 627, clip]{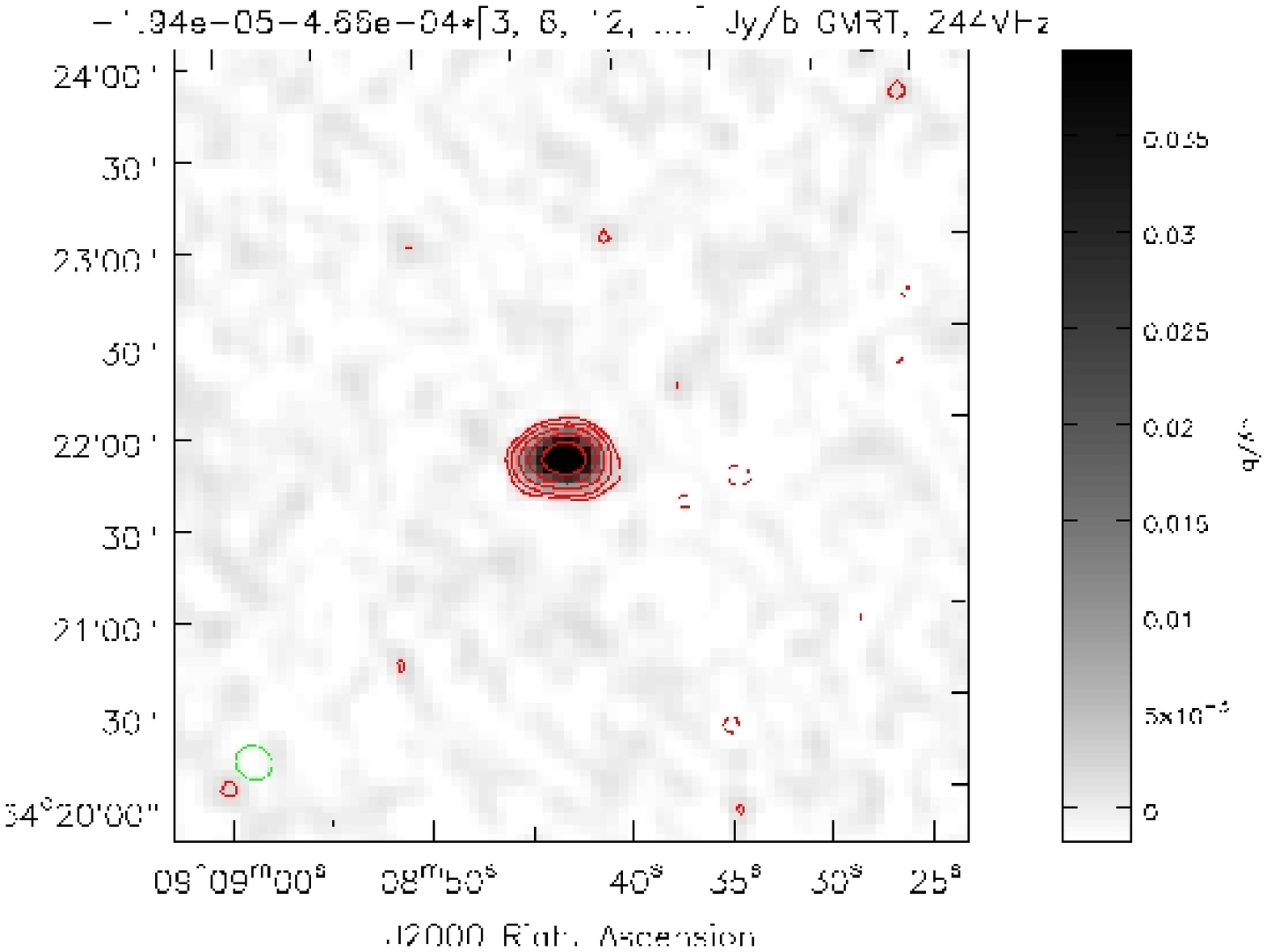}
 }
 \hbox{\hskip 0.9cm GMRT091133+630623 \hskip 3.2cm GMRT091308+633945 \hskip 3.2cm GMRT091328+640209}
 \hbox{
  \includegraphics[angle=0, totalheight=1.8in, viewport=19 212 573 627, clip]{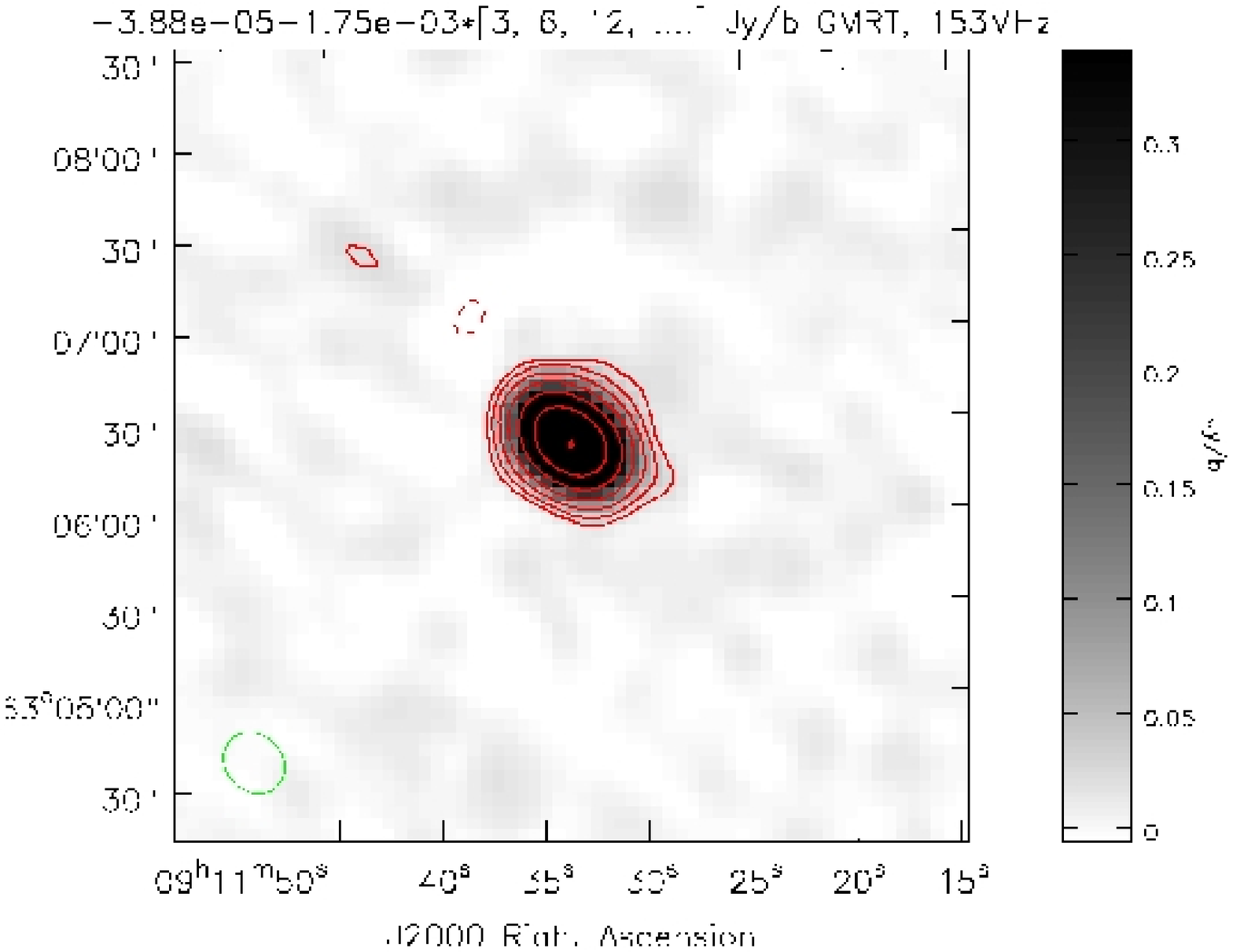}
  \includegraphics[angle=0, totalheight=1.8in, viewport=19 212 573 627, clip]{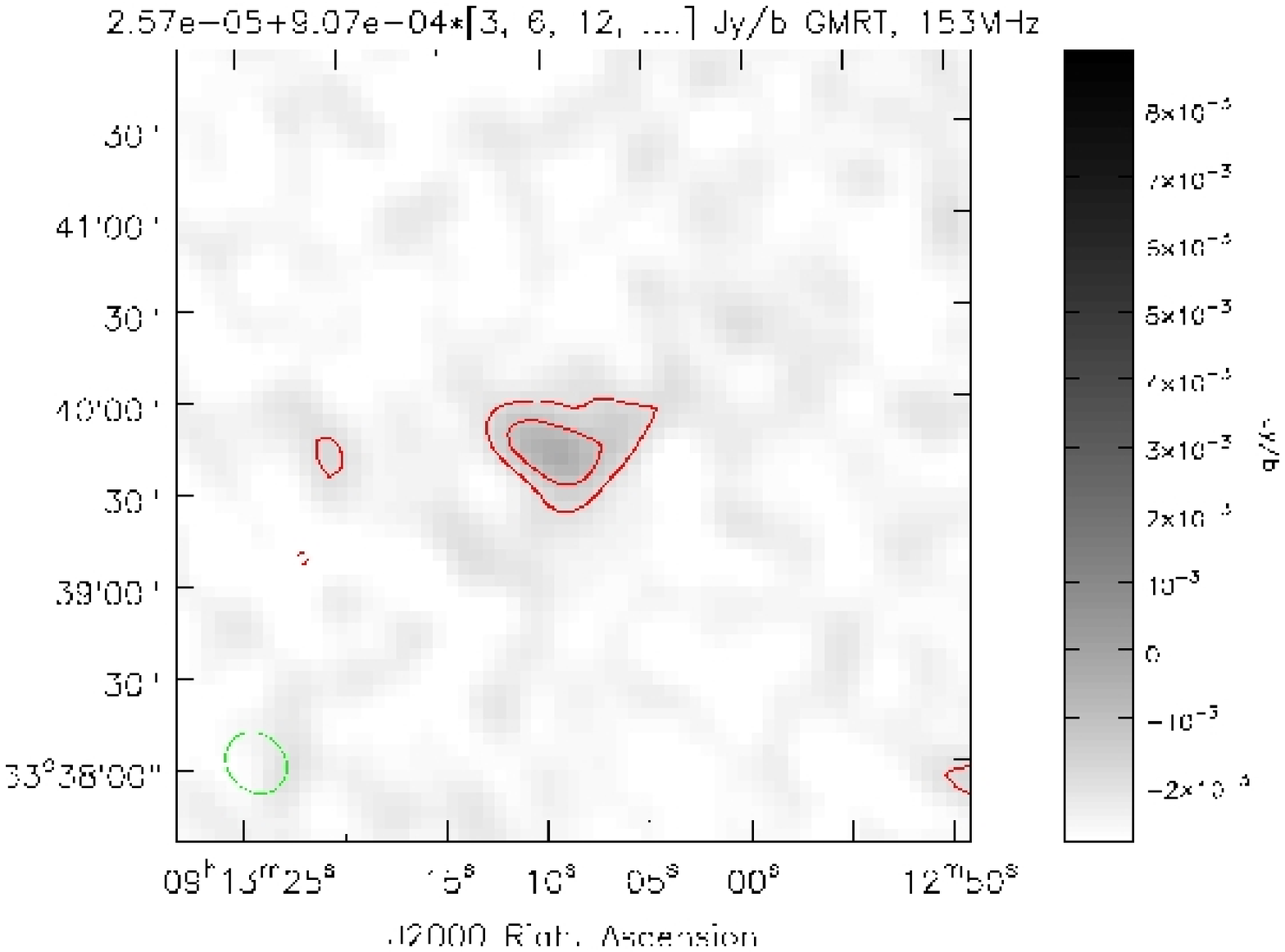}
  \includegraphics[angle=0, totalheight=1.8in, viewport=19 212 573 627, clip]{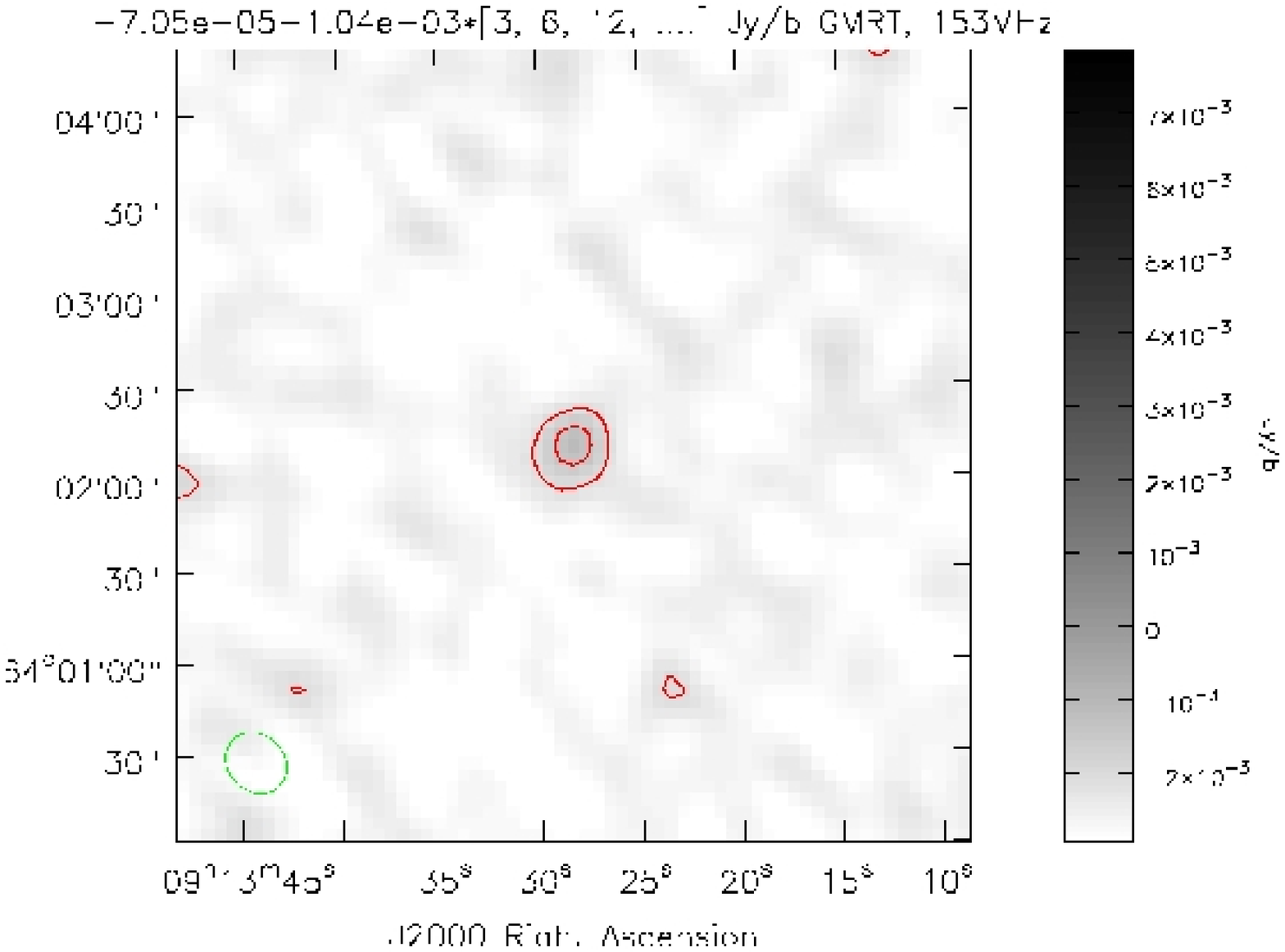}
 }
 \hbox{
  \includegraphics[angle=0, totalheight=1.8in, viewport=19 212 573 627, clip]{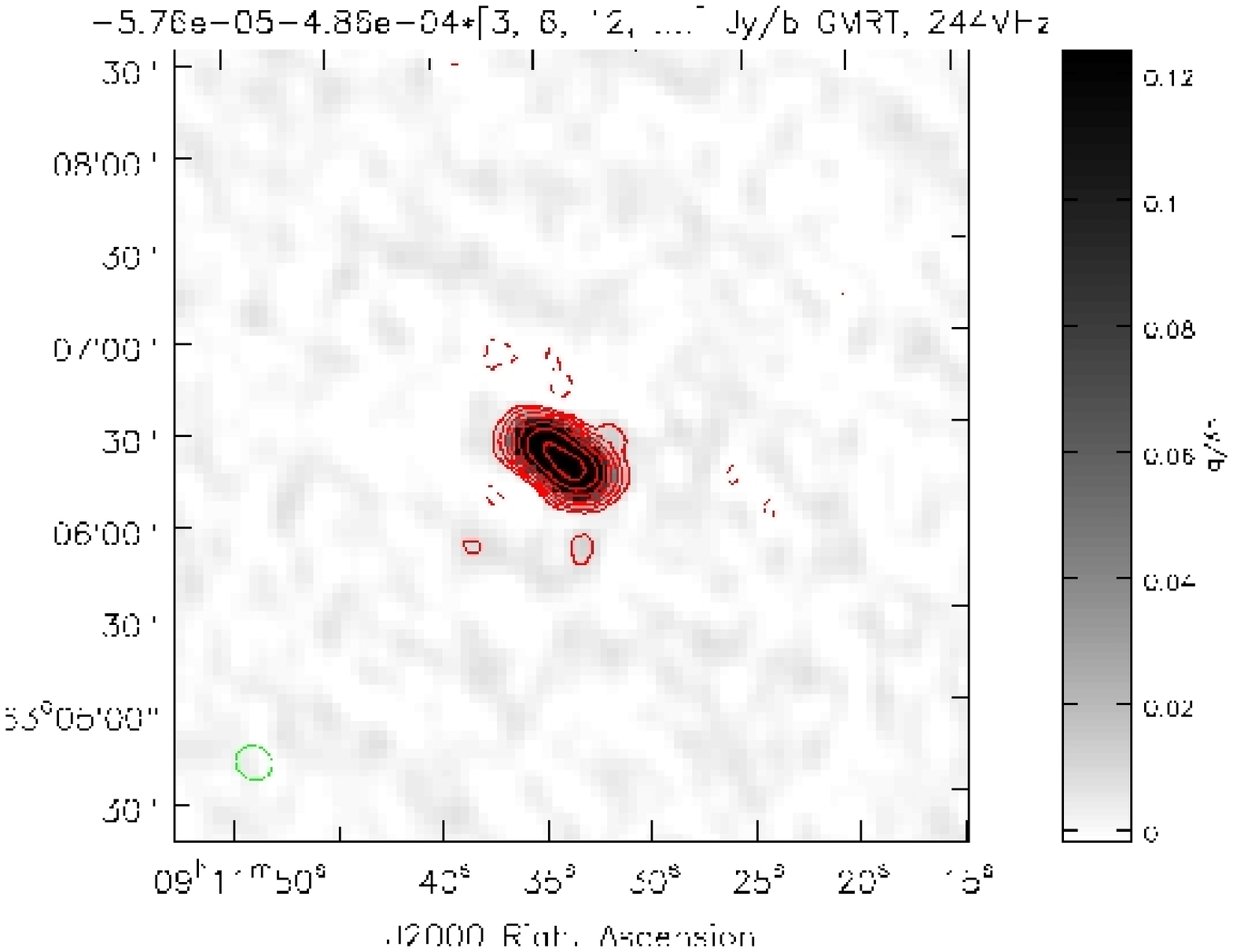}
  \includegraphics[angle=0, totalheight=1.8in, viewport=19 212 573 627, clip]{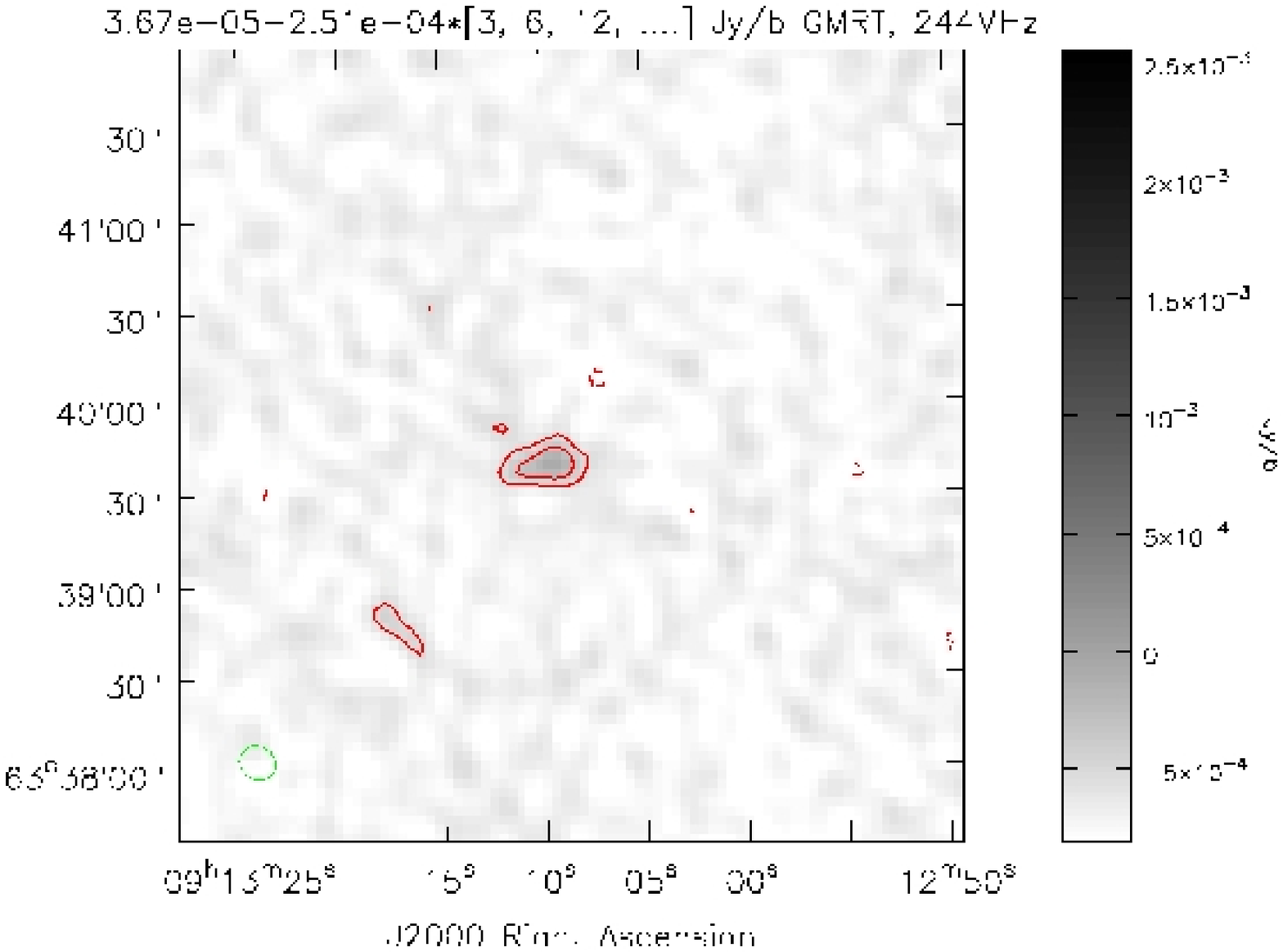}
  \includegraphics[angle=0, totalheight=1.8in, viewport=19 212 573 627, clip]{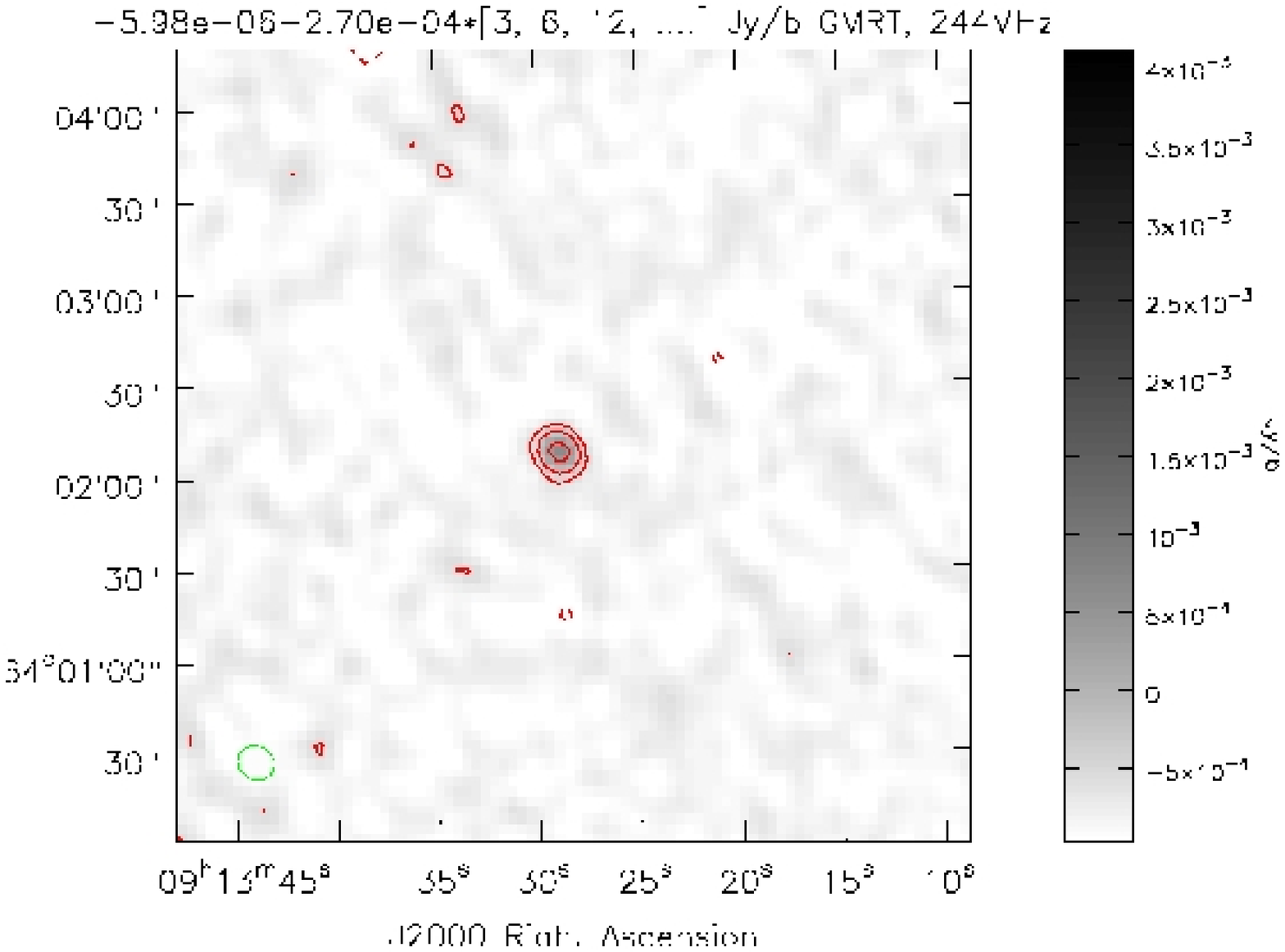}
 }
}
\caption{The GMRT images of the very steep spectrum objects at 153 and 244 MHz. The images of the 
sources at 153 MHz are in the first and third rows, while the corresponding images of the same sources at
244 MHz are in the second and fourth rows. Contour levels: mean$+$rms$\times$(n) in units of Jy beam$^{-1}$.}
\label{0916p6348_f0:fig:lfqima}
\end{center}
\end{figure*}

\begin{figure*}
\begin{center}
\vbox{
 \hbox{\hskip 0.9cm GMRT091734+640001 \hskip 3.2cm GMRT091744+630229 \hskip 3.2cm GMRT091845+642854}
 \hbox{
  \includegraphics[angle=0, totalheight=1.8in, viewport=19 212 573 627, clip]{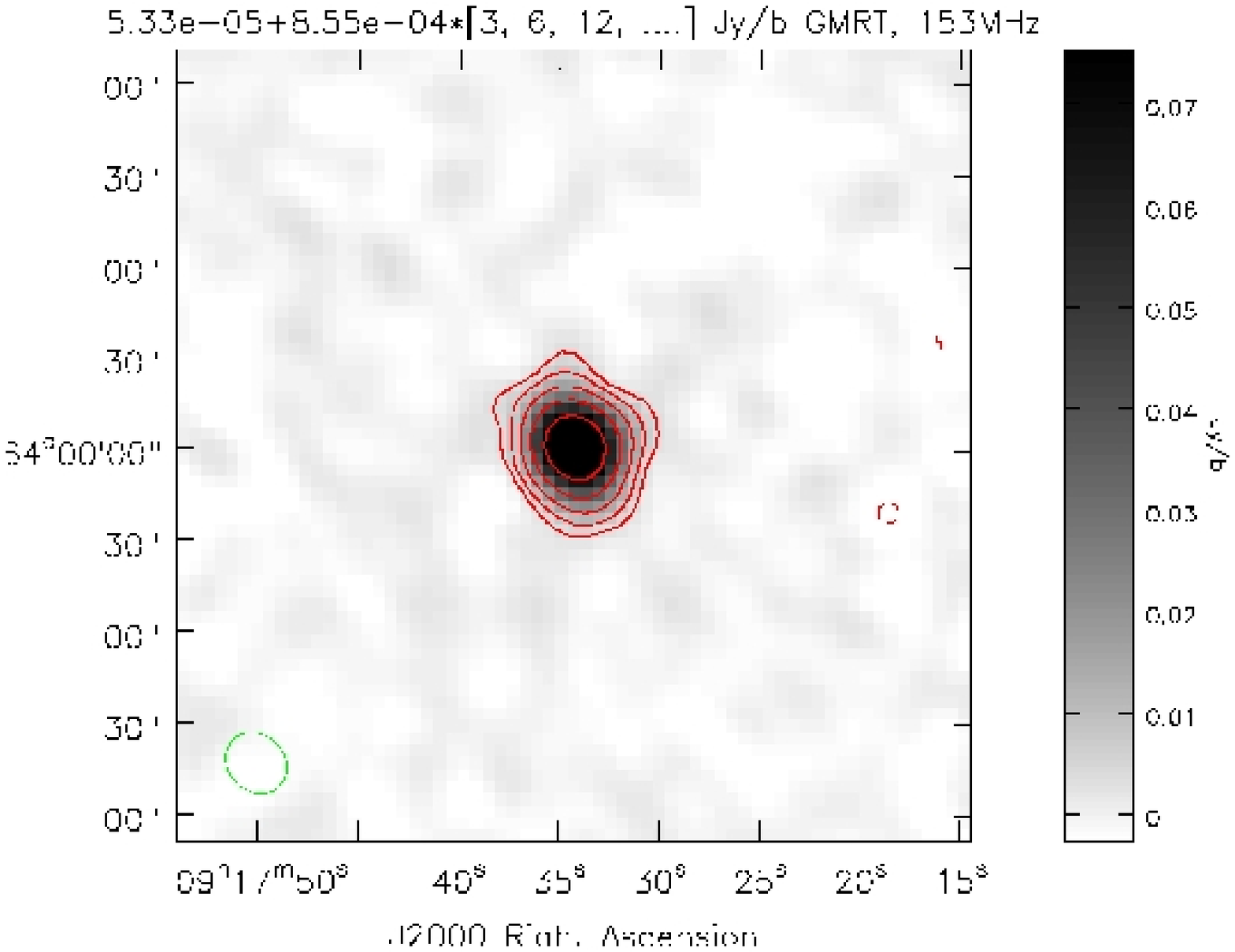}
  \includegraphics[angle=0, totalheight=1.8in, viewport=19 212 573 627, clip]{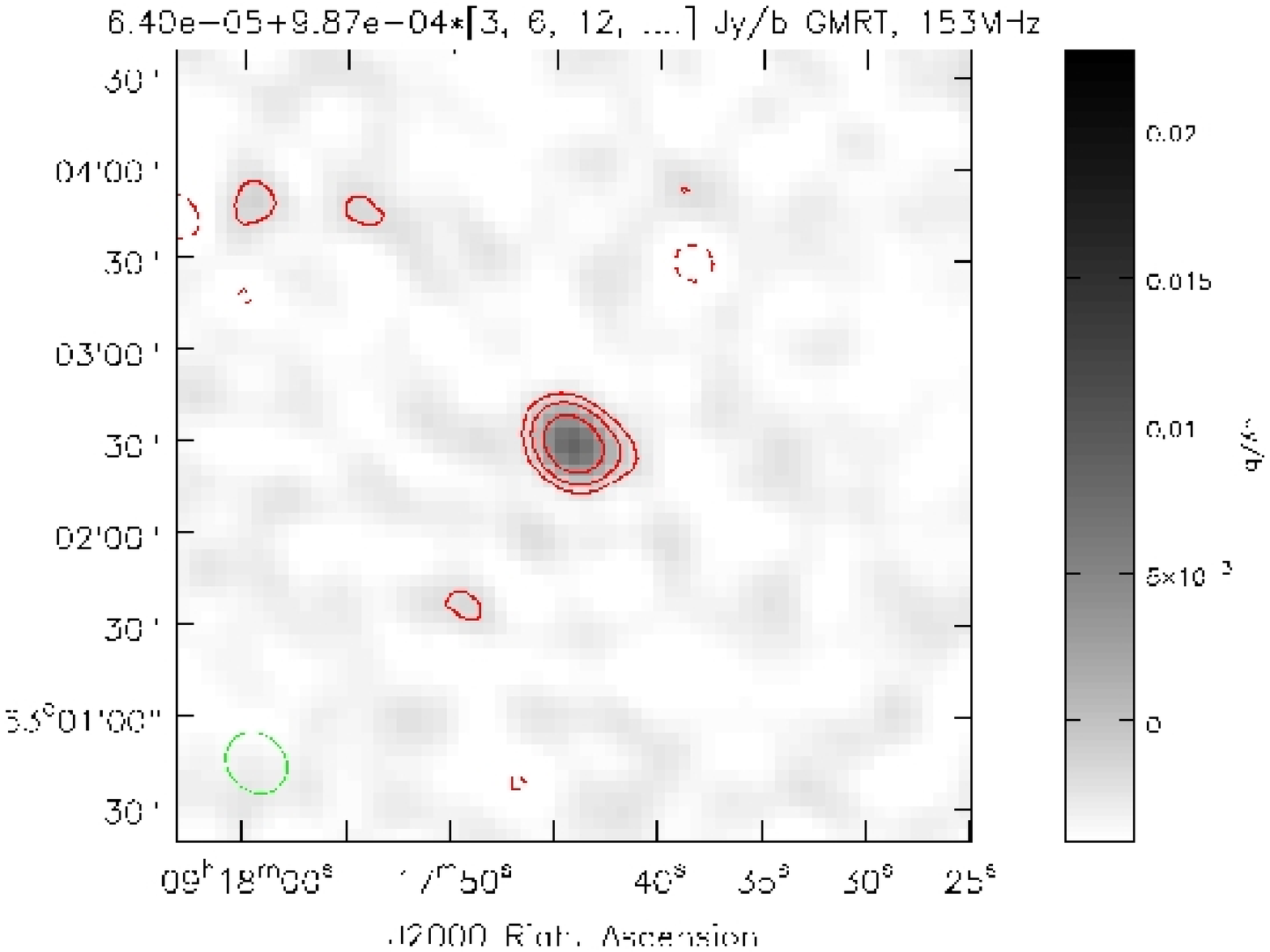}
  \includegraphics[angle=0, totalheight=1.8in, viewport=19 212 573 627, clip]{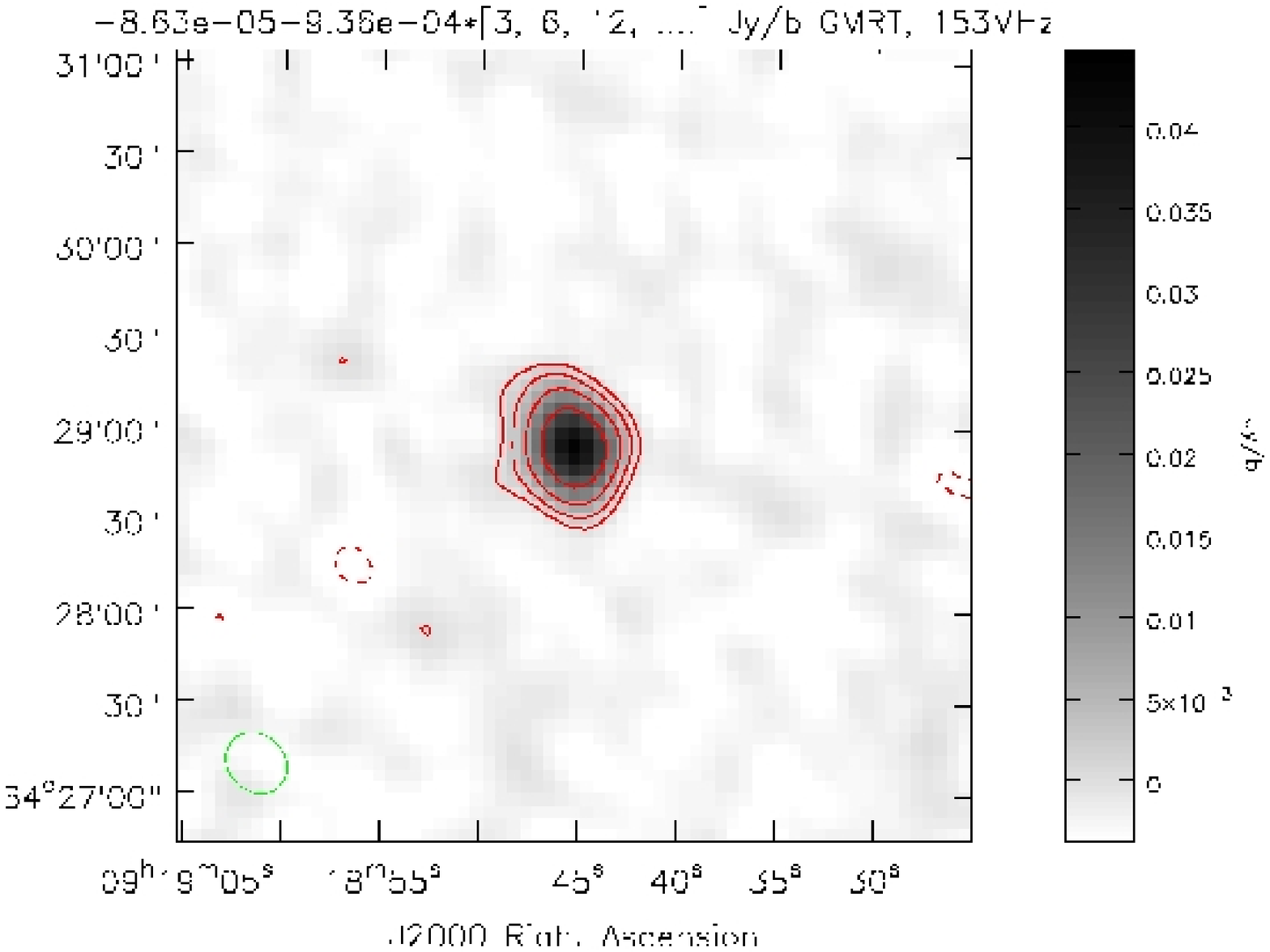}
 }
 \hbox{
  \includegraphics[angle=0, totalheight=1.8in, viewport=19 212 573 627, clip]{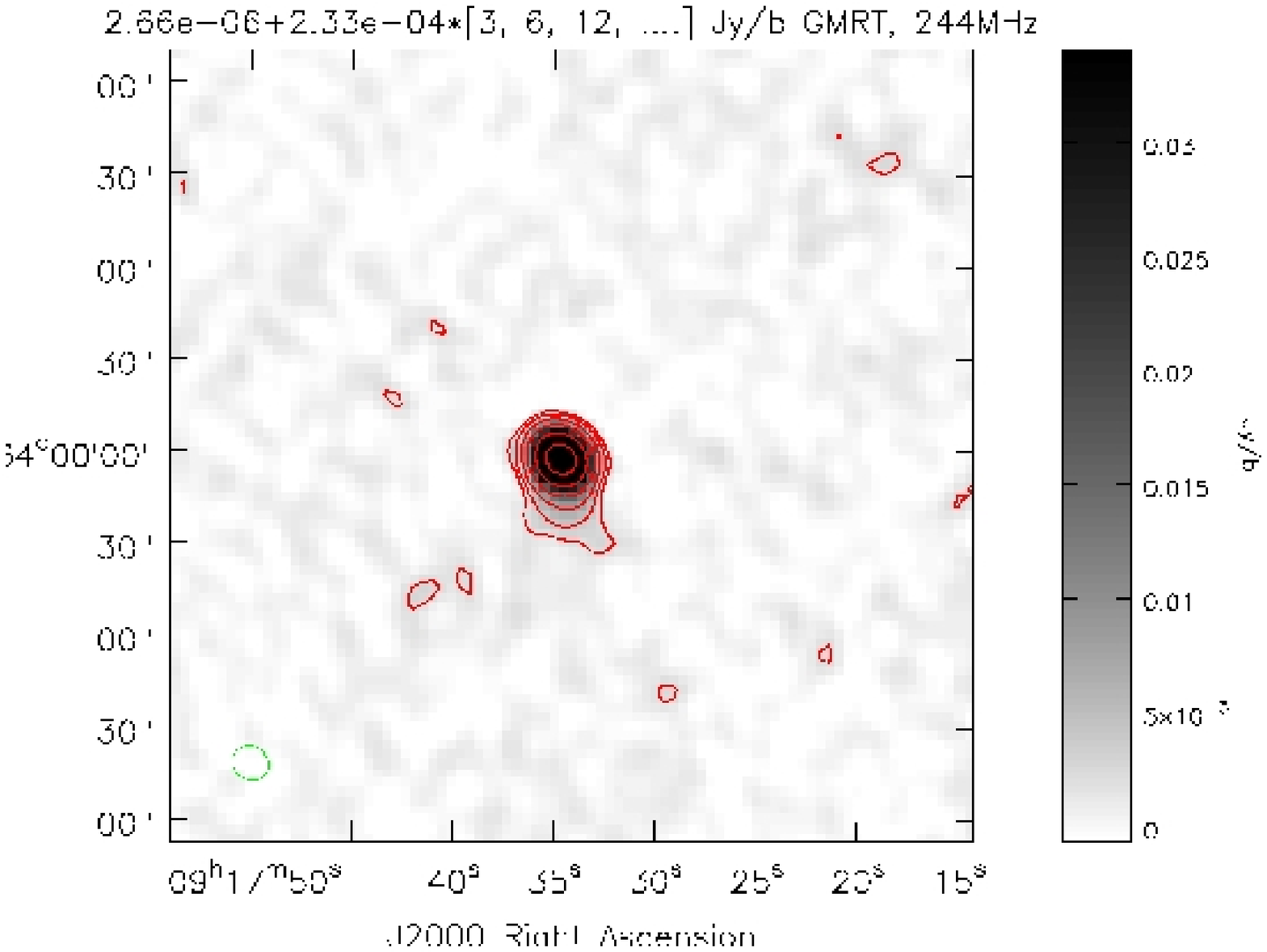}
  \includegraphics[angle=0, totalheight=1.8in, viewport=19 212 573 627, clip]{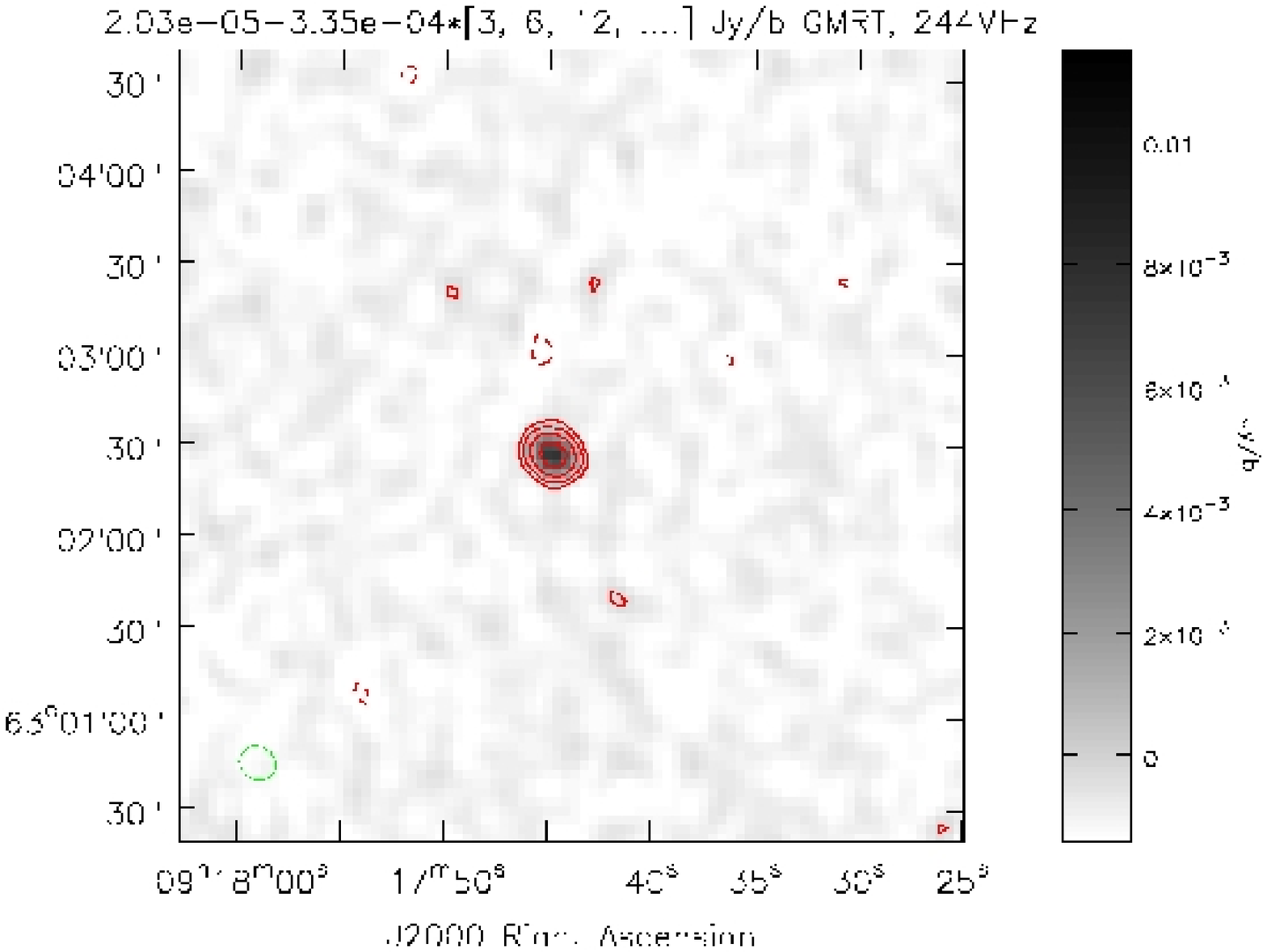}
  \includegraphics[angle=0, totalheight=1.8in, viewport=19 212 573 627, clip]{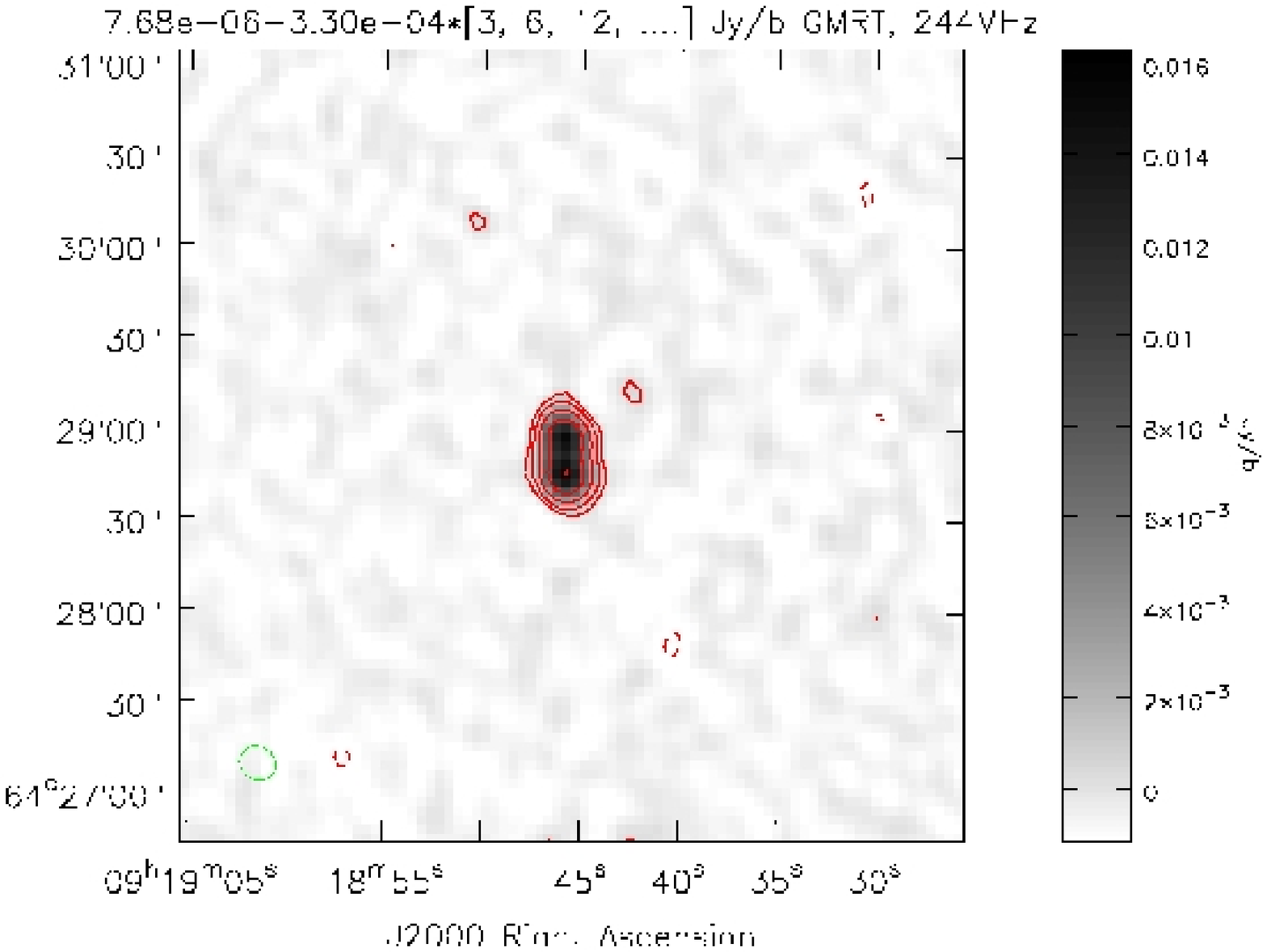}
 }
 \hbox{\hskip 0.9cm GMRT092004+633626 \hskip 3.2cm GMRT092009+641550 \hskip 3.2cm GMRT092131+633939}
 \hbox{
  \includegraphics[angle=0, totalheight=1.8in, viewport=19 212 573 627, clip]{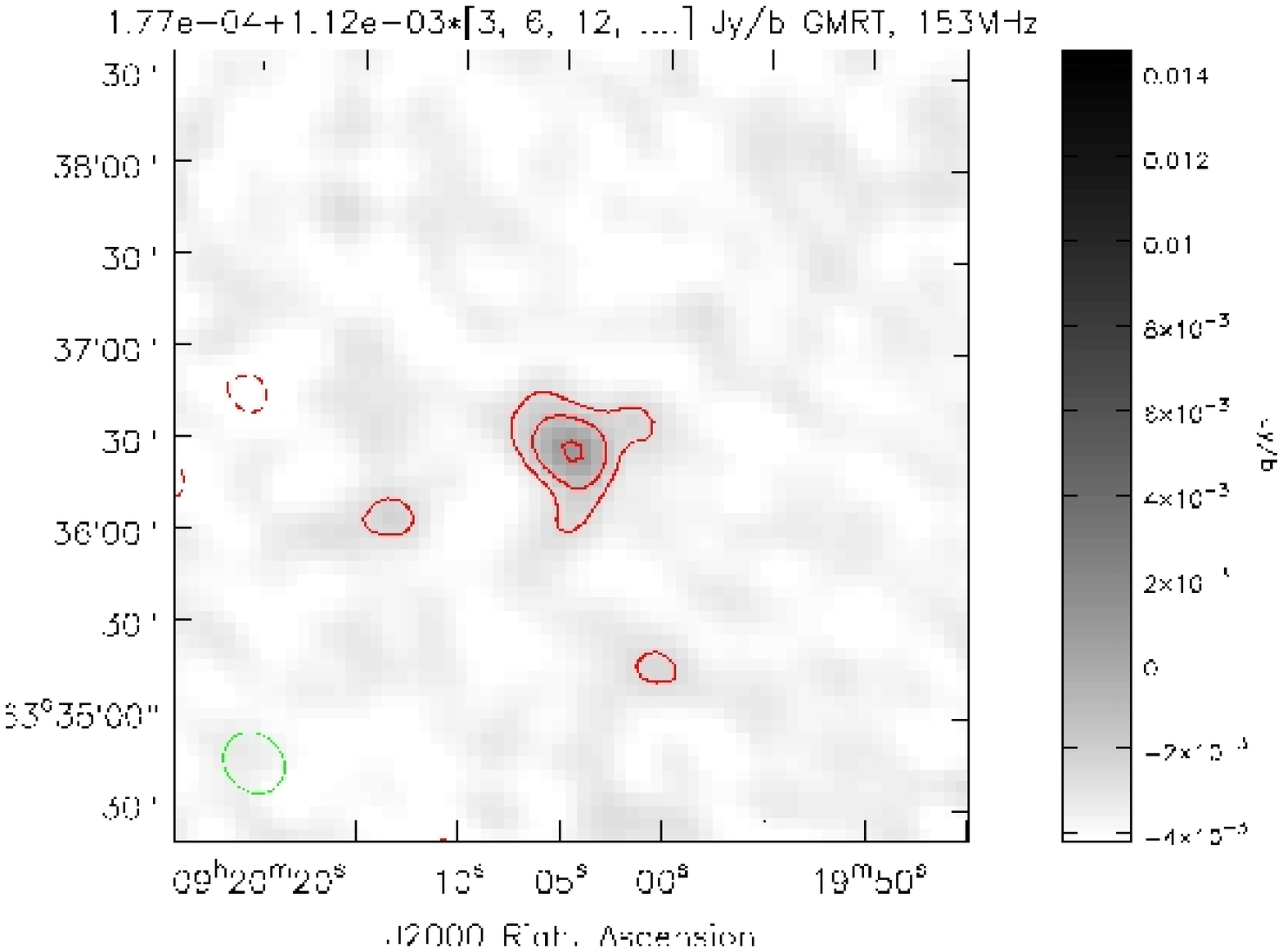}
  \includegraphics[angle=0, totalheight=1.8in, viewport=19 212 573 627, clip]{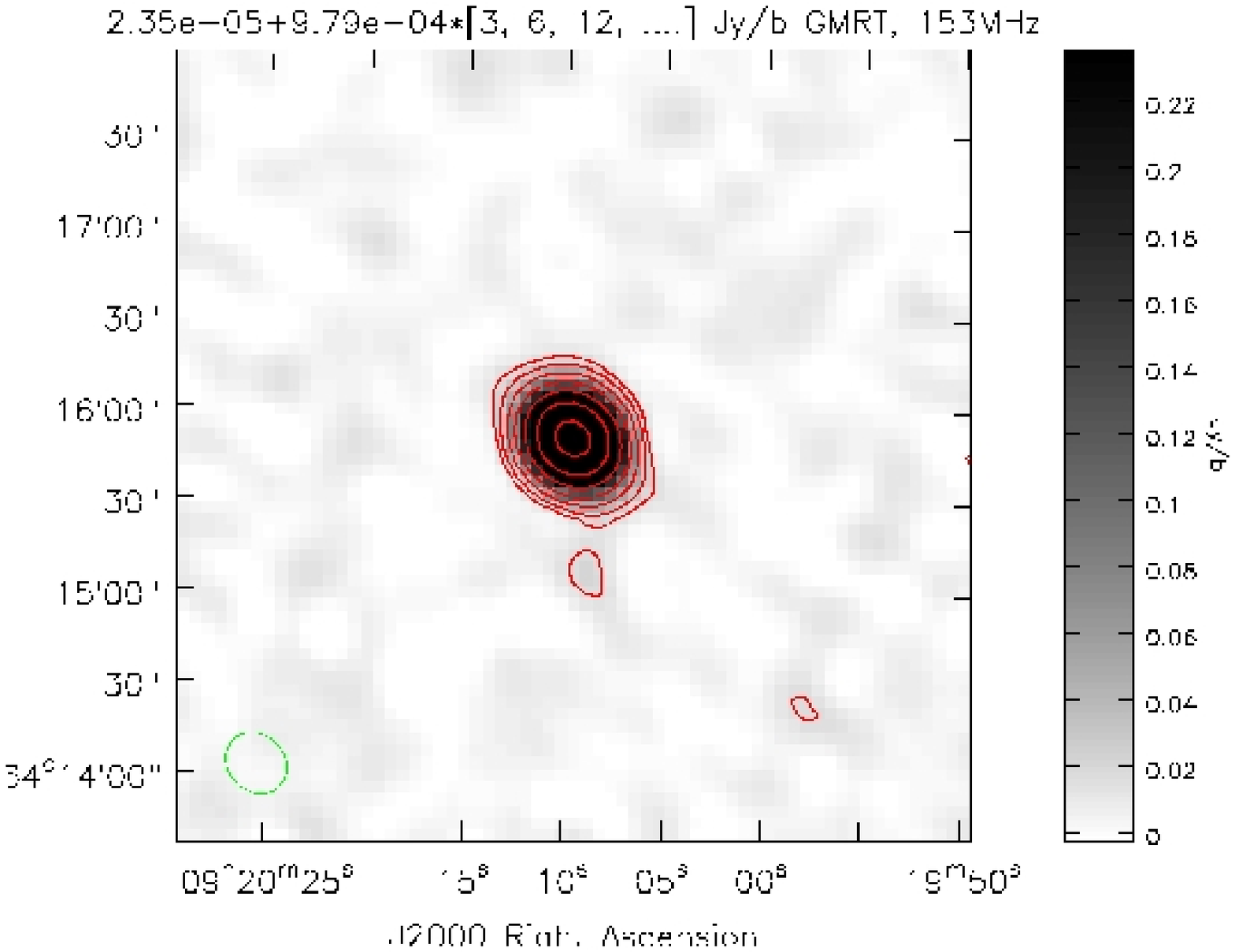}
  \includegraphics[angle=0, totalheight=1.8in, viewport=19 212 573 627, clip]{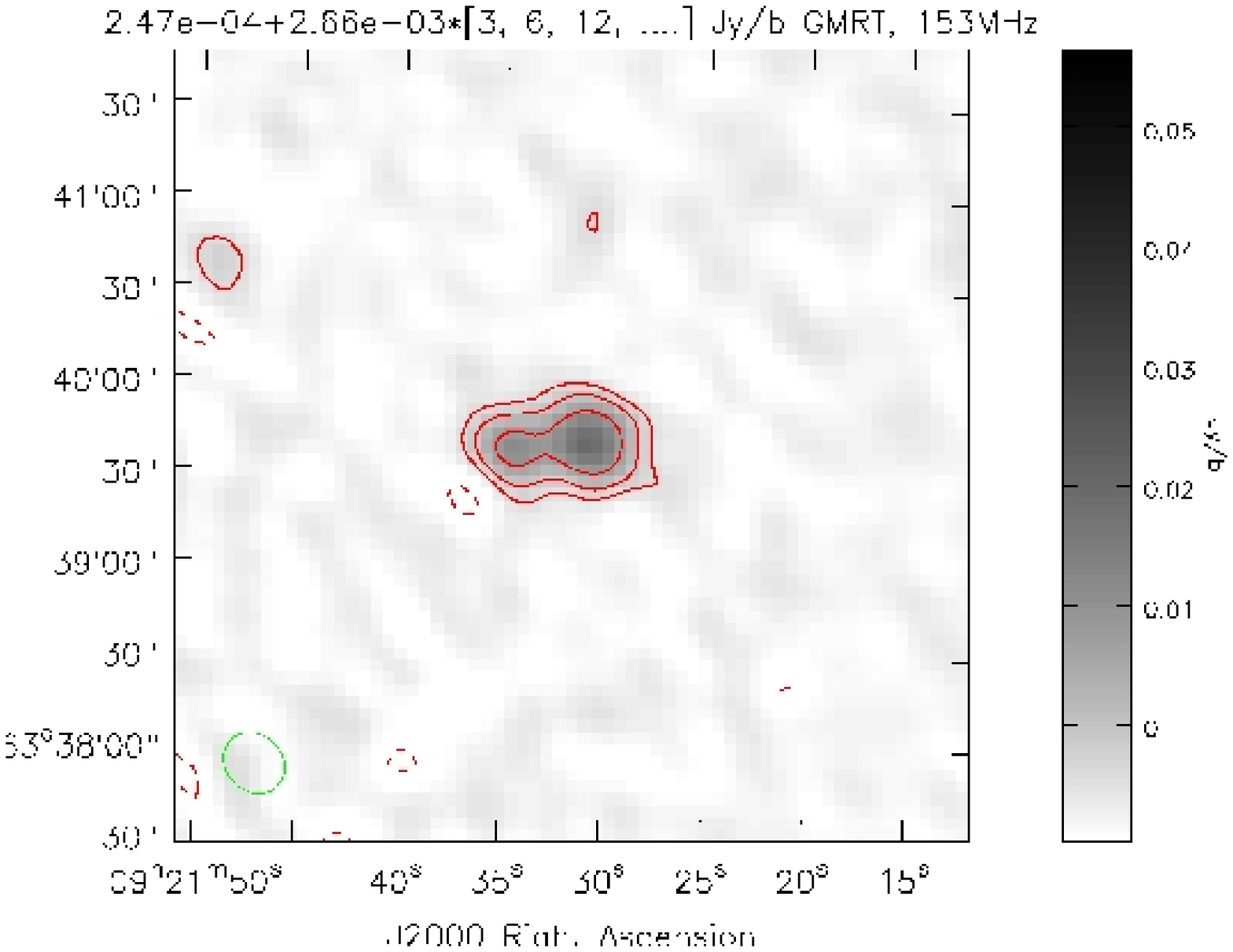}
 }
 \hbox{
  \includegraphics[angle=0, totalheight=1.8in, viewport=19 212 573 627, clip]{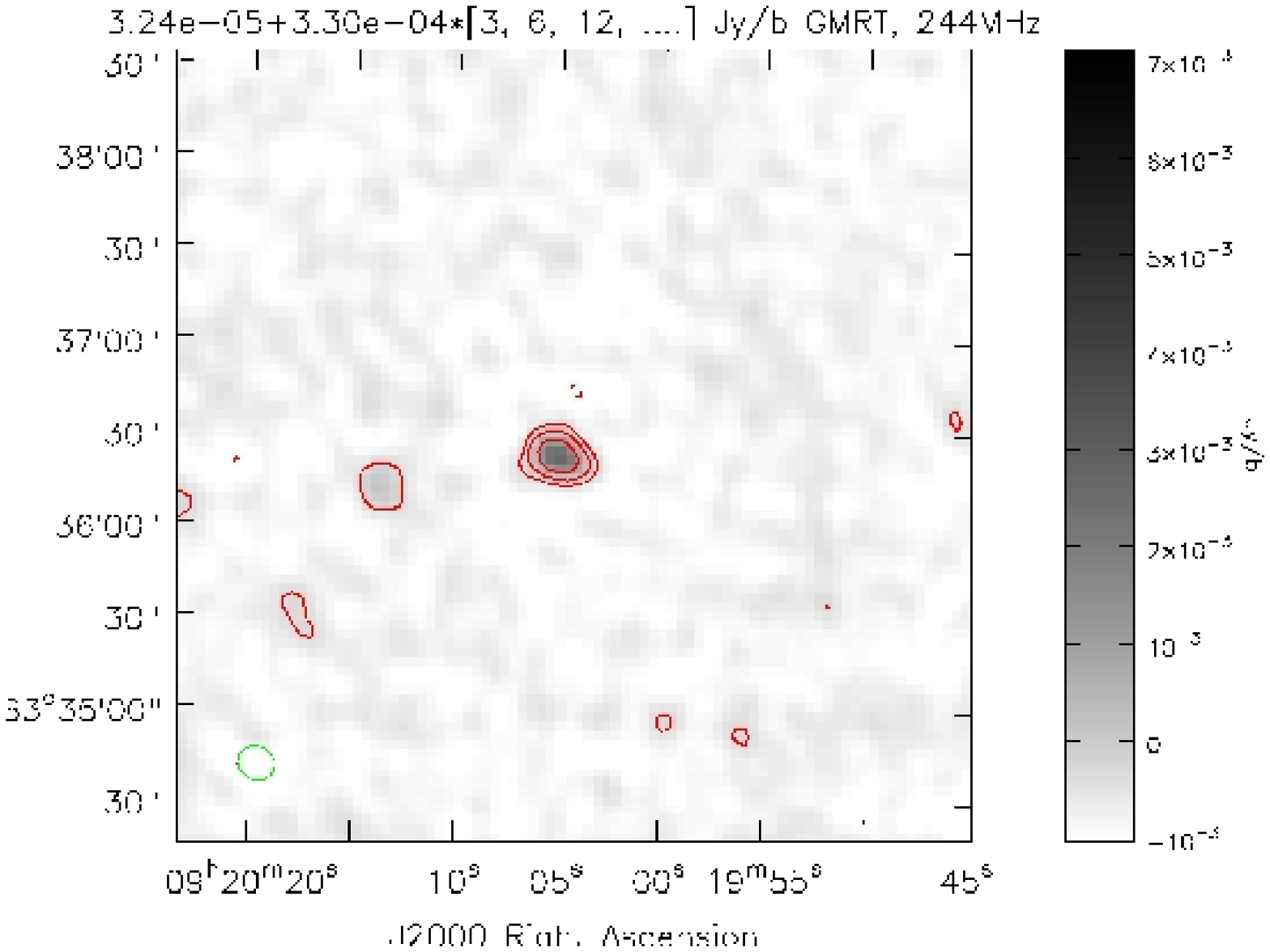}
  \includegraphics[angle=0, totalheight=1.8in, viewport=19 212 573 627, clip]{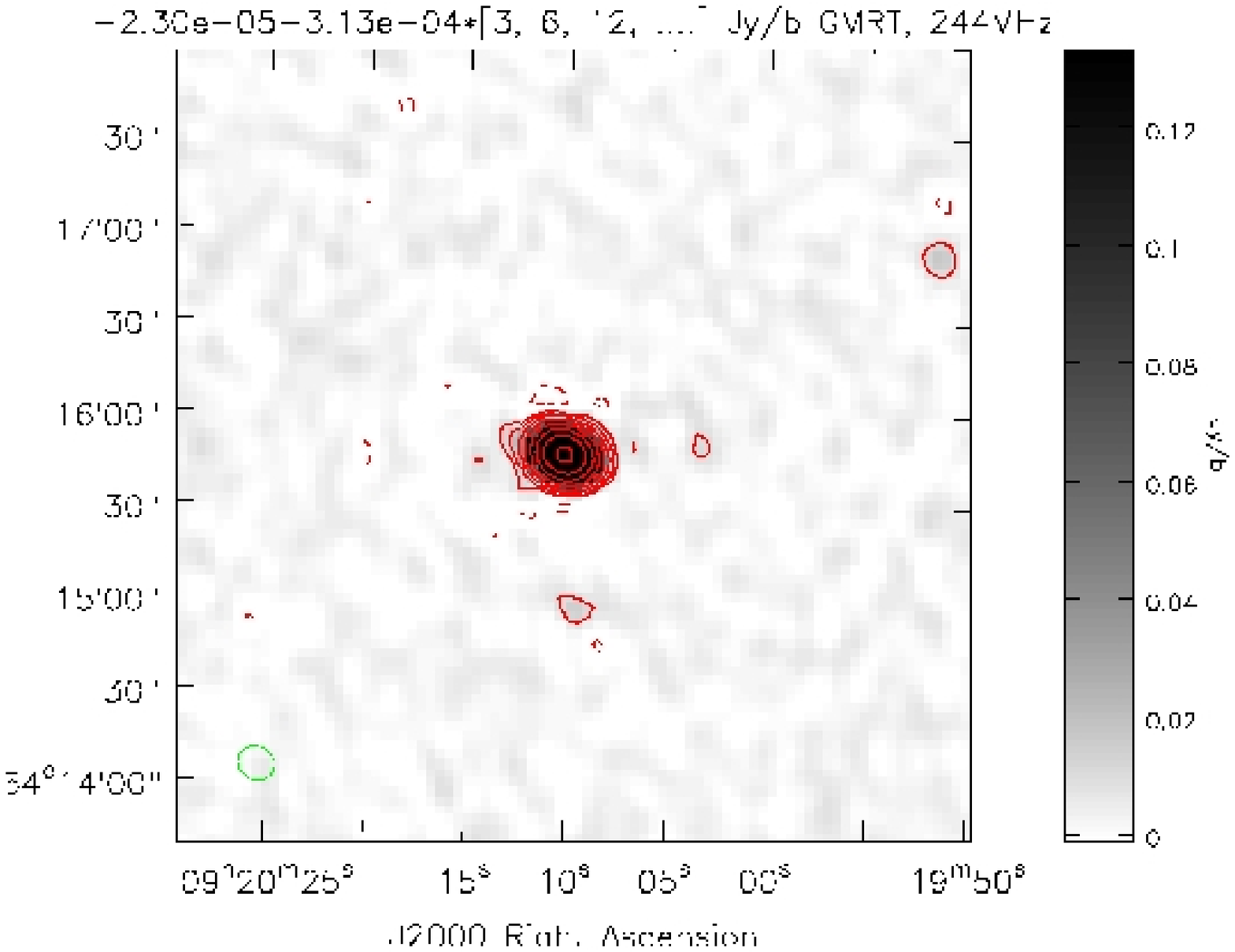}
  \includegraphics[angle=0, totalheight=1.8in, viewport=19 212 573 627, clip]{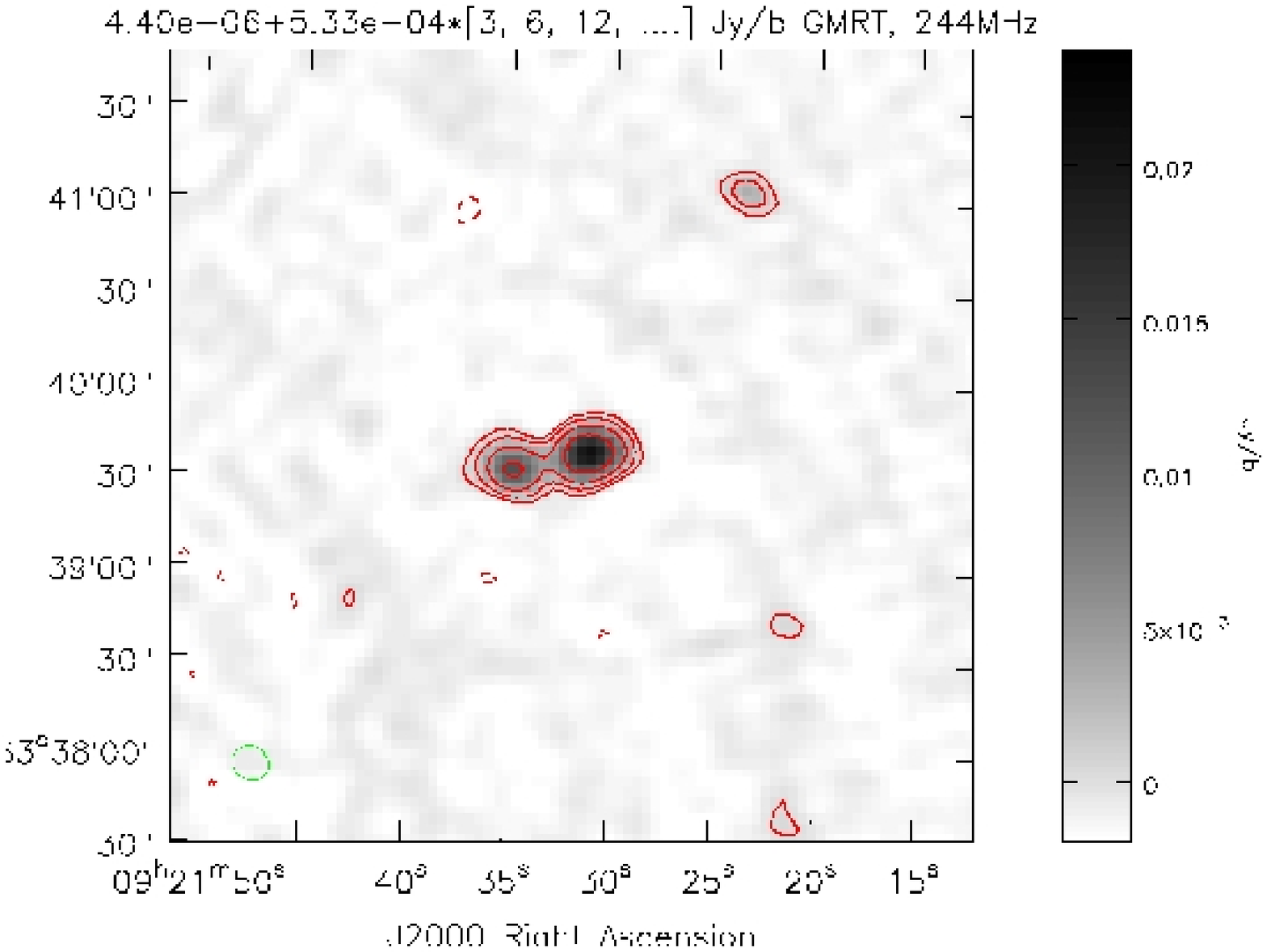}
 }
}
\contcaption{ }
\label{0916p6348_f0:fig:lfqimb}
\end{center}
\end{figure*}

\begin{figure*}
\begin{center}
\vbox{
 \hbox{\hskip 0.9cm GMRT092332+633611 \hskip 3.2cm GMRT092830+634748}
 \hbox{
  \includegraphics[angle=0, totalheight=1.8in, viewport=19 212 573 627, clip]{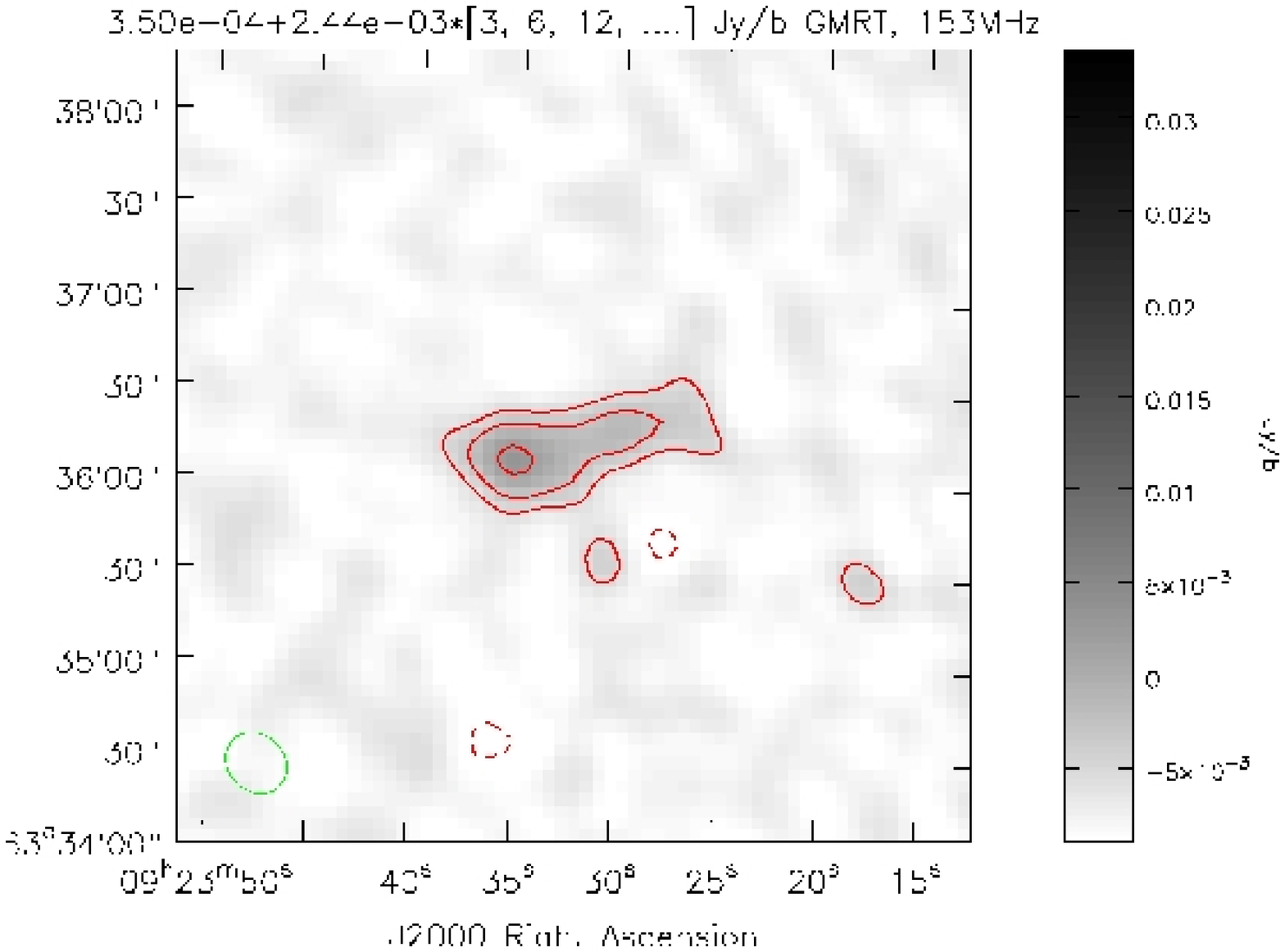}
  \includegraphics[angle=0, totalheight=1.8in, viewport=19 212 573 627, clip]{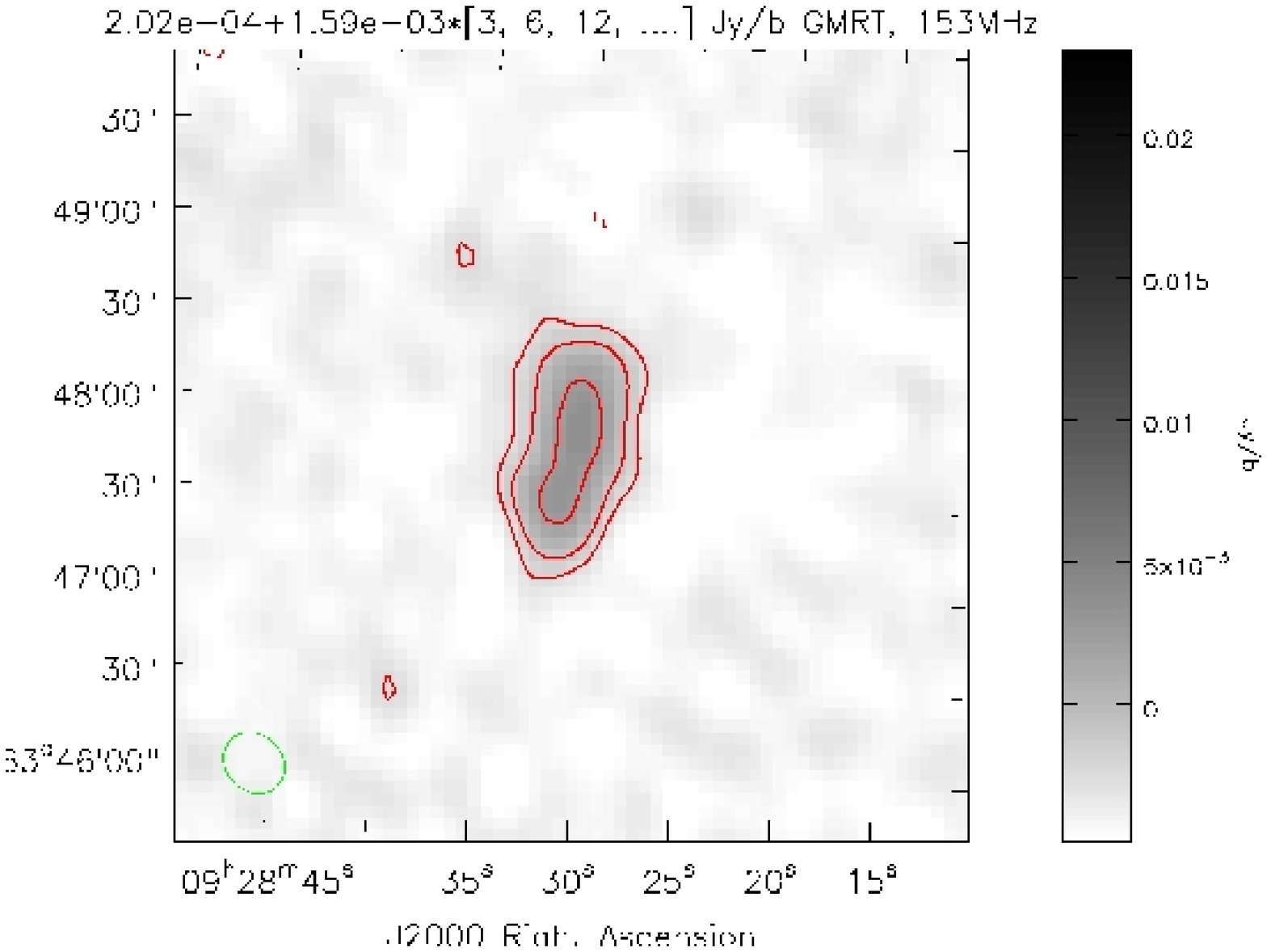}
 }
 \hbox{
  \includegraphics[angle=0, totalheight=1.8in, viewport=19 212 573 627, clip]{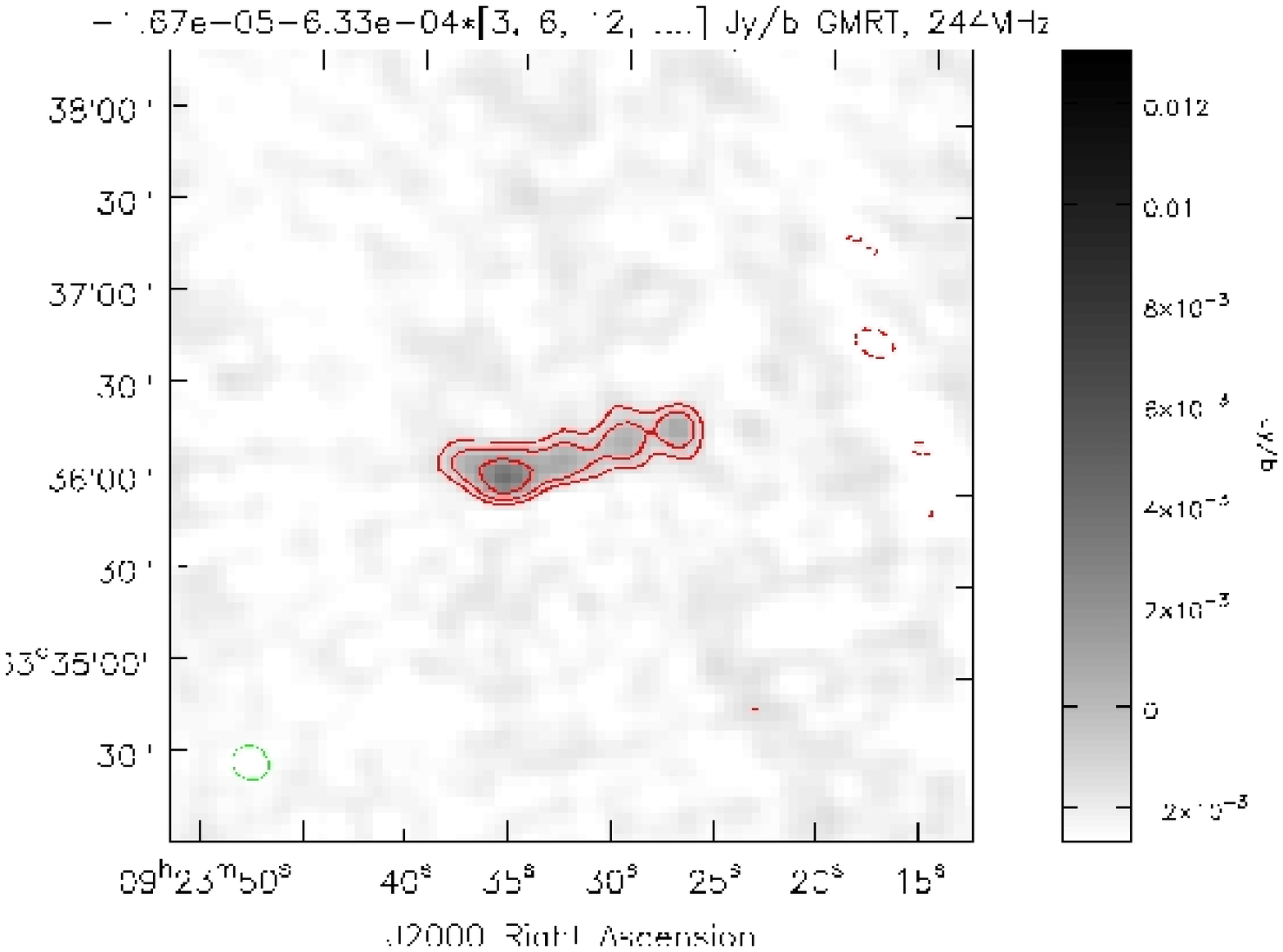}
  \includegraphics[angle=0, totalheight=1.8in, viewport=19 212 573 627, clip]{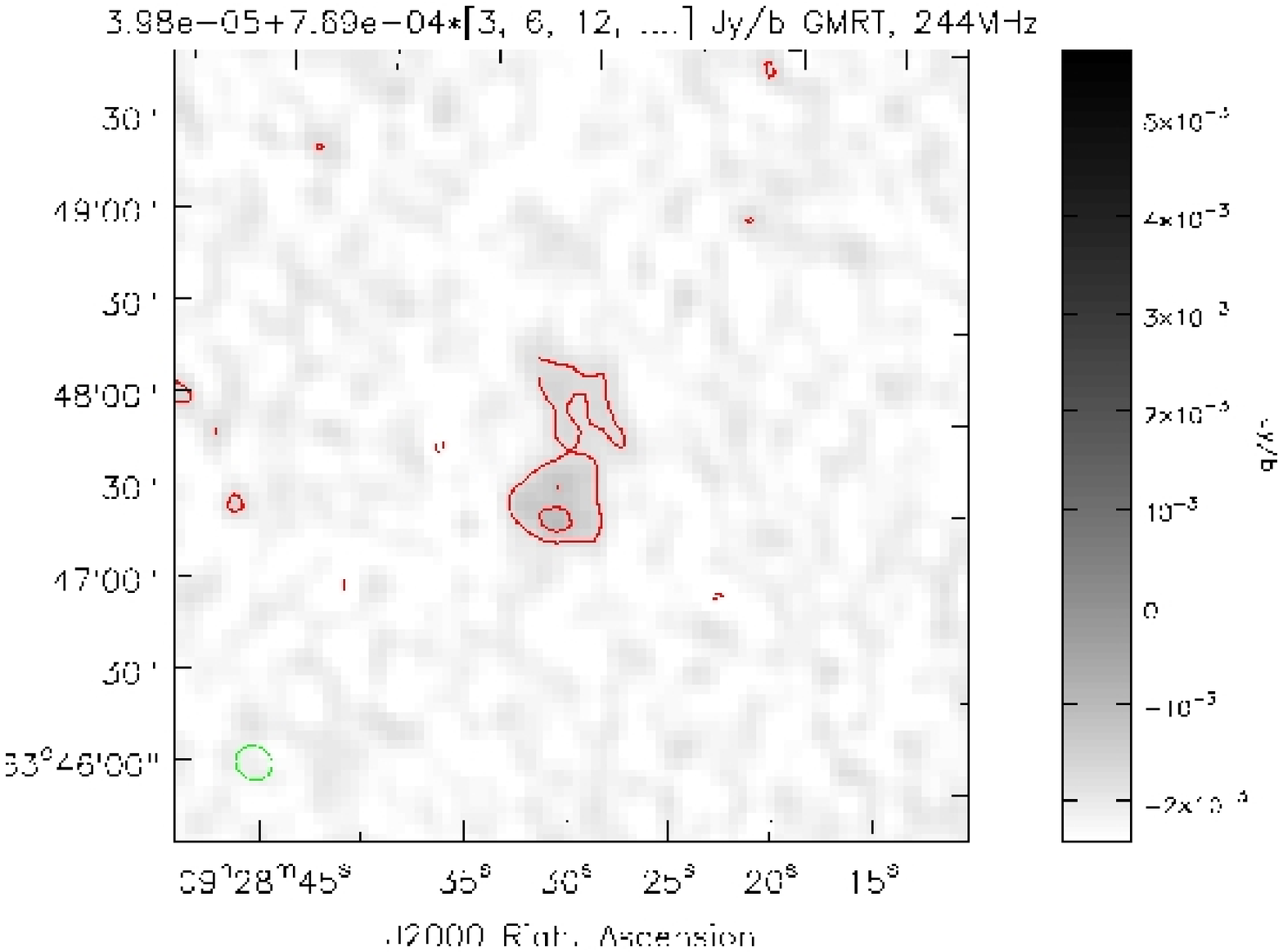}
 }
}
\contcaption{}
\label{0916p6348_f0:fig:lfqimc}
\end{center}
\end{figure*}


\subsection{Very steep spectrum sources}
The relic lobes of radio sources from an earlier cycle of activity which are 
no longer being fed with energy from the nucleus are expected to have a 
very steep radio spectrum due to radiative losses. For a source with a typical 
spectral index 
of 0.8, the expected spectral index beyond the break due to synchrotron ageing
is expected to be about 
1.3 \citep{1970ranp.book.....P, 2007A+A...470..875P}.
The 14 objects from our list with at least three measurements, and 
spectral indices $>$1.3 are listed in Table~\ref{0916p6348_f0:table:ss}. We
have confirmed the steep spectra of these sources by convolving the higher
resolution images to those of the lower resolution ones. This shows
that the number of very steep spectrum sources comprises only $\sim$3.7 per
cent of our entire list of 374 objects. There have been earlier searches for 
such steep spectrum sources with the primary objective of finding high-redshift
radio galaxies \citep[e.g.][]{1994A+AS..108...79R, 2000A+AS..143..303D}. These
studies also indicate that very steep spectrum sources are rather rare. For 
example, in the study by \cite{2000A+AS..143..303D}, only 0.5 per cent of a
complete low-frequency selected sample is found to have a spectral index $>$1.3.
Our detection rate is somewhat higher because our pointing centre is close to two
clusters of galaxies. Radio galaxies in clusters  are known  to have 
steeper spectra \citep[e.g.][]{1983AuJPh..36..101S, 1985A+A...148..323R, 1984PASAu...5..516S},
possibly due to confinement of the radio emitting plasma by the intracluster
medium. Also, clusters are known to often harbour steep-spectrum relics and
halos of emission 
(cf. \citeauthor{2008A+A...484..327V}~\citeyear{2008A+A...484..327V} and references therein).

These sources exhibit
structures ranging from single sources to double-lobed and diffuse extended
sources. The GMRT images at 153 and 244 MHz of six of these sources are shown
in Fig.~\ref{0916p6348_f0:fig:lfqima}.
We have examined the optical fields of each of the 14 sources in the
Sloan Digital Sky Survey \citep{2008ApJS..175..297A}. We have tried to find optical counterparts 
of these sources as well attempt to identify any cluster of galaxies that might be
associated with the radio sources. We have plotted the positions of the galaxies
from the SDSS on the radio images, and have described it as a possible identification 
if the optical object lies within the radio source and is within a few arcsec of 
the radio centroid. 

Of the 14 sources, three (GMRT091308+633945, GMRT092009+641550 and GMRT092131+633939) 
have possible identifications, and three of the remaining sources 
(GMRT091734+640001, GMRT091845+0642854, GMRT092830+0634748)
are in the directions of clusters of galaxies.  These are described briefly here.

GMRT091308+633945: This is a single point source at 610 MHz. The nearest optical galaxy
 from the SDSS catalogue is J091309.5+633943.0, which is 0.9 arcsec from radio source at 610 MHz.
 Its r magnitude is 22 and its estimated redshift ranges from 0.1 to 0.77.

GMRT091734+640001: This source has an extension toward the south with an overall 
size of 19.2 arcsec at 610 MHz. The object is at distance of 10.9 arcmin from the cluster Abell 0764 
cluster which is centered at 09$^{\rm h}$17$^{\rm m}$09.7$^{\rm s}$, 63$^{\circ}$49$^{\prime}$26.4$^{\prime\prime}$.
The cluster diameter is about 28 arcmin and is at 
z=0.166 with an x-ray flux density in the 0.1-2.4keV band  of 0.904$\times10^{-15}$W/m$^2$, as listed in 
the BAX (Base de Donn\'ees Amas de Galaxies X) X-Ray Galaxy Clusters and Groups Catalog~\citep{2004A+A...424.1097S}.

GMRT091845+0642854: This is an double source with an angular size of 19.9 arcsec at 244 MHz.
 The object is at a distance of 11.2 arcmin from the Zwicky cluster \citep{2003AJ....125.2064G} ZW0916+6448 centered at 
09$^{\rm h}$20$^{\rm m}$11.3$^{\rm s}$, 64$^{\circ}$35$^{\prime}$16.8$^{\prime\prime}$. The cluster radius is 
0.2$^\circ$.

GMRT092009+641550: This is a single source with a deconvolved size  
of 3.92 arcsec from the FIRST catalogue. The nearest optical galaxy from the SDSS catalogue 
is J092010.8+641548.5, which is located about 2.7 arcsec from the FIRST position, has an r 
magnitude of 22.71 and an estimated photometric redshift of 0.85$\pm$0.67. 

GMRT092131+633939: This radio object is an asymmetric double with an angular size of 
53.6 arcsec at 244 MHz. The nearest optical galaxy from the SDSS catalogue is J092133.7+633930.7, 
which is 9.9 arcsec from the centroid and lies between the two radio components. Its r magnitude is 21.52
and its estimated redshift ranges from 0.42 to 0.59.

GMRT092830+0634748: This is a diffuse source with deconvolved size of 13.9 arcsec at 244 MHz.
The object is at a distance of 8.2 arcmin from the Zwicky cluster \citep{2003AJ....125.2064G} ZW0923+6407 centered at 
09$^{\rm h}$27$^{\rm m}$41.0$^{\rm s}$, 63$^{\circ}$53$^{\prime}$56.4$^{\prime\prime}$. The cluster radius is 
0.5$^\circ$.

\section{CONCLUDING REMARKS}
We have presented deep multifrequency GMRT observations 
at 153, 244, 610 and 1260 MHz of a field centered on J0916+6348. 

We do not find any unambiguous evidence of emission from an
earlier cycle of activity in a list of 374 sources,
suggesting that such activity is rare even in relatively 
deep low-frequency observations. However, the median size of
our objects is expected to be less than about a 100 kpc, 
which may be partly responsible for the non-detection of
fossil lobes from an earlier cycle of AGN activity.

By combining our flux density measurements with those of
WENSS, NVSS and FIRST we find a median 
spectral index of $\sim$0.8, which is similar to that of
the stronger 3CR sources, but appears steeper 
than theoretical expectations of the injection spectral index.

We identify 14 very steep-spectrum sources 
with a steep spectral index $\geq$1.3. We find three of these
sources to be identified with an optical galaxy visible in 
the SDSS, while another three are in the directions of clusters
of galaxies. One of these three,  
GMRT092830+0634748, is a diffuse extended source and could
be a cluster relic.

\section*{Acknowledgments}
We thank the anonymous referee for his/her helpful comments and
the staff of the GMRT who have made these observations possible. 
GMRT is run by the National Centre for Radio Astrophysics of the Tata Institute of Fundamental Research. 
This research has made use of the NASA/IPAC Extragalactic Database (NED) which is operated by 
the Jet Propulsion Laboratory, California Institute of Technology, under contract with the National 
Aeronautics and Space Administration.

\bibliography{../bib/ref.bib} 
\bibliographystyle{./mn2e.bst}

\begin{table*}
\setcounter{table}{1}
\caption{The Table for the entire list of sources with spectral index information (available in online version only).}
\label{0916p6348_f0:table:ot}
\begin{tabular}{l r r r r r r r r r r l}
\hline
\multicolumn{1}{c}{Source name} & \multicolumn{1}{c}{RA} & \multicolumn{1}{c}{DEC} & \multicolumn{1}{c}{Dis} & \multicolumn{1}{c}{S$_{153}$} & \multicolumn{1}{c}{S$_{244}$}
                    & \multicolumn{1}{c}{S$_{330}$} & \multicolumn{1}{c}{S$_{610}$} & \multicolumn{1}{c}{S$_{1260}$} & \multicolumn{1}{c}{S$_{1400}$}
                    & \multicolumn{1}{c}{$\alpha$} &\multicolumn{1}{c}{Class} \\
                    & \multicolumn{1}{c}{hh:mm:ss.s} & \multicolumn{1}{c}{dd:mm:ss.s} & \multicolumn{1}{c}{deg.} & \multicolumn{1}{c}{mJy} & \multicolumn{1}{c}{mJy}
                    & \multicolumn{1}{c}{mJy} & \multicolumn{1}{c}{mJy} & \multicolumn{1}{c}{mJy} & \multicolumn{1}{c}{mJy} &  &        \\
\multicolumn{1}{c}{(1)} & \multicolumn{1}{c}{(2)} & \multicolumn{1}{c}{(3)} & \multicolumn{1}{c}{(4)} &  \multicolumn{1}{c}{(5)} & \multicolumn{1}{c}{(6)} & \multicolumn{1}{c}{(7)} & \multicolumn{1}{c}{(8)} & \multicolumn{1}{c}{(9)} & \multicolumn{1}{c}{(10)} & \multicolumn{1}{c}{(11)} & \multicolumn{1}{c}{(12)} \\
\hline
GMRT090312+0640001 & 09:03:12.9 & 64:00:01.1 &   1.48 &   59.6 &   32.4 &   13.0 &    $-$ &    $-$ &   11.8 &   0.77 &      S \\ 
GMRT090336+0633813 & 09:03:37.0 & 63:38:13.6 &   1.45 &  418.2 &  235.7 &  176.0 &    $-$ &    $-$ &   49.1 &   0.99 &      S \\ 
GMRT090353+0633416 & 09:03:53.7 & 63:34:16.6 &   1.43 &   16.2 &    6.1 &    $-$ &    $-$ &    $-$ &    3.0 &   0.71 &      S \\ 
GMRT090357+0640848 & 09:03:57.8 & 64:08:48.2 &   1.43 &  427.0 &  202.4 &  139.0 &    $-$ &    $-$ &   25.1 &   1.31 &      S \\ 
GMRT090402+0632412 & 09:04:02.9 & 63:24:12.9 &   1.45 &   36.3 &   19.3 &   18.0 &    $-$ &    $-$ &    9.6 &   0.58 &      S \\ 
\\ 
GMRT090403+0641148 & 09:04:03.1 & 64:11:48.3 &   1.43 &   17.6 &    9.2 &    $-$ &    $-$ &    $-$ &    3.1 &   0.76 &      S \\ 
GMRT090413+0634054 & 09:04:13.3 & 63:40:54.4 &   1.37 &   58.3 &   32.0 &   29.0 &    $-$ &    $-$ &    7.6 &   0.91 &      S \\ 
GMRT090421+0641502 & 09:04:21.9 & 64:15:02.2 &   1.41 &   23.8 &   18.3 &    $-$ &    $-$ &    $-$ &    6.9 &   0.56 &      S \\ 
GMRT090425+0640650 & 09:04:25.8 & 64:06:50.8 &   1.37 &   48.6 &   28.6 &   16.0 &    $-$ &    $-$ &    8.5 &   0.81 &      S \\ 
GMRT090444+0634149 & 09:04:44.7 & 63:41:49.4 &   1.31 &   13.6 &    7.2 &    $-$ &    $-$ &    $-$ &    1.4 &   1.02 &      S \\ 
\\ 
GMRT090455+0632924 & 09:04:55.7 & 63:29:24.0 &   1.33 &  108.3 &   59.8 &   44.0 &    $-$ &    $-$ &   10.7 &   1.05 &      S \\ 
GMRT090459+0642958 & 09:04:59.7 & 64:29:58.6 &   1.44 &    7.0 &    7.8 &    $-$ &    $-$ &    $-$ &    6.9 &   0.02 &      S \\ 
GMRT090510+0640149 & 09:05:10.6 & 64:01:49.6 &   1.27 &  209.1 &  127.2 &   73.0 &    $-$ &    $-$ &   25.8 &   0.98 &      S \\ 
GMRT090518+0634333 & 09:05:18.1 & 63:43:33.6 &   1.25 &   68.6 &   47.6 &   34.0 &    $-$ &    $-$ &   12.5 &   0.78 &      S \\ 
GMRT090524+0632512 & 09:05:24.1 & 63:25:12.4 &   1.30 &   19.0 &   11.4 &    $-$ &    $-$ &    $-$ &   25.8 &  -0.21 &      S \\ 
\\ 
GMRT090531+0635941 & 09:05:31.3 & 63:59:41.9 &   1.23 &   38.7 &   26.5 &    $-$ &    $-$ &    $-$ &   11.3 &   0.55 &      S \\ 
GMRT090534+0635639 & 09:05:34.3 & 63:56:39.6 &   1.22 &  165.3 &  109.3 &   79.0 &    $-$ &    $-$ &   30.4 &   0.78 &      S \\ 
GMRT090539+0642121 & 09:05:39.3 & 64:21:21.0 &   1.32 &   33.9 &   22.4 &   15.0 &    $-$ &    $-$ &    7.3 &   0.70 &      S \\ 
GMRT090550+0631447 & 09:05:50.2 & 63:14:47.2 &   1.32 &   18.4 &   14.2 &    $-$ &    $-$ &    $-$ &    3.3 &   0.78 &      S \\ 
GMRT090604+0642026 & 09:06:04.4 & 64:20:26.5 &   1.27 &   21.2 &   15.8 &    $-$ &    $-$ &    $-$ &    4.1 &   0.75 &      S \\ 
\\ 
GMRT090607+0634848 & 09:06:07.3 & 63:48:48.5 &   1.15 &  692.6 &  414.1 &  263.0 &    $-$ &    $-$ &   91.3 &   0.97 &      D \\ 
GMRT090614+0630148 & 09:06:14.7 & 63:01:48.6 &   1.39 &   38.6 &   17.5 &    $-$ &    $-$ &    $-$ &    3.6 &   1.06 &      S \\ 
GMRT090621+0644304 & 09:06:21.6 & 64:43:04.2 &   1.44 &   19.7 &   16.6 &    $-$ &    $-$ &    $-$ &    3.8 &   0.76 &      S \\ 
GMRT090624+0631218 & 09:06:24.1 & 63:12:18.5 &   1.28 &  423.7 &  355.9 &  284.0 &    $-$ &    $-$ &   82.5 &   0.72 &      S \\ 
GMRT090641+0643742 & 09:06:41.4 & 64:37:42.8 &   1.36 &   52.9 &   23.7 &   17.0 &    $-$ &    $-$ &    7.8 &   0.87 &      S \\ 
\\ 
GMRT090645+0635220 & 09:06:46.0 & 63:52:20.7 &   1.08 &   60.2 &   39.8 &   15.0 &    $-$ &    $-$ &    8.8 &   0.92 &      S \\ 
GMRT090645+0640829 & 09:06:45.5 & 64:08:29.2 &   1.13 &    6.9 &    4.2 &    $-$ &    $-$ &    $-$ &    2.0 &   0.54 &      S \\ 
GMRT090655+0642508 & 09:06:55.8 & 64:25:08.2 &   1.22 &   32.3 &   14.8 &    $-$ &    $-$ &    $-$ &    2.6 &   1.13 &      S \\ 
GMRT090657+0633704 & 09:06:57.1 & 63:37:04.8 &   1.08 &   21.8 &   15.3 &    $-$ &    $-$ &    $-$ &    4.5 &   0.71 &      S \\ 
GMRT090704+0641405 & 09:07:04.5 & 64:14:05.7 &   1.13 &   10.0 &    6.4 &    $-$ &    $-$ &    $-$ &    4.8 &   0.30 &      S \\ 
\\ 
GMRT090707+0632649 & 09:07:07.6 & 63:26:49.6 &   1.11 &   81.4 &   49.8 &   24.0 &    $-$ &    $-$ &   10.0 &   0.99 &      S \\ 
GMRT090714+0630308 & 09:07:14.3 & 63:03:08.6 &   1.29 &   36.6 &   25.8 &    $-$ &    $-$ &    $-$ &    5.6 &   0.85 &      S \\ 
GMRT090729+0640339 & 09:07:29.7 & 64:03:39.9 &   1.03 &   12.6 &    6.0 &    $-$ &    $-$ &    $-$ &    2.9 &   0.63 &      T \\ 
GMRT090731+0643627 & 09:07:31.2 & 64:36:27.0 &   1.27 &   66.7 &   47.3 &   34.0 &    $-$ &    $-$ &    9.1 &   0.90 &      S \\ 
GMRT090733+0642226 & 09:07:33.2 & 64:22:26.0 &   1.14 &  138.2 &   84.6 &   53.0 &    $-$ &    $-$ &   20.6 &   0.88 &      D \\ 
\\ 
GMRT090740+0642941 & 09:07:41.0 & 64:29:41.4 &   1.19 &   53.3 &   30.4 &    $-$ &    $-$ &    $-$ &    2.8 &   1.33 &      D \\ 
GMRT090742+0641012 & 09:07:42.1 & 64:10:12.3 &   1.04 &   33.5 &   25.5 &    $-$ &    $-$ &    $-$ &   19.4 &   0.23 &      S \\ 
GMRT090745+0631759 & 09:07:45.4 & 63:17:59.1 &   1.10 &   28.4 &   23.8 &   16.0 &    $-$ &    $-$ &    5.5 &   0.76 &      S \\ 
GMRT090750+0642030 & 09:07:50.2 & 64:20:30.0 &   1.10 &   41.6 &   35.4 &   27.0 &    $-$ &    $-$ &   12.4 &   0.56 &      S \\ 
GMRT090756+0641856 & 09:07:56.3 & 64:18:56.2 &   1.08 &   14.6 &    6.6 &    $-$ &    $-$ &    $-$ &    2.6 &   0.75 &      S \\ 
\\ 
GMRT090808+0624820 & 09:08:08.5 & 62:48:20.9 &   1.38 &  273.2 &  144.9 &   87.0 &    $-$ &    $-$ &   25.3 &   1.12 &      D \\ 
GMRT090816+0625249 & 09:08:16.1 & 62:52:49.5 &   1.31 &   17.9 &   10.1 &    $-$ &    $-$ &    $-$ &    3.1 &   0.78 &      S \\ 
GMRT090834+0633533 & 09:08:34.5 & 63:35:33.1 &   0.91 &   20.7 &   17.6 &    $-$ &    $-$ &    $-$ &    4.6 &   0.69 &      S \\ 
GMRT090838+0632257 & 09:08:38.7 & 63:22:57.2 &   0.98 &   28.7 &   19.2 &    $-$ &    $-$ &    $-$ &    3.3 &   0.98 &      S \\ 
GMRT090839+0635954 & 09:08:39.6 & 63:59:54.5 &   0.89 &  173.3 &  134.2 &   85.0 &    $-$ &    $-$ &   30.3 &   0.81 &      S \\ 
\\ 
GMRT090842+0641138 & 09:08:42.4 & 64:11:38.2 &   0.95 &   27.1 &   14.0 &    $-$ &    $-$ &    $-$ &    2.1 &   1.15 &      S \\ 
GMRT090842+0642154 & 09:08:42.4 & 64:21:54.2 &   1.03 &  102.3 &   57.2 &   32.0 &    $-$ &    $-$ &    6.0 &   1.31 &      S \\ 
GMRT090846+0624817 & 09:08:46.5 & 62:48:17.9 &   1.33 &  319.8 &  159.8 &  120.0 &    $-$ &    $-$ &   28.5 &   1.11 &      D \\ 
GMRT090847+0635629 & 09:08:47.3 & 63:56:29.5 &   0.87 &   73.1 &   51.8 &   31.0 &    $-$ &    $-$ &   14.3 &   0.76 &      S \\ 
GMRT090848+0630901 & 09:08:48.5 & 63:09:02.0 &   1.09 &  549.0 &  350.2 &  250.0 &    $-$ &    $-$ &   58.9 &   1.01 &      S \\ 
\hline \hline
\end{tabular}

\end{table*}
\begin{table*}
\contcaption{}
\begin{tabular}{l r r r r r r r r r r l}
\hline
\multicolumn{1}{c}{Source name} & \multicolumn{1}{c}{RA} & \multicolumn{1}{c}{DEC} & \multicolumn{1}{c}{Dis} & \multicolumn{1}{c}{S$_{153}$} & \multicolumn{1}{c}{S$_{244}$}
                    & \multicolumn{1}{c}{S$_{330}$} & \multicolumn{1}{c}{S$_{610}$} & \multicolumn{1}{c}{S$_{1260}$} & \multicolumn{1}{c}{S$_{1400}$}
                    & \multicolumn{1}{c}{$\alpha$} &\multicolumn{1}{c}{Class} \\
                    & \multicolumn{1}{c}{hh:mm:ss.s} & \multicolumn{1}{c}{dd:mm:ss.s} & \multicolumn{1}{c}{deg.} & \multicolumn{1}{c}{mJy} & \multicolumn{1}{c}{mJy}
                    & \multicolumn{1}{c}{mJy} & \multicolumn{1}{c}{mJy} & \multicolumn{1}{c}{mJy} & \multicolumn{1}{c}{mJy} &  &        \\
\multicolumn{1}{c}{(1)} & \multicolumn{1}{c}{(2)} & \multicolumn{1}{c}{(3)} & \multicolumn{1}{c}{(4)} &  \multicolumn{1}{c}{(5)} & \multicolumn{1}{c}{(6)} & \multicolumn{1}{c}{(7)} & \multicolumn{1}{c}{(8)} & \multicolumn{1}{c}{(9)} & \multicolumn{1}{c}{(10)} & \multicolumn{1}{c}{(11)} & \multicolumn{1}{c}{(12)} \\
\hline
GMRT090859+0640830 & 09:08:59.4 & 64:08:30.8 &   0.90 &   17.2 &   11.4 &    $-$ &    $-$ &    $-$ &    3.7 &   0.69 &      S \\ 
GMRT090859+0645054 & 09:08:59.9 & 64:50:54.6 &   1.33 &  147.7 &   88.6 &   53.0 &    $-$ &    $-$ &   19.0 &   0.96 &      S \\ 
GMRT090904+0640923 & 09:09:04.2 & 64:09:23.6 &   0.90 &   22.2 &   15.4 &    $-$ &    $-$ &    $-$ &    3.3 &   0.86 &      S \\ 
GMRT090906+0635117 & 09:09:06.7 & 63:51:17.4 &   0.83 &   24.3 &   18.1 &    $-$ &    $-$ &    $-$ &    6.1 &   0.62 &      D \\ 
GMRT090909+0623744 & 09:09:09.4 & 62:37:44.6 &   1.44 &   37.3 &   18.0 &    $-$ &    $-$ &    $-$ &    5.2 &   0.87 &      S \\ 
\\ 
GMRT090911+0643524 & 09:09:11.6 & 64:35:24.2 &   1.13 &  455.3 &  288.5 &  197.0 &    $-$ &    $-$ &   64.9 &   0.91 &      S \\ 
GMRT090913+0642731 & 09:09:13.3 & 64:27:31.2 &   1.04 &   29.2 &   18.3 &    $-$ &    $-$ &    $-$ &    3.9 &   0.91 &      S \\ 
GMRT090918+0635958 & 09:09:18.0 & 63:59:58.1 &   0.83 & 1092.2 &  710.9 &  465.0 &    $-$ &    $-$ &  159.6 &   0.91 &   Cplx \\ 
GMRT090922+0631604 & 09:09:22.3 & 63:16:04.3 &   0.97 &   46.4 &   33.9 &   25.0 &    $-$ &    $-$ &    7.7 &   0.82 &      D \\ 
GMRT090923+0650100 & 09:09:23.8 & 65:01:00.3 &   1.44 &  302.9 &  167.1 &  133.0 &    $-$ &    $-$ &   37.1 &   0.96 &      S \\ 
\\ 
GMRT090926+0634214 & 09:09:27.0 & 63:42:14.1 &   0.80 &   72.3 &   50.8 &   18.0 &    $-$ &    $-$ &    7.2 &   1.11 &      D \\ 
GMRT090936+0634214 & 09:09:36.4 & 63:42:14.0 &   0.78 &    5.6 &    6.2 &    $-$ &    $-$ &    $-$ &   12.4 &  -0.37 &      S \\ 
GMRT090952+0644549 & 09:09:52.4 & 64:45:49.9 &   1.21 &   81.0 &   60.5 &   32.0 &    $-$ &    $-$ &   13.8 &   0.83 &      D \\ 
GMRT090952+0650625 & 09:09:52.2 & 65:06:25.0 &   1.49 &   19.8 &   12.0 &    $-$ &    $-$ &    $-$ &    3.7 &   0.75 &      S \\ 
GMRT091002+0625229 & 09:10:02.4 & 62:52:29.3 &   1.18 &  124.7 &   66.0 &   60.0 &    $-$ &    $-$ &   10.5 &   1.09 &      D \\ 
\\ 
GMRT091007+0633546 & 09:10:08.0 & 63:35:46.5 &   0.74 &   16.5 &   12.4 &    $-$ &    $-$ &    $-$ &    1.1 &   1.25 &      S \\ 
GMRT091011+0632832 & 09:10:12.0 & 63:28:32.4 &   0.78 &   14.4 &    9.0 &    $-$ &    $-$ &    $-$ &    2.1 &   0.86 &      S \\ 
GMRT091012+0625309 & 09:10:12.6 & 62:53:09.0 &   1.16 &   96.4 &   55.1 &    $-$ &    $-$ &    $-$ &   11.8 &   0.95 &      D \\ 
GMRT091012+0645718 & 09:10:12.9 & 64:57:18.9 &   1.34 &   49.2 &   29.9 &   17.0 &    $-$ &    $-$ &    8.4 &   0.82 &      S \\ 
GMRT091016+0634249 & 09:10:16.0 & 63:42:50.0 &   0.70 &   35.5 &   24.6 &    $-$ &    $-$ &    $-$ &    5.8 &   0.82 &      D \\ 
\\ 
GMRT091018+0633718 & 09:10:18.4 & 63:37:18.5 &   0.72 &   27.9 &   15.7 &    $-$ &    $-$ &    $-$ &    7.8 &   0.55 &      S \\ 
GMRT091019+0625558 & 09:10:19.4 & 62:55:58.2 &   1.12 &   14.1 &    8.1 &    $-$ &    $-$ &    $-$ &    2.4 &   0.79 &      S \\ 
GMRT091021+0623435 & 09:10:21.1 & 62:34:35.2 &   1.41 &  255.3 &  121.0 &   84.0 &    $-$ &    $-$ &   17.2 &   1.24 &      S \\ 
GMRT091023+0635150 & 09:10:23.4 & 63:51:50.8 &   0.69 &  354.9 &  248.7 &  171.0 &    $-$ &    $-$ &   48.7 &   0.91 &      S \\ 
GMRT091051+0625408 & 09:10:51.6 & 62:54:08.9 &   1.11 &   10.6 &    7.6 &    $-$ &    $-$ &    $-$ &    1.5 &   0.88 &      S \\ 
\\ 
GMRT091053+0641930 & 09:10:53.9 & 64:19:30.8 &   0.81 &  676.5 &  447.7 &    $-$ &    $-$ &    $-$ &   72.5 &   1.00 &      D \\ 
GMRT091056+0644151 & 09:10:56.9 & 64:41:51.0 &   1.08 & 1907.8 & 1285.5 &  864.0 &    $-$ &    $-$ &  253.5 &   0.94 &      S \\ 
GMRT091059+0640107 & 09:10:59.7 & 64:01:07.7 &   0.65 &  248.3 &  184.3 &  119.0 &    $-$ &    $-$ &   45.1 &   0.79 &      S \\ 
GMRT091103+0630312 & 09:11:03.5 & 63:03:12.2 &   0.97 &   26.3 &   21.9 &    $-$ &    $-$ &    $-$ &    9.8 &   0.45 &      S \\ 
GMRT091105+0642013 & 09:11:05.6 & 64:20:13.6 &   0.80 &  426.6 &  297.5 &    $-$ &    $-$ &    $-$ &   52.9 &   0.94 &      S \\ 
\\ 
GMRT091117+0630930 & 09:11:17.9 & 63:09:30.2 &   0.87 &   14.1 &   16.1 &    $-$ &    $-$ &    $-$ &   13.0 &   0.06 &      S \\ 
GMRT091120+0632133 & 09:11:20.5 & 63:21:33.6 &   0.73 &   34.9 &   29.3 &    $-$ &    $-$ &    $-$ &    6.4 &   0.78 &      S \\ 
GMRT091122+0640813 & 09:11:22.3 & 64:08:14.0 &   0.66 &   33.3 &   22.5 &    $-$ &    $-$ &    $-$ &    4.4 &   0.92 &      S \\ 
GMRT091126+0642320 & 09:11:26.2 & 64:23:20.7 &   0.81 &   20.6 &   14.5 &    $-$ &    $-$ &    $-$ &    1.5 &   1.19 &      S \\ 
GMRT091133+0630623 & 09:11:33.4 & 63:06:23.5 &   0.89 &  436.8 &  222.1 &  118.0 &    $-$ &    $-$ &   11.1 &   1.66 &      D \\ 
\\ 
GMRT091135+0641609 & 09:11:35.3 & 64:16:09.8 &   0.72 &   16.3 &   16.5 &    $-$ &    $-$ &    $-$ &    4.9 &   0.57 &      S \\ 
GMRT091136+0651054 & 09:11:36.6 & 65:10:54.7 &   1.48 &  387.9 &  185.0 &  139.0 &    $-$ &    $-$ &   41.2 &   1.05 &      D \\ 
GMRT091139+0623918 & 09:11:39.9 & 62:39:18.7 &   1.27 &   34.5 &   24.5 &    $-$ &    $-$ &    $-$ &    9.8 &   0.56 &      S \\ 
GMRT091153+0633955 & 09:11:53.7 & 63:39:55.7 &   0.54 &    6.6 &    5.3 &    $-$ &    3.7 &    $-$ &    2.8 &   0.39 &      S \\ 
GMRT091154+0644346 & 09:11:54.0 & 64:43:46.4 &   1.06 &   67.2 &   50.2 &   37.0 &    $-$ &    $-$ &   12.5 &   0.77 &      S \\ 
\\ 
GMRT091157+0632128 & 09:11:57.6 & 63:21:28.3 &   0.68 &  665.9 &  489.7 &  354.0 &    $-$ &    $-$ &  122.9 &   0.77 &      S \\ 
GMRT091201+0632720 & 09:12:01.1 & 63:27:20.7 &   0.61 &  167.9 &  107.5 &   57.0 &    $-$ &    $-$ &   15.3 &   1.12 &      S \\ 
GMRT091205+0630913 & 09:12:05.7 & 63:09:13.2 &   0.82 &   76.2 &   47.5 &   30.0 &    $-$ &    $-$ &    7.9 &   1.04 &      S \\ 
GMRT091212+0634541 & 09:12:12.5 & 63:45:41.9 &   0.49 &  147.4 &  106.8 &   65.0 &   56.2 &    $-$ &   23.1 &   0.80 &      S \\ 
GMRT091223+0635840 & 09:12:23.6 & 63:58:40.2 &   0.49 &   24.2 &   17.4 &    $-$ &   10.9 &    $-$ &    5.9 &   0.62 &      S \\ 
\\ 
GMRT091238+0633330 & 09:12:38.3 & 63:33:30.5 &   0.50 &   10.0 &    8.6 &    $-$ &    4.8 &    $-$ &    3.6 &   0.49 &      D \\ 
GMRT091238+0650419 & 09:12:38.7 & 65:04:19.6 &   1.34 &   18.8 &   10.3 &    $-$ &    $-$ &    $-$ &    4.8 &   0.59 &      S \\ 
GMRT091246+0634932 & 09:12:46.1 & 63:49:32.1 &   0.42 &  488.0 &  298.5 &  169.0 &  111.9 &    $-$ &   36.9 &   1.15 &      D \\ 
GMRT091250+0623803 & 09:12:50.4 & 62:38:04.0 &   1.24 &   83.3 &   55.8 &   42.0 &    $-$ &    $-$ &   17.6 &   0.71 &      S \\ 
GMRT091252+0633051 & 09:12:52.5 & 63:30:51.4 &   0.50 &   84.5 &   59.7 &   42.0 &   28.6 &    $-$ &   12.7 &   0.84 &      S \\ 
\hline \hline
\end{tabular}

\end{table*}
\begin{table*}
\contcaption{}
\begin{tabular}{l r r r r r r r r r r l}
\hline
\multicolumn{1}{c}{Source name} & \multicolumn{1}{c}{RA} & \multicolumn{1}{c}{DEC} & \multicolumn{1}{c}{Dis} & \multicolumn{1}{c}{S$_{153}$} & \multicolumn{1}{c}{S$_{244}$}
                    & \multicolumn{1}{c}{S$_{330}$} & \multicolumn{1}{c}{S$_{610}$} & \multicolumn{1}{c}{S$_{1260}$} & \multicolumn{1}{c}{S$_{1400}$}
                    & \multicolumn{1}{c}{$\alpha$} &\multicolumn{1}{c}{Class} \\
                    & \multicolumn{1}{c}{hh:mm:ss.s} & \multicolumn{1}{c}{dd:mm:ss.s} & \multicolumn{1}{c}{deg.} & \multicolumn{1}{c}{mJy} & \multicolumn{1}{c}{mJy}
                    & \multicolumn{1}{c}{mJy} & \multicolumn{1}{c}{mJy} & \multicolumn{1}{c}{mJy} & \multicolumn{1}{c}{mJy} &  &        \\
\multicolumn{1}{c}{(1)} & \multicolumn{1}{c}{(2)} & \multicolumn{1}{c}{(3)} & \multicolumn{1}{c}{(4)} &  \multicolumn{1}{c}{(5)} & \multicolumn{1}{c}{(6)} & \multicolumn{1}{c}{(7)} & \multicolumn{1}{c}{(8)} & \multicolumn{1}{c}{(9)} & \multicolumn{1}{c}{(10)} & \multicolumn{1}{c}{(11)} & \multicolumn{1}{c}{(12)} \\
\hline
GMRT091256+0625901 & 09:12:56.7 & 62:59:01.1 &   0.92 &   22.1 &   17.2 &    $-$ &    $-$ &    $-$ &    4.8 &   0.70 &      S \\ 
GMRT091257+0624148 & 09:12:57.4 & 62:41:48.9 &   1.18 &   59.1 &   33.4 &   21.0 &    $-$ &    $-$ &   10.8 &   0.78 &      S \\ 
GMRT091259+0631651 & 09:12:59.3 & 63:16:51.2 &   0.66 &    7.7 &    4.3 &    $-$ &    $-$ &    $-$ &    0.7 &   1.07 &      S \\ 
GMRT091303+0631808 & 09:13:03.8 & 63:18:08.9 &   0.64 &   12.2 &    8.6 &    $-$ &    $-$ &    $-$ &    2.9 &   0.65 &      S \\ 
GMRT091306+0630428 & 09:13:06.7 & 63:04:28.9 &   0.83 &    6.8 &    5.0 &    $-$ &    $-$ &    $-$ &    1.3 &   0.75 &      S \\ 
\\ 
GMRT091308+0633945 & 09:13:08.5 & 63:39:45.8 &   0.41 &   19.1 &    4.9 &    $-$ &    0.7 &    $-$ &    $-$ &   2.37 &      S \\ 
GMRT091311+0645419 & 09:13:11.9 & 64:54:19.4 &   1.16 &  715.6 &  351.2 &  234.0 &    $-$ &    $-$ &   59.7 &   1.18 &   Cplx \\ 
GMRT091313+0635126 & 09:13:14.0 & 63:51:26.4 &   0.37 &   23.1 &   16.4 &    $-$ &    8.0 &    $-$ &    4.1 &   0.78 &      D \\ 
GMRT091325+0633032 & 09:13:25.6 & 63:30:32.4 &   0.46 &   54.3 &   42.0 &   23.0 &   20.1 &    $-$ &    9.2 &   0.80 &      S \\ 
GMRT091326+0642810 & 09:13:26.6 & 64:28:10.9 &   0.75 &   59.1 &   43.1 &   27.0 &    $-$ &    $-$ &    8.9 &   0.87 &      S \\ 
\\ 
GMRT091328+0640209 & 09:13:28.4 & 64:02:09.1 &   0.41 &    7.6 &    3.9 &    $-$ &    1.1 &    $-$ &    $-$ &   1.37 &      S \\ 
GMRT091335+0625203 & 09:13:35.2 & 62:52:03.1 &   0.99 &   14.9 &    7.7 &    $-$ &    $-$ &    $-$ &    3.4 &   0.64 &      S \\ 
GMRT091336+0631857 & 09:13:36.0 & 63:18:57.8 &   0.59 &   70.2 &   45.5 &   22.0 &    $-$ &    $-$ &   13.4 &   0.78 &   Cplx \\ 
GMRT091337+0630408 & 09:13:37.5 & 63:04:08.7 &   0.81 &  109.4 &   66.3 &   37.0 &    $-$ &    $-$ &   11.1 &   1.07 &      S \\ 
GMRT091342+0630731 & 09:13:42.6 & 63:07:31.7 &   0.75 &    6.0 &    4.9 &    $-$ &    $-$ &    $-$ &    1.4 &   0.67 &      S \\ 
\\ 
GMRT091343+0632751 & 09:13:43.5 & 63:27:51.9 &   0.46 &   31.4 &   23.9 &    $-$ &   12.1 &    $-$ &    5.6 &   0.77 &      S \\ 
GMRT091346+0644249 & 09:13:46.5 & 64:42:49.2 &   0.96 &  256.7 &  162.2 &  109.0 &    $-$ &    $-$ &   37.2 &   0.89 &      S \\ 
GMRT091347+0632240 & 09:13:47.0 & 63:22:40.5 &   0.53 &   53.5 &   37.8 &   18.0 &   22.4 &    $-$ &   12.1 &   0.66 &      S \\ 
GMRT091348+0623508 & 09:13:48.1 & 62:35:08.8 &   1.26 &   14.2 &    7.7 &    $-$ &    $-$ &    $-$ &    4.4 &   0.50 &      S \\ 
GMRT091349+0641603 & 09:13:49.9 & 64:16:03.7 &   0.55 &  170.9 &  111.0 &   52.0 &    $-$ &    $-$ &   17.3 &   1.09 &      D \\ 
\\ 
GMRT091352+0625805 & 09:13:52.1 & 62:58:05.2 &   0.89 &   40.3 &   25.5 &    $-$ &    $-$ &    $-$ &    5.8 &   0.87 &      S \\ 
GMRT091353+0642301 & 09:13:53.5 & 64:23:01.2 &   0.65 &  117.8 &   75.5 &   38.0 &    $-$ &    $-$ &   12.1 &   1.07 &      S \\ 
GMRT091354+0645157 & 09:13:54.6 & 64:51:57.8 &   1.10 &   10.5 &    6.2 &    $-$ &    $-$ &    $-$ &    2.4 &   0.65 &      S \\ 
GMRT091358+0640801 & 09:13:58.7 & 64:08:01.0 &   0.44 &   13.2 &   10.7 &    $-$ &    6.6 &    $-$ &    2.2 &   0.74 &      S \\ 
GMRT091401+0632103 & 09:14:01.4 & 63:21:03.0 &   0.53 &   13.6 &   11.6 &    $-$ &    6.2 &    $-$ &    2.8 &   0.70 &      S \\ 
\\ 
GMRT091402+0635652 & 09:14:02.6 & 63:56:52.4 &   0.32 &  430.2 &  288.2 &  183.0 &  139.4 &   81.0 &   57.0 &   0.85 &      D \\ 
GMRT091402+0651235 & 09:14:02.7 & 65:12:35.2 &   1.43 &  538.0 &  236.6 &  205.0 &    $-$ &    $-$ &   65.8 &   0.99 &      S \\ 
GMRT091403+0625000 & 09:14:03.9 & 62:50:00.4 &   1.01 &    4.4 &    3.9 &    $-$ &    $-$ &    $-$ &    0.7 &   0.84 &      S \\ 
GMRT091409+0631548 & 09:14:09.2 & 63:15:48.8 &   0.60 &    5.6 &    4.9 &    $-$ &    $-$ &    $-$ &    1.5 &   0.61 &      S \\ 
GMRT091410+0643332 & 09:14:10.4 & 64:33:33.0 &   0.80 &   12.2 &   11.6 &    $-$ &    $-$ &    $-$ &    3.0 &   0.66 &      S \\ 
\\ 
GMRT091415+0640324 & 09:14:15.0 & 64:03:24.8 &   0.36 &  129.8 &  100.8 &   67.0 &   64.5 &    $-$ &   33.2 &   0.59 &      D \\ 
GMRT091417+0623840 & 09:14:17.1 & 62:38:40.0 &   1.19 &    7.5 &    4.7 &    $-$ &    $-$ &    $-$ &    3.1 &   0.37 &      S \\ 
GMRT091424+0640613 & 09:14:24.1 & 64:06:13.6 &   0.39 &  101.3 &   91.5 &   57.0 &   68.9 &    $-$ &   37.4 &   0.41 &      S \\ 
GMRT091435+0632643 & 09:14:35.0 & 63:26:43.6 &   0.42 &   11.9 &   10.2 &    $-$ &    7.0 &    $-$ &    3.7 &   0.50 &      S \\ 
GMRT091437+0645522 & 09:14:37.1 & 64:55:22.9 &   1.14 &   33.7 &   24.9 &   16.0 &    $-$ &    $-$ &    8.6 &   0.63 &      S \\ 
\\ 
GMRT091440+0631316 & 09:14:40.7 & 63:13:16.7 &   0.62 &   24.2 &   12.4 &    $-$ &    $-$ &    $-$ &    3.2 &   0.90 &      T \\ 
GMRT091445+0641414 & 09:14:45.3 & 64:14:14.4 &   0.48 &  996.8 &  756.4 &  511.0 &  407.5 &    $-$ &  174.4 &   0.75 &      S \\ 
GMRT091446+0633757 & 09:14:46.9 & 63:37:57.7 &   0.26 &   64.2 &   52.0 &    $-$ &   25.8 &    $-$ &   15.4 &   0.66 &      T \\ 
GMRT091446+0642706 & 09:14:46.2 & 64:27:07.0 &   0.68 &  179.1 &  115.1 &   56.0 &    $-$ &    $-$ &   15.3 &   1.16 &      S \\ 
GMRT091446+0644244 & 09:14:46.1 & 64:42:44.3 &   0.93 &   34.5 &   25.8 &   20.0 &    $-$ &    $-$ &    7.4 &   0.70 &      S \\ 
\\ 
GMRT091449+0651507 & 09:14:49.4 & 65:15:07.5 &   1.46 &   13.3 &    9.7 &    $-$ &    $-$ &    $-$ &    5.2 &   0.41 &      S \\ 
GMRT091452+0622452 & 09:14:52.2 & 62:24:52.3 &   1.40 &   42.1 &   18.2 &   17.0 &    $-$ &    $-$ &    4.8 &   0.96 &      S \\ 
GMRT091508+0633507 & 09:15:08.2 & 63:35:07.5 &   0.27 &   85.0 &   60.2 &   40.0 &   28.2 &    $-$ &   15.3 &   0.79 &      D \\ 
GMRT091513+0624944 & 09:15:13.7 & 62:49:44.0 &   0.99 &    6.4 &    6.6 &    $-$ &    $-$ &    $-$ &    6.2 &   0.02 &      S \\ 
GMRT091515+0651008 & 09:15:15.4 & 65:10:08.1 &   1.37 &   49.4 &   24.8 &   13.0 &    $-$ &    $-$ &    8.4 &   0.82 &      S \\ 
\\ 
GMRT091518+0650114 & 09:15:18.3 & 65:01:14.5 &   1.23 &   67.9 &   46.2 &   21.0 &    $-$ &    $-$ &   11.3 &   0.85 &      S \\ 
GMRT091520+0630119 & 09:15:20.4 & 63:01:19.2 &   0.79 &  276.9 &  175.6 &  122.0 &    $-$ &    $-$ &   34.6 &   0.95 &      D \\ 
GMRT091520+0632754 & 09:15:20.7 & 63:27:54.1 &   0.36 &   60.7 &   46.8 &    $-$ &   18.2 &    $-$ &    7.8 &   0.93 &      S \\ 
GMRT091520+0635953 & 09:15:20.3 & 63:59:53.6 &   0.24 &  284.2 &  178.6 &  102.0 &   66.8 &   34.2 &   23.1 &   1.07 &      D \\ 
GMRT091521+0635558 & 09:15:21.5 & 63:55:58.1 &   0.19 &   11.6 &    7.3 &    $-$ &    3.4 &    1.7 &    $-$ &   0.90 &      S \\ 
\hline \hline
\end{tabular}

\end{table*}
\begin{table*}
\contcaption{}
\begin{tabular}{l r r r r r r r r r r l}
\hline
\multicolumn{1}{c}{Source name} & \multicolumn{1}{c}{RA} & \multicolumn{1}{c}{DEC} & \multicolumn{1}{c}{Dis} & \multicolumn{1}{c}{S$_{153}$} & \multicolumn{1}{c}{S$_{244}$}
                    & \multicolumn{1}{c}{S$_{330}$} & \multicolumn{1}{c}{S$_{610}$} & \multicolumn{1}{c}{S$_{1260}$} & \multicolumn{1}{c}{S$_{1400}$}
                    & \multicolumn{1}{c}{$\alpha$} &\multicolumn{1}{c}{Class} \\
                    & \multicolumn{1}{c}{hh:mm:ss.s} & \multicolumn{1}{c}{dd:mm:ss.s} & \multicolumn{1}{c}{deg.} & \multicolumn{1}{c}{mJy} & \multicolumn{1}{c}{mJy}
                    & \multicolumn{1}{c}{mJy} & \multicolumn{1}{c}{mJy} & \multicolumn{1}{c}{mJy} & \multicolumn{1}{c}{mJy} &  &        \\
\multicolumn{1}{c}{(1)} & \multicolumn{1}{c}{(2)} & \multicolumn{1}{c}{(3)} & \multicolumn{1}{c}{(4)} &  \multicolumn{1}{c}{(5)} & \multicolumn{1}{c}{(6)} & \multicolumn{1}{c}{(7)} & \multicolumn{1}{c}{(8)} & \multicolumn{1}{c}{(9)} & \multicolumn{1}{c}{(10)} & \multicolumn{1}{c}{(11)} & \multicolumn{1}{c}{(12)} \\
\hline
GMRT091523+0624511 & 09:15:23.8 & 62:45:11.5 &   1.06 &   10.0 &    5.6 &    $-$ &    $-$ &    $-$ &    1.2 &   0.95 &      S \\ 
GMRT091526+0625436 & 09:15:26.6 & 62:54:36.7 &   0.90 &   16.6 &    8.8 &    $-$ &    $-$ &    $-$ &    2.7 &   0.80 &      S \\ 
GMRT091528+0632939 & 09:15:28.4 & 63:29:39.3 &   0.33 &  662.4 &  445.9 &  337.0 &  186.9 &    $-$ &   77.5 &   0.95 &      S \\ 
GMRT091531+0630722 & 09:15:31.6 & 63:07:22.0 &   0.69 &    9.5 &   11.0 &    $-$ &    $-$ &    $-$ &    3.5 &   0.49 &      S \\ 
GMRT091533+0642008 & 09:15:33.2 & 64:20:08.9 &   0.54 &   11.8 &    7.6 &    $-$ &    $-$ &    $-$ &    4.2 &   0.44 &      D \\ 
\\ 
GMRT091540+0650940 & 09:15:40.7 & 65:09:40.5 &   1.36 &  228.2 &  123.6 &   92.0 &    $-$ &    $-$ &   35.4 &   0.86 &      S \\ 
GMRT091543+0625659 & 09:15:43.8 & 62:56:60.0 &   0.86 &    7.1 &    3.8 &    $-$ &    $-$ &    $-$ &    3.3 &   0.30 &      S \\ 
GMRT091543+0631112 & 09:15:43.3 & 63:11:12.8 &   0.62 &    6.5 &    5.7 &    $-$ &    $-$ &    $-$ &    0.8 &   0.99 &      S \\ 
GMRT091544+0643404 & 09:15:44.2 & 64:34:04.2 &   0.77 &    7.5 &    6.8 &    $-$ &    $-$ &    $-$ &    2.3 &   0.55 &      S \\ 
GMRT091544+0645228 & 09:15:44.3 & 64:52:28.6 &   1.08 &   71.6 &   43.5 &   18.0 &    $-$ &    $-$ &    5.8 &   1.19 &      S \\ 
\\ 
GMRT091549+0634409 & 09:15:49.8 & 63:44:09.6 &   0.11 &    8.2 &    2.7 &    $-$ &    1.6 &    0.5 &    $-$ &   1.21 &      S \\ 
GMRT091550+0631545 & 09:15:50.3 & 63:15:45.9 &   0.55 &   45.4 &   29.6 &   21.0 &    $-$ &    $-$ &    8.0 &   0.79 &      S \\ 
GMRT091552+0635541 & 09:15:52.3 & 63:55:41.0 &   0.15 &   49.8 &   32.5 &   13.0 &   14.1 &    5.9 &    7.5 &   0.91 &      T \\ 
GMRT091554+0623927 & 09:15:54.3 & 62:39:27.0 &   1.15 &    6.1 &    2.9 &    $-$ &    $-$ &    $-$ &    1.1 &   0.75 &      S \\ 
GMRT091603+0623034 & 09:16:03.4 & 62:30:34.7 &   1.29 &   18.1 &    9.1 &    $-$ &    $-$ &    $-$ &    2.8 &   0.82 &      S \\ 
\\ 
GMRT091612+0641110 & 09:16:12.1 & 64:11:10.0 &   0.39 &   36.4 &   28.0 &   21.0 &   16.0 &    $-$ &    7.6 &   0.68 &      S \\ 
GMRT091614+0651445 & 09:16:14.6 & 65:14:45.8 &   1.44 &   38.6 &    7.5 &    $-$ &    $-$ &    $-$ &    3.3 &   1.07 &      S \\ 
GMRT091616+0623915 & 09:16:16.7 & 62:39:15.6 &   1.15 &   10.4 &    7.0 &    $-$ &    $-$ &    $-$ &    2.7 &   0.60 &      S \\ 
GMRT091616+0643009 & 09:16:16.0 & 64:30:09.5 &   0.70 &    5.0 &    3.6 &    $-$ &    $-$ &    $-$ &    2.5 &   0.29 &      S \\ 
GMRT091617+0641327 & 09:16:17.8 & 64:13:27.4 &   0.42 &   15.1 &    7.6 &    $-$ &    3.2 &    $-$ &    2.1 &   0.92 &      S \\ 
\\ 
GMRT091618+0634929 & 09:16:18.2 & 63:49:29.5 &   0.04 &  123.1 &   94.2 &    $-$ &   43.5 &   20.4 &   19.5 &   0.84 &      S \\ 
GMRT091619+0631840 & 09:16:19.1 & 63:18:40.5 &   0.49 &   76.1 &   51.4 &   27.0 &   24.7 &    $-$ &   14.4 &   0.77 &      S \\ 
GMRT091621+0622445 & 09:16:21.8 & 62:24:45.0 &   1.39 &   26.1 &   15.1 &    $-$ &    $-$ &    $-$ &    7.1 &   0.57 &      S \\ 
GMRT091621+0651021 & 09:16:21.5 & 65:10:21.5 &   1.37 &   28.3 &   17.3 &   18.0 &    $-$ &    $-$ &    5.6 &   0.71 &      S \\ 
GMRT091623+0640017 & 09:16:23.9 & 64:00:17.4 &   0.20 &    7.0 &   13.6 &    $-$ &   14.5 &   12.4 &    9.8 &  -0.11 &      S \\ 
\\ 
GMRT091626+0631416 & 09:16:26.7 & 63:14:16.4 &   0.57 &  159.6 &  120.3 &   85.0 &    $-$ &    $-$ &   33.1 &   0.72 &      D \\ 
GMRT091627+0624041 & 09:16:27.3 & 62:40:41.4 &   1.12 &   12.0 &    9.2 &    $-$ &    $-$ &    $-$ &    3.9 &   0.51 &      D \\ 
GMRT091631+0634820 & 09:16:32.0 & 63:48:20.0 &   0.01 & 1212.2 &  826.8 &  523.0 &  347.4 &    $-$ &  143.6 &   0.96 &   Cplx \\ 
GMRT091632+0640438 & 09:16:32.2 & 64:04:38.7 &   0.27 &   21.1 &   10.3 &    $-$ &    6.0 &    2.7 &    2.5 &   0.92 &      S \\ 
GMRT091633+0643528 & 09:16:33.2 & 64:35:28.1 &   0.79 &  263.1 &  154.6 &   71.0 &    $-$ &    $-$ &   20.6 &   1.21 &      D \\ 
\\ 
GMRT091638+0625110 & 09:16:38.6 & 62:51:10.5 &   0.95 &    7.5 &    5.2 &    $-$ &    $-$ &    $-$ &    4.9 &   0.16 &      S \\ 
GMRT091639+0641738 & 09:16:39.6 & 64:17:38.9 &   0.49 &    6.0 &    4.9 &    $-$ &    4.0 &    $-$ &    2.1 &   0.43 &      S \\ 
GMRT091647+0633008 & 09:16:47.6 & 63:30:08.4 &   0.30 &   16.9 &   10.6 &    $-$ &    6.4 &    1.8 &    3.3 &   0.84 &      S \\ 
GMRT091658+0643948 & 09:16:58.1 & 64:39:48.4 &   0.86 &  147.8 &   96.0 &   62.0 &    $-$ &    $-$ &   15.9 &   1.02 &      S \\ 
GMRT091705+0635514 & 09:17:05.0 & 63:55:14.7 &   0.13 &   12.7 &    7.2 &    $-$ &    2.5 &    0.8 &    $-$ &   1.25 &      S \\ 
\\ 
GMRT091706+0642227 & 09:17:06.4 & 64:22:27.7 &   0.57 &   24.9 &   15.4 &    $-$ &    $-$ &    $-$ &    2.4 &   1.06 &      S \\ 
GMRT091709+0632430 & 09:17:09.0 & 63:24:30.8 &   0.40 &  180.7 &  127.4 &   90.0 &   66.9 &    $-$ &   39.4 &   0.70 &      D \\ 
GMRT091711+0641810 & 09:17:11.6 & 64:18:10.8 &   0.50 &    4.2 &    3.3 &    $-$ &    1.1 &    $-$ &    $-$ &   0.97 &      S \\ 
GMRT091717+0632106 & 09:17:18.0 & 63:21:06.3 &   0.46 &   10.7 &   13.8 &    $-$ &    5.2 &    $-$ &    1.9 &   0.80 &      S \\ 
GMRT091719+0641734 & 09:17:19.9 & 64:17:34.7 &   0.50 &   11.2 &    8.6 &    $-$ &    6.0 &    $-$ &    3.7 &   0.48 &      S \\ 
\\ 
GMRT091726+0622734 & 09:17:27.0 & 62:27:34.4 &   1.35 &   13.7 &    6.9 &    $-$ &    $-$ &    $-$ &    2.4 &   0.76 &      S \\ 
GMRT091728+0633100 & 09:17:28.5 & 63:31:00.0 &   0.30 &    6.8 &    5.6 &    $-$ &    2.1 &    $-$ &    $-$ &   0.87 &      S \\ 
GMRT091729+0642227 & 09:17:29.6 & 64:22:27.5 &   0.58 &   62.9 &   43.2 &   29.0 &    $-$ &    $-$ &   10.9 &   0.80 &      D \\ 
GMRT091729+0645744 & 09:17:29.5 & 64:57:44.3 &   1.16 &   22.2 &   15.5 &    $-$ &    $-$ &    $-$ &    4.3 &   0.74 &      S \\ 
GMRT091732+0623932 & 09:17:32.4 & 62:39:32.6 &   1.15 &   26.2 &    9.5 &    $-$ &    $-$ &    $-$ &    4.5 &   0.75 &      S \\ 
\\ 
GMRT091733+0650126 & 09:17:33.4 & 65:01:26.5 &   1.23 &  656.7 &  354.4 &    $-$ &    $-$ &    $-$ &   48.1 &   1.19 &      S \\ 
GMRT091734+0640000 & 09:17:34.2 & 64:00:01.0 &   0.23 &   98.2 &   50.2 &   26.0 &   13.1 &    3.1 &    3.8 &   1.54 &      S \\ 
GMRT091735+0633739 & 09:17:35.8 & 63:37:39.6 &   0.21 &    6.1 &    5.8 &    $-$ &    2.2 &    0.6 &    $-$ &   1.07 &      S \\ 
GMRT091736+0640522 & 09:17:36.2 & 64:05:22.5 &   0.31 &   22.8 &   12.0 &    $-$ &    5.7 &    2.4 &    2.2 &   1.03 &      S \\ 
GMRT091741+0650258 & 09:17:41.0 & 65:02:58.5 &   1.25 &  472.9 &  343.0 &    $-$ &    $-$ &    $-$ &   82.9 &   0.79 &      D \\ 
\hline \hline
\end{tabular}

\end{table*}
\begin{table*}
\contcaption{}
\begin{tabular}{l r r r r r r r r r r l}
\hline
\multicolumn{1}{c}{Source name} & \multicolumn{1}{c}{RA} & \multicolumn{1}{c}{DEC} & \multicolumn{1}{c}{Dis} & \multicolumn{1}{c}{S$_{153}$} & \multicolumn{1}{c}{S$_{244}$}
                    & \multicolumn{1}{c}{S$_{330}$} & \multicolumn{1}{c}{S$_{610}$} & \multicolumn{1}{c}{S$_{1260}$} & \multicolumn{1}{c}{S$_{1400}$}
                    & \multicolumn{1}{c}{$\alpha$} &\multicolumn{1}{c}{Class} \\
                    & \multicolumn{1}{c}{hh:mm:ss.s} & \multicolumn{1}{c}{dd:mm:ss.s} & \multicolumn{1}{c}{deg.} & \multicolumn{1}{c}{mJy} & \multicolumn{1}{c}{mJy}
                    & \multicolumn{1}{c}{mJy} & \multicolumn{1}{c}{mJy} & \multicolumn{1}{c}{mJy} & \multicolumn{1}{c}{mJy} &  &        \\
\multicolumn{1}{c}{(1)} & \multicolumn{1}{c}{(2)} & \multicolumn{1}{c}{(3)} & \multicolumn{1}{c}{(4)} &  \multicolumn{1}{c}{(5)} & \multicolumn{1}{c}{(6)} & \multicolumn{1}{c}{(7)} & \multicolumn{1}{c}{(8)} & \multicolumn{1}{c}{(9)} & \multicolumn{1}{c}{(10)} & \multicolumn{1}{c}{(11)} & \multicolumn{1}{c}{(12)} \\
\hline
GMRT091743+0635046 & 09:17:43.3 & 63:50:46.8 &   0.13 &    3.6 &    5.5 &    $-$ &    2.3 &    0.7 &    $-$ &   0.78 &      S \\ 
GMRT091744+0630229 & 09:17:44.0 & 63:02:29.5 &   0.77 &   20.6 &   11.5 &    $-$ &    $-$ &    $-$ &    1.1 &   1.34 &      S \\ 
GMRT091745+0625955 & 09:17:45.1 & 62:59:55.3 &   0.81 &   15.5 &    9.1 &    $-$ &    $-$ &    $-$ &    1.8 &   0.97 &      S \\ 
GMRT091748+0622852 & 09:17:48.5 & 62:28:52.1 &   1.33 &   25.9 &   16.9 &   18.0 &    $-$ &    $-$ &    5.6 &   0.67 &      S \\ 
GMRT091749+0631617 & 09:17:49.5 & 63:16:17.5 &   0.55 &   27.0 &   19.6 &    $-$ &    $-$ &    $-$ &    3.8 &   0.89 &      S \\ 
\\ 
GMRT091750+0651521 & 09:17:50.7 & 65:15:21.0 &   1.46 &  274.6 &  132.2 &  113.0 &    $-$ &    $-$ &   37.5 &   0.91 &      D \\ 
GMRT091757+0625445 & 09:17:57.6 & 62:54:45.2 &   0.90 &  153.8 &   88.7 &   58.0 &    $-$ &    $-$ &   11.5 &   1.18 &      D \\ 
GMRT091800+0631340 & 09:18:00.4 & 63:13:40.0 &   0.60 &   27.5 &   20.4 &    $-$ &    $-$ &    $-$ &    2.1 &   1.18 &      D \\ 
GMRT091802+0645918 & 09:18:02.9 & 64:59:18.9 &   1.20 &   53.1 &   28.8 &    $-$ &    $-$ &    $-$ &    5.9 &   0.99 &      S \\ 
GMRT091803+0634156 & 09:18:03.1 & 63:41:56.6 &   0.19 &   11.8 &   11.1 &    $-$ &    5.4 &    2.0 &    3.0 &   0.75 &      S \\ 
\\ 
GMRT091807+0640831 & 09:18:07.0 & 64:08:31.6 &   0.38 &   57.3 &   39.0 &    $-$ &   11.7 &    $-$ &    4.1 &   1.20 &      T \\ 
GMRT091810+0633919 & 09:18:10.1 & 63:39:19.8 &   0.23 &   11.8 &   11.2 &    $-$ &    4.8 &    $-$ &    1.7 &   0.87 &      S \\ 
GMRT091812+0645112 & 09:18:12.4 & 64:51:12.9 &   1.07 & 2315.2 & 1317.2 &  851.0 &    $-$ &    $-$ &  197.6 &   1.17 &      D \\ 
GMRT091815+0633724 & 09:18:15.2 & 63:37:24.9 &   0.26 &   33.1 &   24.2 &    $-$ &   12.5 &    3.9 &    7.0 &   0.83 &      S \\ 
GMRT091827+0641031 & 09:18:27.2 & 64:10:31.6 &   0.43 &   18.8 &   22.2 &   16.0 &   36.3 &    $-$ &   21.9 &  -0.17 &      S \\ 
\\ 
GMRT091841+0635253 & 09:18:41.2 & 63:52:53.0 &   0.24 &   24.3 &   21.1 &    $-$ &   11.5 &    5.6 &    6.2 &   0.67 &      S \\ 
GMRT091842+0650229 & 09:18:42.1 & 65:02:29.4 &   1.26 &  215.8 &  126.3 &  117.0 &    $-$ &    $-$ &   40.6 &   0.75 &      S \\ 
GMRT091845+0642854 & 09:18:45.5 & 64:28:54.4 &   0.72 &   62.4 &   32.9 &   17.0 &    $-$ &    $-$ &    2.5 &   1.48 &      S \\ 
GMRT091846+0634654 & 09:18:46.5 & 63:46:54.2 &   0.24 &  260.5 &  170.8 &  112.0 &   69.2 &   21.8 &   30.8 &   1.05 &      S \\ 
GMRT091847+0625241 & 09:18:47.5 & 62:52:41.6 &   0.96 &   13.6 &   13.3 &    $-$ &    $-$ &    $-$ &    7.7 &   0.27 &      S \\ 
\\ 
GMRT091853+0644403 & 09:18:53.2 & 64:44:03.8 &   0.96 &   24.2 &   18.6 &    $-$ &    $-$ &    $-$ &    3.3 &   0.91 &      S \\ 
GMRT091903+0644834 & 09:19:03.0 & 64:48:34.2 &   1.04 &   59.2 &   35.1 &   21.0 &    $-$ &    $-$ &    6.2 &   1.04 &      S \\ 
GMRT091914+0623359 & 09:19:14.6 & 62:33:59.1 &   1.27 &   68.1 &   32.5 &    $-$ &    $-$ &    $-$ &   10.6 &   0.82 &      D \\ 
GMRT091914+0634340 & 09:19:14.2 & 63:43:40.2 &   0.30 &    6.5 &    3.9 &    $-$ &    1.9 &    $-$ &    $-$ &   0.87 &      S \\ 
GMRT091929+0641436 & 09:19:29.4 & 64:14:36.0 &   0.54 &   16.3 &   14.8 &    $-$ &    $-$ &    $-$ &    3.1 &   0.77 &      S \\ 
\\ 
GMRT091930+0642537 & 09:19:30.3 & 64:25:37.4 &   0.70 &  175.7 &  123.7 &   89.0 &    $-$ &    $-$ &   32.8 &   0.77 &      S \\ 
GMRT091938+0622856 & 09:19:38.9 & 62:28:56.3 &   1.36 &   30.8 &   19.4 &   14.0 &    $-$ &    $-$ &    8.2 &   0.60 &      S \\ 
GMRT091946+0640544 & 09:19:46.8 & 64:05:44.7 &   0.46 &   14.5 &    8.1 &    $-$ &    4.5 &    $-$ &    2.1 &   0.84 &      S \\ 
GMRT091947+0650953 & 09:19:47.1 & 65:09:53.5 &   1.40 &  250.1 &  143.5 &  118.0 &    $-$ &    $-$ &   42.1 &   0.81 &      S \\ 
GMRT091948+0625832 & 09:19:48.6 & 62:58:32.8 &   0.90 & 1643.9 & 1062.9 &  788.0 &    $-$ &    $-$ &  241.0 &   0.89 &      D \\ 
\\ 
GMRT091957+0643231 & 09:19:57.4 & 64:32:31.5 &   0.83 &  124.5 &   75.9 &   38.0 &    $-$ &    $-$ &   11.6 &   1.11 &      S \\ 
GMRT091957+0643800 & 09:19:57.9 & 64:38:00.2 &   0.91 &   25.8 &   17.6 &    $-$ &    $-$ &    $-$ &    9.2 &   0.45 &   Cplx \\ 
GMRT092002+0631414 & 09:20:02.3 & 63:14:14.2 &   0.68 &  202.1 &  131.5 &   83.0 &    $-$ &    $-$ &   24.6 &   0.97 &      D \\ 
GMRT092002+0634606 & 09:20:02.6 & 63:46:06.7 &   0.38 &   22.1 &   19.3 &    $-$ &    8.5 &    $-$ &    5.2 &   0.69 &      S \\ 
GMRT092004+0633626 & 09:20:04.4 & 63:36:26.5 &   0.43 &   22.5 &    8.6 &    $-$ &    2.9 &    $-$ &    $-$ &   1.49 &      S \\ 
\\ 
GMRT092009+0641550 & 09:20:09.6 & 64:15:50.1 &   0.60 &  256.5 &  145.9 &   94.0 &    $-$ &    $-$ &   14.4 &   1.30 &      S \\ 
GMRT092016+0632357 & 09:20:16.1 & 63:23:57.1 &   0.57 &   19.0 &   14.0 &    $-$ &    $-$ &    $-$ &    3.3 &   0.79 &      S \\ 
GMRT092020+0644259 & 09:20:20.5 & 64:42:59.5 &   1.00 &   31.5 &   21.4 &    $-$ &    $-$ &    $-$ &    3.9 &   0.95 &      S \\ 
GMRT092022+0630258 & 09:20:22.9 & 63:02:58.0 &   0.86 &   13.0 &   11.0 &    $-$ &    $-$ &    $-$ &    3.6 &   0.59 &      S \\ 
GMRT092022+0635143 & 09:20:22.5 & 63:51:43.6 &   0.42 &   93.9 &   66.7 &   36.0 &   31.1 &    $-$ &   15.7 &   0.82 &      S \\ 
\\ 
GMRT092025+0624607 & 09:20:25.7 & 62:46:07.3 &   1.12 &   36.9 &   21.3 &   16.0 &    $-$ &    $-$ &    5.4 &   0.87 &      S \\ 
GMRT092030+0635437 & 09:20:30.4 & 63:54:37.3 &   0.44 &    5.9 &    3.1 &    $-$ &    1.4 &    $-$ &    $-$ &   1.05 &      S \\ 
GMRT092033+0633921 & 09:20:34.0 & 63:39:21.8 &   0.46 &   10.5 &    8.2 &    $-$ &    3.4 &    $-$ &    2.6 &   0.69 &      S \\ 
GMRT092033+0650452 & 09:20:33.4 & 65:04:52.6 &   1.35 &   27.9 &   20.3 &    $-$ &    $-$ &    $-$ &    6.1 &   0.69 &      S \\ 
GMRT092034+0643829 & 09:20:35.0 & 64:38:29.9 &   0.94 &   61.4 &   41.1 &   19.0 &    $-$ &    $-$ &    9.8 &   0.87 &      S \\ 
\\ 
GMRT092036+0630453 & 09:20:36.0 & 63:04:53.7 &   0.85 &   15.7 &   10.4 &    $-$ &    $-$ &    $-$ &    4.5 &   0.55 &      S \\ 
GMRT092040+0623255 & 09:20:40.3 & 62:32:55.8 &   1.34 &   13.8 &   12.0 &    $-$ &    $-$ &    $-$ &    6.0 &   0.38 &      S \\ 
GMRT092045+0630309 & 09:20:45.9 & 63:03:09.3 &   0.88 &   23.2 &   18.9 &   14.0 &    $-$ &    $-$ &    6.0 &   0.62 &      S \\ 
GMRT092056+0640008 & 09:20:57.0 & 64:00:08.2 &   0.52 &   13.8 &   11.2 &    $-$ &    4.0 &    $-$ &    $-$ &   0.91 &      S \\ 
GMRT092057+0632047 & 09:20:57.1 & 63:20:47.3 &   0.67 &    8.1 &   12.9 &    $-$ &    $-$ &    $-$ &    6.7 &   0.16 &      S \\ 
\hline \hline
\end{tabular}

\end{table*}
\begin{table*}
\contcaption{}
\begin{tabular}{l r r r r r r r r r r l}
\hline
\multicolumn{1}{c}{Source name} & \multicolumn{1}{c}{RA} & \multicolumn{1}{c}{DEC} & \multicolumn{1}{c}{Dis} & \multicolumn{1}{c}{S$_{153}$} & \multicolumn{1}{c}{S$_{244}$}
                    & \multicolumn{1}{c}{S$_{330}$} & \multicolumn{1}{c}{S$_{610}$} & \multicolumn{1}{c}{S$_{1260}$} & \multicolumn{1}{c}{S$_{1400}$}
                    & \multicolumn{1}{c}{$\alpha$} &\multicolumn{1}{c}{Class} \\
                    & \multicolumn{1}{c}{hh:mm:ss.s} & \multicolumn{1}{c}{dd:mm:ss.s} & \multicolumn{1}{c}{deg.} & \multicolumn{1}{c}{mJy} & \multicolumn{1}{c}{mJy}
                    & \multicolumn{1}{c}{mJy} & \multicolumn{1}{c}{mJy} & \multicolumn{1}{c}{mJy} & \multicolumn{1}{c}{mJy} &  &        \\
\multicolumn{1}{c}{(1)} & \multicolumn{1}{c}{(2)} & \multicolumn{1}{c}{(3)} & \multicolumn{1}{c}{(4)} &  \multicolumn{1}{c}{(5)} & \multicolumn{1}{c}{(6)} & \multicolumn{1}{c}{(7)} & \multicolumn{1}{c}{(8)} & \multicolumn{1}{c}{(9)} & \multicolumn{1}{c}{(10)} & \multicolumn{1}{c}{(11)} & \multicolumn{1}{c}{(12)} \\
\hline
GMRT092106+0623042 & 09:21:06.8 & 62:30:42.9 &   1.39 &   55.4 &   28.5 &   25.0 &    $-$ &    $-$ &   11.1 &   0.72 &      S \\ 
GMRT092111+0641445 & 09:21:11.3 & 64:14:45.6 &   0.67 &   69.8 &   41.5 &    $-$ &    $-$ &    $-$ &   12.4 &   0.77 &      S \\ 
GMRT092118+0645237 & 09:21:18.3 & 64:52:37.2 &   1.19 &  958.3 &  599.3 &  452.0 &    $-$ &    $-$ &  134.9 &   0.90 &      S \\ 
GMRT092123+0643404 & 09:21:23.6 & 64:34:04.4 &   0.93 &  100.5 &   79.4 &   31.0 &    $-$ &    $-$ &   52.1 &   0.32 &   Cplx \\ 
GMRT092131+0633939 & 09:21:31.9 & 63:39:39.6 &   0.57 &  114.2 &   60.5 &   40.0 &    $-$ &    $-$ &    4.0 &   1.50 &      D \\ 
\\ 
GMRT092138+0623054 & 09:21:38.4 & 62:30:54.9 &   1.41 &  151.3 &   76.1 &   77.0 &    $-$ &    $-$ &   32.2 &   0.69 &      S \\ 
GMRT092143+0641520 & 09:21:43.6 & 64:15:20.8 &   0.72 &  320.9 &  184.0 &  141.0 &    $-$ &    $-$ &   60.6 &   0.77 &   Cplx \\ 
GMRT092145+0623528 & 09:21:45.1 & 62:35:28.8 &   1.34 &  163.6 &   83.3 &   83.0 &    $-$ &    $-$ &   22.2 &   0.88 &      S \\ 
GMRT092149+0640311 & 09:21:49.5 & 64:03:11.3 &   0.63 &   36.5 &   25.1 &    $-$ &    $-$ &    $-$ &    6.3 &   0.79 &      D \\ 
GMRT092150+0634702 & 09:21:50.9 & 63:47:02.0 &   0.58 &   56.9 &   47.2 &   27.0 &    $-$ &    $-$ &   11.0 &   0.77 &      S \\ 
\\ 
GMRT092209+0651030 & 09:22:09.7 & 65:10:30.5 &   1.50 &  474.2 &  193.0 &  168.0 &    $-$ &    $-$ &   32.4 &   1.24 &      S \\ 
GMRT092219+0633655 & 09:22:19.5 & 63:36:55.4 &   0.66 & 3351.8 & 2174.2 & 1486.0 &    $-$ &    $-$ &  310.7 &   1.05 &      S \\ 
GMRT092219+0650556 & 09:22:20.0 & 65:05:56.4 &   1.44 &  232.8 &   98.7 &   93.0 &    $-$ &    $-$ &   17.9 &   1.15 &      D \\ 
GMRT092228+0623756 & 09:22:28.3 & 62:37:56.4 &   1.35 &   12.1 &    7.7 &    $-$ &    $-$ &    $-$ &    2.1 &   0.78 &      S \\ 
GMRT092234+0625237 & 09:22:34.1 & 62:52:37.0 &   1.14 &   58.4 &   35.6 &   17.0 &    $-$ &    $-$ &   11.9 &   0.75 &      S \\ 
\\ 
GMRT092234+0640556 & 09:22:34.7 & 64:05:56.9 &   0.72 &  201.0 &  119.8 &   80.0 &    $-$ &    $-$ &   37.7 &   0.78 &      T \\ 
GMRT092235+0623518 & 09:22:35.3 & 62:35:18.9 &   1.39 &   15.2 &    7.4 &    $-$ &    $-$ &    $-$ &    2.0 &   0.89 &      S \\ 
GMRT092251+0630618 & 09:22:51.7 & 63:06:18.9 &   0.99 &   59.5 &   39.1 &   22.0 &    $-$ &    $-$ &    8.1 &   0.93 &      S \\ 
GMRT092255+0645259 & 09:22:55.1 & 64:52:59.7 &   1.28 &  112.6 &   54.5 &   39.0 &    $-$ &    $-$ &   12.0 &   1.03 &      S \\ 
GMRT092302+0632925 & 09:23:02.9 & 63:29:25.0 &   0.78 &   38.5 &   29.1 &   39.0 &    $-$ &    $-$ &   23.4 &   0.20 &      S \\ 
\\ 
GMRT092307+0632853 & 09:23:07.4 & 63:28:53.0 &   0.79 &   16.6 &   17.7 &    $-$ &    $-$ &    $-$ &   23.4 &  -0.16 &      S \\ 
GMRT092309+0623246 & 09:23:09.2 & 62:32:46.9 &   1.46 &    9.1 &    6.6 &    $-$ &    $-$ &    $-$ &    3.6 &   0.41 &      S \\ 
GMRT092311+0632429 & 09:23:11.7 & 63:24:29.2 &   0.83 &   31.9 &   18.8 &    $-$ &    $-$ &    $-$ &    5.4 &   0.79 &      S \\ 
GMRT092312+0633140 & 09:23:12.1 & 63:31:40.5 &   0.78 &  158.7 &   94.3 &   72.0 &    $-$ &    $-$ &   17.5 &   1.00 &      D \\ 
GMRT092318+0643531 & 09:23:18.0 & 64:35:31.3 &   1.08 &   30.2 &   11.7 &    $-$ &    $-$ &    $-$ &    5.6 &   0.72 &      D \\ 
\\ 
GMRT092323+0630847 & 09:23:23.9 & 63:08:47.0 &   1.00 &   91.2 &   56.4 &   39.0 &    $-$ &    $-$ &   14.3 &   0.85 &      S \\ 
GMRT092324+0630643 & 09:23:24.9 & 63:06:43.5 &   1.03 &  133.8 &   74.1 &   51.0 &    $-$ &    $-$ &   14.2 &   1.03 &      D \\ 
GMRT092332+0632817 & 09:23:32.8 & 63:28:17.2 &   0.84 &  760.3 &  461.8 &  319.0 &    $-$ &    $-$ &   81.0 &   1.03 &      T \\ 
GMRT092332+0633611 & 09:23:32.4 & 63:36:11.9 &   0.80 &   89.0 &   48.6 &   38.0 &    $-$ &    $-$ &    4.6 &   1.31 &   Cplx \\ 
GMRT092344+0641122 & 09:23:44.1 & 64:11:22.3 &   0.87 &    8.5 &    5.0 &    $-$ &    $-$ &    $-$ &    1.2 &   0.86 &      S \\ 
\\ 
GMRT092355+0625336 & 09:23:55.2 & 62:53:36.4 &   1.23 & 1130.4 &  785.9 &  763.0 &    $-$ &    $-$ &  299.6 &   0.58 &      S \\ 
GMRT092401+0645058 & 09:24:01.3 & 64:50:58.5 &   1.32 &  275.7 &  141.8 &  119.0 &    $-$ &    $-$ &   37.3 &   0.91 &      D \\ 
GMRT092401+0645641 & 09:24:01.7 & 64:56:41.4 &   1.40 &   16.2 &   11.4 &    $-$ &    $-$ &    $-$ &    3.2 &   0.73 &      S \\ 
GMRT092403+0623912 & 09:24:03.4 & 62:39:12.6 &   1.42 &   59.1 &   33.9 &   19.0 &    $-$ &    $-$ &   10.2 &   0.82 &      D \\ 
GMRT092408+0641424 & 09:24:08.9 & 64:14:24.8 &   0.94 &   15.8 &   21.8 &   16.0 &    $-$ &    $-$ &   15.2 &   0.06 &      S \\ 
\\ 
GMRT092412+0632324 & 09:24:12.4 & 63:23:24.2 &   0.94 &   25.8 &    9.7 &    $-$ &    $-$ &    $-$ &    2.8 &   0.98 &      S \\ 
GMRT092412+0643911 & 09:24:12.8 & 64:39:11.4 &   1.19 &    9.2 &    8.4 &    $-$ &    $-$ &    $-$ &    2.9 &   0.54 &      S \\ 
GMRT092412+0645922 & 09:24:12.3 & 64:59:22.7 &   1.44 &   12.8 &    6.3 &    $-$ &    $-$ &    $-$ &    3.1 &   0.60 &      S \\ 
GMRT092428+0632912 & 09:24:28.6 & 63:29:12.7 &   0.93 &   11.1 &    7.7 &    $-$ &    $-$ &    $-$ &    4.3 &   0.41 &      S \\ 
GMRT092428+0635917 & 09:24:28.6 & 63:59:18.0 &   0.89 &   48.0 &   16.6 &    $-$ &    $-$ &    $-$ &    3.2 &   1.21 &      D \\ 
\\ 
GMRT092434+0635751 & 09:24:34.8 & 63:57:51.0 &   0.89 &   16.1 &   11.4 &    $-$ &    $-$ &    $-$ &    1.5 &   1.08 &      S \\ 
GMRT092441+0634926 & 09:24:41.8 & 63:49:26.6 &   0.89 &   21.7 &   12.8 &    $-$ &    $-$ &    $-$ &    4.6 &   0.69 &      S \\ 
GMRT092459+0643337 & 09:24:59.4 & 64:33:37.5 &   1.19 &   46.1 &   22.3 &   18.0 &    $-$ &    $-$ &    8.5 &   0.76 &      S \\ 
GMRT092501+0635941 & 09:25:01.8 & 63:59:41.6 &   0.95 &   21.8 &   13.7 &    $-$ &    $-$ &    $-$ &    3.9 &   0.77 &      S \\ 
GMRT092526+0632708 & 09:25:26.4 & 63:27:08.6 &   1.04 &  122.9 &   74.9 &   47.0 &    $-$ &    $-$ &   21.3 &   0.82 &      S \\ 
\\ 
GMRT092530+0640832 & 09:25:30.2 & 64:08:33.0 &   1.04 &    9.7 &   14.5 &    $-$ &    $-$ &    $-$ &    4.6 &   0.41 &      S \\ 
GMRT092540+0644841 & 09:25:40.3 & 64:48:41.0 &   1.41 &   41.4 &   16.0 &    $-$ &    $-$ &    $-$ &    5.0 &   0.93 &      S \\ 
GMRT092549+0631414 & 09:25:49.5 & 63:14:14.4 &   1.17 &  106.5 &   63.8 &   56.0 &    $-$ &    $-$ &   15.2 &   0.87 &      D \\ 
GMRT092551+0645158 & 09:25:51.6 & 64:51:58.7 &   1.46 &   53.8 &   26.5 &   33.0 &    $-$ &    $-$ &    5.5 &   0.98 &      S \\ 
GMRT092608+0632559 & 09:26:08.7 & 63:25:59.6 &   1.12 &  122.2 &   75.1 &   37.0 &    $-$ &    $-$ &   15.9 &   0.97 &      S \\ 
\hline \hline
\end{tabular}

\end{table*}
\begin{table*}
\contcaption{}
\begin{tabular}{l r r r r r r r r r r l}
\hline
\multicolumn{1}{c}{Source name} & \multicolumn{1}{c}{RA} & \multicolumn{1}{c}{DEC} & \multicolumn{1}{c}{Dis} & \multicolumn{1}{c}{S$_{153}$} & \multicolumn{1}{c}{S$_{244}$}
                    & \multicolumn{1}{c}{S$_{330}$} & \multicolumn{1}{c}{S$_{610}$} & \multicolumn{1}{c}{S$_{1260}$} & \multicolumn{1}{c}{S$_{1400}$}
                    & \multicolumn{1}{c}{$\alpha$} &\multicolumn{1}{c}{Class} \\
                    & \multicolumn{1}{c}{hh:mm:ss.s} & \multicolumn{1}{c}{dd:mm:ss.s} & \multicolumn{1}{c}{deg.} & \multicolumn{1}{c}{mJy} & \multicolumn{1}{c}{mJy}
                    & \multicolumn{1}{c}{mJy} & \multicolumn{1}{c}{mJy} & \multicolumn{1}{c}{mJy} & \multicolumn{1}{c}{mJy} &  &        \\
\multicolumn{1}{c}{(1)} & \multicolumn{1}{c}{(2)} & \multicolumn{1}{c}{(3)} & \multicolumn{1}{c}{(4)} &  \multicolumn{1}{c}{(5)} & \multicolumn{1}{c}{(6)} & \multicolumn{1}{c}{(7)} & \multicolumn{1}{c}{(8)} & \multicolumn{1}{c}{(9)} & \multicolumn{1}{c}{(10)} & \multicolumn{1}{c}{(11)} & \multicolumn{1}{c}{(12)} \\
\hline
GMRT092635+0633938 & 09:26:35.4 & 63:39:38.8 &   1.12 &   34.1 &   19.2 &    $-$ &    $-$ &    $-$ &    4.0 &   0.96 &      S \\ 
GMRT092639+0632523 & 09:26:39.2 & 63:25:23.0 &   1.18 &   28.3 &   15.8 &    $-$ &    $-$ &    $-$ &    4.3 &   0.84 &      S \\ 
GMRT092643+0635212 & 09:26:43.3 & 63:52:12.7 &   1.12 &   64.0 &   40.1 &   25.0 &    $-$ &    $-$ &   12.3 &   0.76 &      S \\ 
GMRT092647+0631454 & 09:26:47.7 & 63:14:54.5 &   1.27 &   46.2 &   22.5 &    $-$ &    $-$ &    $-$ &    6.8 &   0.85 &      S \\ 
GMRT092652+0641938 & 09:26:52.4 & 64:19:38.0 &   1.24 & 4665.0 & 3030.9 & 2606.0 &    $-$ &    $-$ & 1093.4 &   0.74 &      S \\ 
\\ 
GMRT092655+0633256 & 09:26:55.3 & 63:32:56.6 &   1.17 &  211.7 &  122.7 &   97.0 &    $-$ &    $-$ &   33.6 &   0.84 &      T \\ 
GMRT092701+0643748 & 09:27:01.0 & 64:37:48.8 &   1.40 &  402.6 &  161.6 &  148.0 &    $-$ &    $-$ &   34.1 &   1.14 &      D \\ 
GMRT092718+0635018 & 09:27:18.3 & 63:50:18.8 &   1.18 &   32.2 &   15.5 &    $-$ &    $-$ &    $-$ &    2.7 &   1.11 &      D \\ 
GMRT092756+0634755 & 09:27:56.8 & 63:47:55.9 &   1.25 &   87.0 &   66.3 &   46.0 &    $-$ &    $-$ &   30.5 &   0.49 &      S \\ 
GMRT092809+0635356 & 09:28:09.7 & 63:53:56.5 &   1.28 &  371.8 &  179.3 &  127.0 &    $-$ &    $-$ &   38.5 &   1.07 &      D \\ 
\\ 
GMRT092819+0632001 & 09:28:19.7 & 63:20:01.3 &   1.39 &   40.6 &   19.2 &   20.0 &    $-$ &    $-$ &    8.8 &   0.66 &      D \\ 
GMRT092830+0634748 & 09:28:30.7 & 63:47:48.8 &   1.32 &   96.8 &   33.1 &   23.0 &    $-$ &    $-$ &    $-$ &   1.95 &      E \\ 
GMRT092830+0641530 & 09:28:30.7 & 64:15:30.2 &   1.38 &   37.4 &   21.9 &    $-$ &    $-$ &    $-$ &    9.7 &   0.59 &      D \\ 
GMRT092857+0635912 & 09:28:57.0 & 63:59:12.9 &   1.37 &  190.3 &   80.5 &   60.0 &    $-$ &    $-$ &   21.1 &   1.02 &      D \\ 
GMRT092858+0634653 & 09:28:58.4 & 63:46:53.5 &   1.37 &   59.6 &   27.5 &   28.0 &    $-$ &    $-$ &    8.5 &   0.86 &      S \\ 
\\ 
GMRT092915+0633058 & 09:29:15.2 & 63:30:59.0 &   1.43 &   95.8 &   42.3 &   28.0 &    $-$ &    $-$ &   11.6 &   0.98 &      S \\ 
GMRT092930+0635029 & 09:29:30.9 & 63:50:29.3 &   1.43 &   95.7 &   43.6 &   34.0 &    $-$ &    $-$ &   11.2 &   0.98 &      S \\ 
\hline \hline
\end{tabular}

\end{table*}

\label{lastpage}

\end{document}